\documentclass{article}
\usepackage{jheppub}
\usepackage[utf8]{inputenc}

\usepackage{amsmath,amsthm}
\usepackage{mathrsfs}
\usepackage{amscd,amssymb, amsfonts, verbatim,subfigure, enumerate}
\usepackage[mathcal]{eucal}
\usepackage[super]{nth}

\newtheorem{theorem}{Theorem}[subsection]
\newtheorem{thm-def}{Theorem/Definition}[theorem]
\newtheorem{proposition}[theorem]{Proposition}
\newtheorem{lemma}[theorem]{Lemma}

\newcommand\NO[1] { \mathopen{:}#1\mathclose{:}}

\newcommand{\half}{\tfrac{1}{2}}

\title{Bootstrapping two-loop QCD amplitudes }
\author{Kevin Costello}
\affiliation{%
    Perimeter Institute for Theoretical Physics 
}
\emailAdd{kcostello@perimeterinstitute.ca}
\renewcommand{\t}{{\sf{t}}}

\newcommand{\sh}{{\sf{h}}} 
\renewcommand{\i}{i}

\newcommand{\eps}{\epsilon}
\newcommand{\g}{\mathfrak{g}}

\newcommand{\what}{\widehat}

\newcommand{\til}{\widetilde}
\newcommand{\mscr}{\mathscr}

\newcommand{\br}{\overline}

\newcommand{\C}{\mathbb C}
\newcommand{\CP}{\mathbb{CP}}
\newcommand{\PT}{\mathbb{PT}}

\newcommand{\Oo}{\mscr O}
\newcommand{\Z}{\mathbb Z}
\newcommand{\defeq}{\overset{\text{def}}{=}}

\newcommand{\op}{\operatorname}

\newcommand{\mbb}{\mathbb}
\newcommand{\mf}{\mathfrak}
\newcommand{\mc}{\mathcal}

\newcommand{\ip}[1]{\left\langle #1 \right\rangle}
\newcommand{\abs}[1]{\left| #1 \right|}

\newcommand{\R}{\mbb R}
\renewcommand{\d}{\mathrm{d}}

\DeclareMathOperator{\Sym}{Sym}

\begin{document}

\abstract{
	Form factors of self-dual gauge theory are equal to correlators of an (extended) celestial chiral algebra. This suggests that these form factors can be computed using the ``bootstrap'' method familiar from 2d CFTs.   The method can also be applied to certain QCD amplitudes, which are built from form-factors of self-dual gauge theory. 
	
	In this paper this bootstrap method is applied to compute two-loop all $+$ QCD amplitudes, for $SU(N)$ gauge theory with certain special matter content. A closed formula is presented for all single-trace amplitudes.

}
\maketitle

\section{Introduction}
The study of scattering amplitudes plays a central role in theoretical physics.  Understanding of scattering amplitudes of $N=4$ gauge theory has seen remarkable progress in the last $15$ years \cite{Arkani-Hamed:2022rwr}, with exact all-order results available for the planar theory.  

Scattering amplitudes of non-supersymmetric theories appear to be considerably more challenging. Two-loop all $+$ amplitudes, which are the focus of this paper, have undergone extensive study in the past twenty years \cite{Bern:2000dn, Badger:2013gxa, Badger:2015lda, Dunbar:2016aux, Dunbar:2016gjb, Dunbar:2017nfy, Dalgleish:2020mof, Dunbar:2016cxp, Kosower:2022bfv, Kosower:2022iju}, with exact results available up to $7$ points and partial amplitudes computed for all points.      

Progress in $N=4$ gauge theory has been made in part  because  of powerful ``bootstrap'' methods \cite{Alday:2010ku, Basso:2013vsa, Caron-Huot:2020bkp} which bypass cumbersome Feynman diagram calculations.

In \cite{Costello:2022wso}, the author and Natalie Paquette suggested a (rather different) bootstrap method for computing form factors of self-dual gauge theories, which are the same as certain QCD amplitudes and form factors.  Our bootstrap method uses $2d$ celestial chiral algebras, related to those studied by Strominger and collaborators \cite{Guevara:2021abz, Strominger:2017zoo,Ball:2021tmb}.  Our technique is purely perturbative.   

Here I apply this method to compute the complete single-trace two-loop all $+$ amplitudes of QCD, with certain special matter content. In these cases,  a simple closed formula for the amplitude is obtained. 

\subsection{Background on self-dual gauge theory}
Let me review the background. Self-dual Yang-Mills theory is the four-dimensional non-unitary gauge theory with Lagrangian
\begin{equation} 
	\int \op{tr} (B \wedge F(A)_-) 
\end{equation}
where $A$ is an ordinary gauge field, and $B \in \Omega^2_-(\R^4,\g)$.  Ordinary Yang-Mills theory is a deformation of the self-dual theory obtained by adding on the term $\half g_{YM}^2 \int \op{tr}(B \wedge B)$.  Adding on this term and integrating out $B$ yields
\begin{equation} 
	-\frac{1}{2 g_{YM}^2} \int \op{tr}( F(A)_- \wedge F(A)_- ) 
\end{equation}
which is the Yang-Mills Lagrangian, up to a topological term $\half \g_{YM}^{-2} \int \op{tr}( F \wedge F)$ which does not contribute in perturbation theory.

The celestial OPE of positive-helicity states in self-dual gauge theory were studied in \cite{Ball:2021tmb} and shown to form a chiral algebra. 

This was generalized in  \cite{Costello:2022wso, Costello:2022upu}, where an extended celestial chiral algebra for self-dual gauge theory including states of both helicity was studied.    We used a twistor perspective, and techniques from twisted holography, to understand the algebra.

We found that for pure gauge theory, the (extended) celestial chiral algebra does not exist at loop level, because of a gauge anomaly on twistor space.   The gauge anomaly can be cancelled by introducing a certain ``axion'' field on space-time: a fourth-order scalar coupled to the topological term in Yang-Mills theory.   This method works for gauge groups $SU(2)$, $SU(3)$, $SO(8)$ or exceptional.   

The key relationship between the chiral algebra and the self-dual gauge theory is that 
\begin{enumerate}
	\item Local operators in the chiral algebra are the same as  states of self-dual gauge theory, in a boost eigenbasis.
	\item OPEs in the chiral algebra are collinear limits of states in self-dual gauge theory.
	\item Conformal blocks of the chiral algebra are local operators in the self-dual gauge theory.
	\item Correlation functions of the chiral algebra in a given conformal block, are scattering amplitudes of the gauge theory in the presence of the corresponding local operator, which is inserted at a point in space-time.    Scattering amplitudes in the presence of a local operator are known as form factors. 
\end{enumerate}
This suggests a natural program:  use the chiral algebra structure to ``bootstrap'' the form factors of self-dual gauge theory.   

This is particularly interesting for the form factors of the operator $\op{tr}(B^2)$. This is because, when we integrate over the position of the operator $\op{tr}(B^2)$, these form factors become ordinary amplitudes of Yang-Mills theory -- albeit with the unusual axion introduced in \cite{Costello:2021bah, Costello:2022wso}.  

The form factor where we insert one copy of the operator $\op{tr}(B^2)$ gives Yang-Mills amplitudes at tree level with $2$ negative helicity states, one loop with $1$ negative helicity state or two loops with only positive helicity states.   

We calculated the tree-level and one loop form factors in \cite{Costello:2022wso, Costello:2022upu}  The goal of this paper is to compute the two-loop form factor. 

Applying these methods beyond two loops (or with different helicity configurations) becomes more difficult, because we would need to deal with two or more insertions of the operator $\op{tr}(B^2)$.  One can think of the form factors with one insertion of $\op{tr}(B^2)$ as being the loop analog of the MHV vertex, and the amplitude obtained using multiple insertions of $\op{tr}(B^2)$ as a loop-level analog of the CSW construction \cite{Cachazo:2004kj}. 

\subsection{Cancelling the anomaly with matter}

One difficulty that has to be overcome is that the methods of \cite{Costello:2022wso, Costello:2022upu} work when we introduce an axion in order to cancel an anomaly on twistor space.  The axion has an unusual fourth-order kinetic term, and QCD amplitudes in the presence of the axion differ from ordinary QCD amplitudes. 

In order to be able to say something about ordinary QCD, in this paper I mostly focus on a different way to cancel the anomaly: by including appropriate matter.  The twistor-space anomaly cancels (and so the chiral algebra exists) when we have a matter representation $R$ where the trace identity
\begin{equation} 
	\op{tr}_{\g}(X^4) - \op{tr}_{R}(X^4) = 0	\label{eqn:trace_identity_strong} 
\end{equation}
holds,  where $X$ is an element of the Lie algebra $\g$.  

This holds, for instance, $SU(2)$ with $N_f = 8$, or $SU(3)$ with $N_f = 9$. These two theories are part of an infinite sequence where \eqref{eqn:trace_identity_strong} holds, which I learned of from Roland Bittleston. The general element of the series has $SU(N)$ gauge group, with matter in 
\begin{equation} 
	8 F \oplus 8 F^\vee \oplus \wedge^2 F \oplus \wedge^2 F^\vee \label{eqn:mattercontent} 
\end{equation}
where $F, F^\vee$ are the fundamental and anti-fundamental.  Here $\wedge^2 F$ is the exterior square of $F$, that is the space of anti-symmetric elements in the tensor product $F \otimes F$.  The representation $\wedge^2 F$ is of dimension $\choose{N}{2}$.   

There is also an interesting set of examples where the anomaly can be cancelled by a mix of matter and axion. This can happen if the identity 
\begin{equation} 
	\op{tr}_{\g}(X^4) - \op{tr}_{R}(X^4) \propto \op{tr}(X^2)^2 
	\label{eqn:trace_identity_mixed} 
\end{equation}
holds. In this paper the main example where the anomaly can be cancelled in this way is for $SU(N)$ gauge theory  with matter
\begin{equation} 
	N F \oplus N F^\vee 
\end{equation}
i.e.\ $N_F = N_c$. 
\subsection{Main results}
The main results of this paper are then the following:
\begin{enumerate} 
	\item I describe the chiral algebra, including one-loop corrections, in the presence of matter. This is a small generalization of the results of \cite{Costello:2022upu}.
	\item The chiral algebra is used to compute the one-loop one minus form factor of the operator $\op{tr}(B^2)$. This is a review of the computation in \cite{Costello:2022upu}, but I also correct a minor error in the published version of \cite{Costello:2022upu}.  Because of an algebra error in the derivation of the formula, we initially stated that the formula for the one-loop one minus form factor holds when the twistor space anomaly is cancelled by the axion. This turns out to not be correct, we need the twistor space anomaly to be cancelled by matter only.  
	\item This one-loop form factor is compared to the known \cite{Bern:1991aq} one-loop one minus QCD amplitude, at $4$ points. Perfect agreement is found as long as the identity \eqref{eqn:trace_identity_strong} holds.  
	\item The chiral algebra is used to compute the two-loop all $+$  form factors of a renormalized version of $\op{tr}(B^2)$, which gives us the two-loop all $+$ single-trace QCD amplitude for $SU(N)$ gauge theory with matter as in equation \eqref{eqn:mattercontent}. 

		The terms which are leading and subleading in $N$ are both proportional to the parity even part of the standard one-loop all $+$ QCD amplitude \cite{Mahlon:1993si,Bern:1993qk}.  
	\item Finally I also calculate the two-loop single-trace all $+$ amplitude for $SU(N)$ gauge theory with $N_F = N$, where the anomaly is cancelled by an axion.  The term which is leading order in $N$ is again proportional to the parity-even part of the standard one-loop all $+$ QCD amplitude.  

		This leading  term can be shown to not depend on the presence of the axion, showing that the planar all $+$ 2-loop QCD amplitude with $N_F = N_c$ is proportional to the parity-even part of the $1$-loop all $+$ amplitude of pure gauge theory. 
\end{enumerate}
The ``renormalized'' version of $\op{tr}(B^2)$ whose form factor we compute is obtained by adding a two-loop counter-term proportional to $\op{tr}(F^2)$.  Since this counter-term is a total derivative, adding this does not affect the amplitude, but it does simplify the form factor computation.  

\subsection{Comparison to previous work}
The QCD amplitudes which arise from self-dual gauge theory by the insertion of one copy of $\op{tr}(B^2)$ are those at tree level with two negative helicity states, one loop with one negative helicity, and two loops with only positive helicity states.

These amplitudes have been very heavily studied over the past several decades, starting with the work of Parke-Taylor \cite{Parke:1986gb} on the tree-level amplitude with two negative helicity states. 

One loop all $+$ amplitudes also play a role in our story. These were first computed independently by Mahlon \cite{Mahlon:1993si} and Bern et al. \cite{Bern:1993qk}.  One loop amplitudes with one negative helicity state were first computed at four points in \cite{Bern:1991aq} and $5$ points in \cite{Bern:1993mq}.  A general formula was found in \cite{Mahlon:1993si}, which was later simplified in \cite{Bern:2005ji}. 

Two loop all $+$ QCD amplitudes have been extensively studied, first at $4$ points in \cite{Bern:2000dn} and at $5$ points  in \cite{Badger:2013gxa, Badger:2015lda, Dunbar:2016aux}.  Six and seven point amplitudes were computed in  \cite{Dunbar:2016gjb, Dunbar:2017nfy, Dalgleish:2020mof}.  The polylogarithmic part at leading colour for arbitrary number of gluons  was determined in \cite{Dunbar:2016cxp}.  New methods and further progress   in computing the rational part of the amplitudes were developed in \cite{Kosower:2022bfv, Kosower:2022iju}. 

The results of this paper are in a sense complementary to most known results.    One important difference in our paper is that the two-loop amplitudes we study are entirely \emph{rational}. While at first this may be surprising, this is explained by our careful choice of matter.  Our matter content is engineered so that the one-loop all $+$ amplitude vanishes.

This special choice of matter ensures that the two-loop amplitudes we study are significantly simpler than the two-loop amplitudes with generic matter, which is why exact results are attainable.  

Since this paper was posted, the two-loop amplitude computed here has been computed in \cite{Dixon:2024mzh} using standard techniques at four points.  In dimensional regularization, the computations agree after subtracting a universal two-loop IR divergence. When using a mass-regulator, \cite{Dixon:2024mzh} found that the amplitude is finite and matches the result of this paper. The work of \cite{Dixon:2024mzh} pushed the method of this paper further by computing the double-trace two-loop all $+$ amplitude.   

In a different direction, our general approach to computing amplitudes has several parallels in the literature. As I have explained, amplitudes in our approach arise as form factors of the operator $\op{tr}(B^2)$ in self-dual gauge theory. Very similar form factors were considered at one loop in \cite{Dixon:2004za}.  From a twistor perspective, which is quite close in spirit to our approach, such form factors at tree level were considered in \cite{Boels:2007qn}, which is related to the CSW approach to amplitudes  \cite{Cachazo:2004kj}. 

\section{Review of the chiral algebra for gauge theory without matter}

Amplitudes and form factors will be described using the spinor-helicity formalism, where massless states are parameterized by a pair $(\lambda_{\alpha}, \til{\lambda}_{\dot{\alpha}})$ of spinors.  The momentum is written in terms of the spinors by
\begin{equation} 
	p_{\alpha \dot{\beta}} = \lambda_{\alpha} \til{\lambda}_{\dot{\alpha}}. 
\end{equation}
By scaling $\lambda,\til{\lambda}$ while leaving $p$ fixed, we can assume the two-component spinor $\lambda$ is 
\begin{equation} 
	\lambda = (1,z). 
\end{equation}
Then $z$ is a coordinate on the $\CP^1$ where our chiral algebra will live. 

In this way, a massless state will be described as a function of $z$ and $\til{\lambda}$, instead of $\lambda$ and $\til{\lambda}$. We use the notation
\begin{equation} 
	J_a[\omega \til{\lambda}](z) \ \ \til{J}_a[\omega \til{\lambda}] (z) 	
\end{equation}
for gauge theory states of positive and negative helicity, where the colour index is $a$ and $\omega$ is the energy. Expanding as a series in $\omega$ we write
\begin{equation} 
	\begin{split} 
		J_a[\omega \til{\lambda}](z) &= \sum \omega^k J_a[k]  (z) \\
		\til{J}_a[\omega \til{\lambda}] (z) &= \sum \omega^k \til{J}[k](z)	 
	\end{split}
\end{equation}
where $J[k]$, $\til{J}[k]$ are homogeneous polynomials of order $k$ in $\til{\lambda}$. We can further expand
\begin{equation} 
	\begin{split} 
		J[k](z) &= \sum_{r + s = k} \frac{1}{r! s!} \til{\lambda}_1^r \til{\lambda}_2^s  J[r,s] (z) \\
		\til{J}[k](z) &= \sum_{r + s = k} \frac{1}{r! s!} \til{\lambda}_1^r \til{\lambda}_2^s  \til{J}[r,s] (z)  
	\end{split}
\end{equation}
For pure self-dual gauge theory (at tree level), the state $J[r,s](z)$ and $\til{J}[r,s](z)$ are generators of the chiral algebra. 

We let\footnote{The convention for the normalization of $[ij]$ has been changed in version 2.  This is to match with the the one-loop splitting function, which was off by a factor of 2 in \cite{Costello:2022upu}.} 
\begin{equation} 
	\begin{split} 
		\ip{ij} &= 2 \pi \i (z_i - z_j) \\
		[ij] &=-  \eps_{\dot{\alpha}\dot{\beta}} \til{\lambda}_i^{\dot \alpha} \til{\lambda}_j^{\dot \beta}.  
	\end{split}
\end{equation}

These states form a chiral algebra on the $z$ plane whose OPEs are collinear limits. At tree level, the OPEs for self dual gauge theory are
\begin{equation} 
	\begin{split} 
		J_a[\til{\lambda}_1] (z_1) J_b[\til{\lambda}_2](z_2) &\sim f_{ab}^c \frac{1}{\ip{12}} J_c[\til{\lambda}_1 + \til{\lambda}_2]  (z_1) 	\\
		J_a[\til{\lambda}_1] (z_1) \til{J}_b[\til{\lambda}_2](z_2) &\sim f_{ab}^c \frac{1}{\ip{12}} \til{J}_c[\til{\lambda}_1 + \til{\lambda}_2]  (z_1).	
	\end{split}
\end{equation}
Here we have absorbed the energy $\omega$ of a state into the normalization  of $ \til{ \lambda} $.

These OPEs are half of the standard splitting amplitudes of QCD \cite{Kosower:1999rx}. Working in the self-dual limit means we only retain splitting amplitudes with angle brackets in the denominator.

\subsection{Including the axion}
 One can ask if there is a chiral algebra which incorporates the loop-level corrections to the splitting amplitudes of QCD, as studied in \cite{Kosower:1999rx}.  We found in \cite{Costello:2022upu} associativity is broken once one-loop corrections are included.  This was traced to a gauge anomaly on twistor space \cite{Costello:2021bah, Costello:2022wso}, and we found that associativity is restored once we include an axion field that cancels the anomaly. Here we will write the states of the chiral algebra including the axion. 

The axion field is a scalar field with a fourth-order kinetic term, and so has the same number of degrees of freedom as two ordinary scalars. This field couples to the topological term $F \wedge F$ of the gauge theory.

In the chiral algebra, there are two towers of states associated to the axion:
\begin{equation} 
	E[m,n](z) \\ F[m,n](z) 
\end{equation}
where for $E[m,n]$, we have $m + n > 0$. 

As before, can arrange these into generating functionals 
\begin{equation} 
	\begin{split} 
		E[\til{\lambda}](z) &= \sum_{r + s > 0} \frac{1}{r! s!} \til{\lambda}_1^r \til{\lambda}_2^s  E[r,s] (z) \\
F[\til{\lambda}](z) &= \sum_{r,s } \frac{1}{r! s!} \til{\lambda}_1^r \til{\lambda}_2^s  F[r,s] (z). 	
	\end{split}
\end{equation}
We also let $E[k]$, $F[k]$ denote the terms in $E[\til{\lambda}]$, $F[\til{\lambda}]$ which are homogeneous of order $k$ in $\til{\lambda}$.

The axion OPEs are
\begin{equation} 
	\begin{split}
		 J_a[r,s](0) E[t,u](z) 
		 \sim & K_{ab} \frac{1}{2 \pi \i z} \frac{(ts - ur)}{t + u}\what{\lambda}_\g \til{J}^b [t+r - 1, s + u -1](0) \\ 
		J_a[r,s](0) F[t,u](z) \sim &  - K_{ab} \what{\lambda}_\g  \frac{1}{2 \pi \i z} \partial_z \til{J}^b[r+t, s + u](0) \\
		 & - K_{ab}  \what{\lambda}_\g \frac{1}{2 \pi \i z^2} (1 + \frac{r + s}{t+u+2}) \til{J}^b[r+t, s+u](0) \\
			 J_a[r,s](0) J_b[t,u](z)  \sim & \what{\lambda}_\g  \frac{1}{2\pi \i z} K_{ab} (ru-st) F[r+t-1,s+u-1] (0)\\
		& -  \what{\lambda}_\g \frac{1}{2\pi \i z} K_{ab} (t+u)  \partial_z E[r+t,s+u](0) \\
		& - \what{\lambda}_\g  \frac{1}{2\pi \i z^2} K_{ab} (r+s+t+u) E[r+t,s+u](0).  \label{eqn:EFope}
\end{split}
\end{equation}

Let us explain the constant $\what{\lambda}_{\g}$. First, we define $\lambda_{\g}$ so that, for $X \in \g$, $\op{Tr}(X^4) = \lambda_{\g}^2 \op{tr}(X^2)^2$, where $\op{Tr}$ means the trace in the adjoint and $\op{tr}$ means the minimal trace (i.e.\ the fundamental for $SU(N)$).   This tensor identity is necessary for the Green-Schwarz mechanism to hold. According to \cite{Okubo:1978qe}, we have 
\begin{equation} 
	\lambda_{\g}^2 = \frac{10 (\sh^\vee)^2 }{\dim \g + 2} 
\end{equation}
where $\sh^\vee$ is the dual Coxeter number, equal to $N_c$ for $SU(N_c)$.  

To account for the normalization of the interaction between the gauge field on twistor space and the field corresponding to the axion, we let
\begin{equation} 
	\what{\lambda}_{\g} = \frac{ \lambda_{\g} } { (2 \pi \i)^{3/2} \sqrt{12} }.
\end{equation}

Since the axion exchange gives the Green-Schwarz mechanism, some of these OPEs should be treated as one-loop effects. We take the convention that in these OPEs, those where the input is an axion are tree level, and those where the output is an axion are one loop. 

\subsection{One-loop correction to the gauge theory OPEs}
Certain one loop corrections to the $JJ$ and $J \til{J}$  OPEs are determined by associativity \cite{Costello:2022upu}\footnote{We use a slightly different convention for the normalization of $[ij]$ than that in \cite{Costello:2022upu} and in the first version of this paper.  This amounts to replacing each occurrence of $[ij]$ by $2[ij]$. }:
\begin{equation}
\begin{split} 
		 	J_a[1](z_1) J_b[1](z_2)
				\sim & \frac{[12]}{\ip{12} }   \frac{1}{ (2\pi \i)^2 8  }	   (f_{ea}^c f_{db}^e  + f_{da}^e f_{eb}^c)   :J_c \til{J}^d :(z_1) \\	
		&	- \frac{[12]}{\ip{12} }    \frac{1}{ (2 \pi \i)^3 48  }   \op{Tr}_{\g }([\t_a,\t_b]\t_c )  \partial_z \til{J}^c (z_1)
			+ \frac{[12] }{\ip{12}^2}  \frac{1}{ (2 \pi \i)^2 24  } \op{Tr}_{\g}([\t_a,\t_b] \t_c )  \til{J}^c(z_1)  \\
	J_a [1] (z_1)  \til{J}^b [1] (z_2)&  \sim  \frac{1}{ (2\pi \i)^2 8  }	 \frac{[12] }{\ip{12} } f_{ac}^e f_{ed}^b   :\til{J}^c \til{J}^d: (z_1)  \label{eqn:oneloopopes_gauge} 
\end{split}
\end{equation}
where we use the notation 
\begin{equation} 
	J \defeq J[0] \ \ \til{J} \defeq \til{J}[0]. 
\end{equation}
The leading pole in the one-loop correction to the OPE matches the part of the one-loop QCD splitting function that comes from self-dual gauge theory \footnote{This means the following. The QCD splitting function is computed by a certain partially off-shell one loop diagrams. Viewing QCD as a deformation of self-dual Yang-Mills, we can extract the diagrams which do not depend on the deformation parameter. These are exactly the diagrams corresponding to a $- \to ++$ splitting function.}.  

The subleading pole is forced on us by associativity -- we know there must be an associative chiral algebra by a formal argument using twistor theory \cite{Costello:2022wso}.  The sub-leading pole involves a normally ordered product, which is a little unusual.  This should be interpreted as follows.  When we study the collinear singularities of form factors of self-dual gauge theory, we will find a term where at $l$ loops, two positive helicity collinear states at $i,j$ can be replaced by one positive and one negative helicity state at $l-1$ loops, multiplied by $[ij]/\ip{ij}$.     In the literature, subleading poles like this one are sometimes referred to as ``non-factorizing contributions'' to the collinear limit. 

Although we know on general grounds that the chiral algebra exists at loop level, we do not have a closed expression for all OPEs.   Remarkably, however, it was shown in  \cite{Fernandez:2024qnu} that all OPEs are determined from the simple OPEs we have discussed by associativity (together with a handful of other natural properties).   If we used supersymmetric gauge theory, all OPEs were computed explicitly in \cite{Zeng:2023qqp}.  

We have been careful to write the one-loop corrections to the OPE in such a way that the Killing form does not appear.  This will be useful when including matter. The only place where the Killing form appears is in the  axion OPEs, where $K_{ab} = \op{tr}(\t_a \t_b)$ is the pairing built from the minimal trace.  

\section{Including matter}
Now let us repeat the analysis using matter. The fields on twistor space corresponding to  fermions on $\R^4$ are given by fermionic fields valued in $\Omega^{0,1}(\PT, \Oo(-1))$ and $\Omega^{0,1}(\PT, \Oo(-3))$.  The fields in $\Oo(-1)$ (respectively, $\Oo(-3)$) correspond to fermions of positive and negative helicity.  

Including matter changes the analysis very little, because the field content on twistor space has changed very little.  Suppose our matter lived in some representation $R$ of $\g$.  To avoid the ABJ anomaly (which is also present as an anomaly on twistor space) we will assume that $R$ is a real representation.  This means that $R$ admits a symmetric invariant non-degenerate pairing. 

The trace identity required to cancel the other anomaly on twistor space is now that
\begin{equation} 
	\op{Tr}_{\g}(X^4) - \op{Tr}_{R} (X^4) = \lambda_{(\g,R)}^2 \op{tr}(X^2)^2 
\end{equation}
for $X \in \g$. On the left hand side we see the difference between the trace in the fundamental and adjoint representations.  


As before, we let
\begin{equation} 
	\what{\lambda}_{\g,R} =  \frac{ \lambda_{\g,R} } { (2 \pi \i)^{3/2} \sqrt{12} }.
\end{equation}
Later we will specialize to the case when $\lambda_{(\g,R)} = 0$. 

To perform our analysis, it is convenient to view the theory with matter as being holomorphic BF theory on twistor space with gauge Lie algebra the super-algebra $\g \oplus \Pi R(-1)$, where $\Pi$ indicates parity shift, and we are twisting $R$ by the bundle $\Oo(-1)$.  From this perspective, the super-gauge field for this super-algebra corresponds to the ordinary gauge field $\mc{A} \in \Omega^{0,1}(\PT,\g)$ and the field in $\Omega^{0,1}(\PT, R \otimes \Oo(-1))$ corresponding to positive helicity fermions. The super-Lagrange multiplier field gives $\mc{B} \in \Omega^{0,1}(\PT, \g \otimes \Oo(-4))$, as well as the field in $\Omega^{0,1}(\PT, \g \otimes \Oo(-3))$ corresponding to negative helicity fermions.

The reason the super-field formalism is helpful is that we can simply borrow all of our previous analysis, but replacing the Lie algebra $\g$ by $\g \oplus \Pi R$.  Let us do this, and write down all the states in the chiral algebra, and their OPEs, including the one-loop correction.   We will use indices $i,j$ for a basis of $R$, and $g_{ai}^j$ for the action of $\g$ on $R$.   The states in the chiral algebra are given in table \ref{table:chiralalgebra_matter}.

\begin{table}	
\begin{tabular}{c |  c |  c |  c | c |c }
	Generator  & Spin & Weight & $SU(2)_+$ representation & Field & Dimension \\
	$J_a[m,n]$, $m,n \ge 0$ & $1-(m+n)/2$   & $(m-n)/2$ & $(m+n)/2$ & $A$ & $-m-n$  \\
	$\til{J}^a[m,n]$, $m,n \ge 0$ & $-1-(m+n)/2$   & $(m-n)/2$ & $(m+n)/2$ & $B$ & $-m-n-2$ \\
	$M_i[m,n]$, $m,n \ge 0$ & $\half - (m+n)/2$ &    $(m-n)/2$ & $(m+n)/2$ & $\psi_i$ & $-m-n-\half$ \\
	$\til{M}^i[m,n]$, $m,n \ge 0$ & $-\half - (m+n)/2$ &    $(m-n)/2$ & $(m+n)/2$ & $\psi^i$ & $-m-n-\tfrac{3}{2}$ \\
	$E[m,n]$, $m+n > 0$ &  $-(m+n)/2$   & $(m-n)/2$ & $(m+n)/2$ & $\rho$ & $-m-n$ \\
	$F[m,n]$, $m,n \ge 0$ & $ -(m+n)/2  $   & $(m-n)/2$ & $(m+n)/2$ & $\rho$ & $-m-n-2$  
\end{tabular}
	\caption{The generators of the 2d chiral algebra including matter. Spin in the table is in the sense of 2d CFT, and captures the transformations under the $SL_2(\C)$ of global conformal symmetries of the chiral algebra plane. Weight refers to charge under the Cartan of the other copy of $SL_2$ inside the (complexified) Lorentz group, which is a flavour symmetry of the chiral algebra.  Dimension refers to how states scale under scaling on $\R^4$.\label{table:chiralalgebra_matter}}
\end{table}

The OPEs with matter become exactly those we wrote above, without matter, except for the following modifications:
\begin{enumerate} 
	\item The index $a$ in equations \eqref{eqn:oneloopopes_gauge} now runs over a basis of $\g \oplus \pi R$.
	\item The appearance of $K_{ab}$, $K^{ab}$ in the OPEs involving the axion now refer to the invariant pairing only on $\g$.
	\item All appearances of $\what{\lambda}_{\g}$ are replaced by $\what{\lambda}_{\g,R}$.   
\end{enumerate}
Choose a basis $e_i$ of $R$ and a dual basis $e^i$ of $R^\vee$,and write $g_{ia}^j = -\ip{e^j , \t_a e_i }$.  Then $g_{ia}^j$ and $f_{ab}^c$ are the structure constants of the super-algebra $\g \oplus \pi R$. Indeed, in this super-algebra,
\begin{equation} 
	[e_i, \t_a] = -t_a (e_i) = g_{ia}^j e_j 
\end{equation}
and
\begin{equation} 
	[e_i, e^j] = g_{ia}^j \t_a 
\end{equation}
Explicitly, the tree-level OPEs are the following:
\begin{equation} 
	\begin{split}
		M_i[\til{\lambda}_1](z_1) J_a[\til{\lambda}_2] (z_2) \sim \frac{1}{\ip{12}} M_j[\til{\lambda}_1 + \til{\lambda}_2] (z_1) g_{ia}^j \\ 
		\til{M}^i[\til{\lambda}_1] (z_1) J_a[\til{\lambda}_2] (z_2) \sim -\frac{1}{\ip{12}} \til{M}^j[\til{\lambda}_1 + \til{\lambda}_2] (z_1) g_{ja}^i \\
		\til{M}^i[\til{\lambda}_1]  (z_1) M_j[\til{\lambda}_2]  (z_2) \sim \frac{1}{\ip{12} } \til{J}^a[\til{\lambda_1} + \til{\lambda}_2] (z_1) g_{ja}^i .
		 \end{split}
\end{equation}

The one-loop OPEs with matter are:
\begin{equation}
	\begin{split} 
		 	J_a[1](z_1) J_b[1](z_2)
				\sim & \frac{[12]}{\ip{12}}   \frac{1}{ (2\pi \i)^2 8  }	   (f_{ea}^c f_{db}^e  + f_{da}^e f_{eb}^c)   :J_c  \til{J}^d  :(z_1) \\
				& + \frac{[12]}{\ip{12}}   \frac{1}{ (2\pi \i)^2 8  }	   (g_{ka}^i g_{jb}^k  + g_{ja}^k g_{kb}^i)   :M_i  \til{M}^j  :(z_1) \\
		&	- \frac{[12]}{\ip{12}}    \frac{1}{ (2 \pi \i)^3 48  }   \op{Tr}_{\g \oplus \pi R}([\t_a,\t_b]\t_c )  \partial_z \til{J}^c (z_1)
			+ \frac{[12]}{\ip{12}^{2}} \frac{1}{ (2 \pi \i)^2 24  } \op{Tr}_{\g \oplus \pi R}([\t_a,\t_b] \t_c )  \til{J}^c(z_1) \\	
		J_a [1] (z_1)  \til{J}^b [1] (z_2)  & \sim  \frac{1}{ (2\pi \i)^2 8  }	 \frac{[12]}{\ip{12}}   f_{ca}^e f_{ed}^b :\til{J}^c  \til{J}^d :    \\
		M_i[1] (z_1) J_b [1](z_2)    
				\sim &- \frac{1}{\ip{12}}   \frac{1}{ (2\pi \i)^2 8  }	   (g_{ie}^j f_{db}^e  + g_{id}^k g_{kb}^j)   :M_j  \til{J}^d  :(z_1) \\
		J_a [1] (z_1)  \til{M}^i [1] (z_2)  & \sim  \frac{1}{ (2\pi \i)^2 8  }	 \frac{1}{\ip{12}}    ( f_{ac}^e g_{je}^i        :\til{J}^c  \til{M}^j :  +  g_{ja}^k g_{kd}^i        :\til{M}^j  \til{J}^d :      )\\
		M_i [1] (z_1)  \til{M}^j [1] (z_2)  & \sim  -\frac{1}{ (2\pi \i)^2 8  }	 \frac{[12]}{\ip{12}}   g_{ic}^k g_{kd}^j :\til{J}^c  \til{J}^d :    
	\end{split}	
\end{equation}

In \cite{Costello:2022upu} we showed that the quantum-corrected OPE is associative only when we include the axion.  The associativity constraint fixes the coefficients of the one-loop OPEs.  The coefficient of the $2 \mapsto 1$ one-loop OPE matches the known one-loop splitting amplitude \cite{Kosower:1999rx}.    In the Appendix we redo the calculation with matter; very little needs to be changed. 

\section{Anomaly cancellation by carefully chosen matter}
There is an important special case of the analysis above.   For some examples of simple group $G$ and matter representation $R$, the constant $\lambda_{G,R} = 0$.  This happens precisely when
\begin{equation} 
	\op{tr}_{\g} (X^4) = \op{tr}_{R} (X^4) \label{eqn:noaxion} 
\end{equation}
As I have already mentioned, examples of such groups and representations as $SU(2)$ with $N_f = 8$ and $SU(3)$ with $N_f = 9$.   

In the case when $\lambda_{\g,R} = 0$, the structure constants derived above still hold. To see this, we note that the chiral algebra can be constructed directly (by the method of Koszul duality \cite{Costello:2022wso}) from the twistor uplift of the theory.  This construction is ultimately in terms of Feynman diagrams.  From this we see that all structure constants of the chiral algebra must be algebraic expressions of $\lambda_{\g,R}$ and of the structure constants of the Lie algebra $\g$ and the representation $R$.   

In particular, any formula for OPEs in the chiral algebra remains valid in the cases when $\lambda_{g,R} = 0$. In those cases, the axion decouples and we can discard it.

One big advantage of cases when the anomaly is cancelled by the matter representation is that results obtained by the chiral algebra method are directly applicable to standard QCD.  We will take advantage of this to compare chiral algebra computations to known results. 

\subsection{Examples of theories with no axion}
Let us quickly explain why \eqref{eqn:noaxion} holds for $SU(2)$, $N_f = 8$.  There is only one quartic invariant function on the Lie algebra.  We can therefore assume that the matrix $X$ in \eqref{eqn:noaxion} is diagonal with eigenvalues $1$ and $-1$.   

The standard convention is that the matter representation consists of $N_f$ fundamental and $N_f$ anti-fundamental. In the case of $SU(2)$, where fundamental and anti-fundamental are the same, this means we have $2 N_f$ fundamental representations. The right hand side of \eqref{eqn:noaxion} is therefore $4 N_f$.  

The eigenvalues of our matrix $X$ in the adjoint are $2$ and $-2$. Therefore the left hand side of \eqref{eqn:noaxion} is $32$, giving us $N_f = 8$ as desired. 

This example is part of an infinite series, with gauge algebra is $\mf{sp}(N)$. Our convention is such that the fundamental representation of $\mf{sp}(N)$ is the symplectic vector space $\C^{2 N}$. The matter representation is
\begin{equation} 
	R = \C^{16} \otimes \C^{2 N} \oplus \wedge^2_0 \C^{2N}  
\end{equation}
Here $\wedge^2_0 \C^{2N}$ indicates the trace-free exterior square, which is the subspace of $\C^{2N} \otimes \C^{2N}$ consisting of two-index tensors which are anti-symmetric and are zero when contracted with the symplectic form. When $N = 1$, this gauge theory and matter content is $SU(2)$ with $N_f = 8$. This system has  $SU(16)$ flavour symmetry.   

It is straightforward to check that the identity \eqref{eqn:noaxion} holds in this case.  To see this, first note that there are two quartic invariant functions on $\mf{sp}(N)$ for $N > 1$. It is enough We need to check that the identity holds when we take $X$ to live in a rank $2$ subspace of the Cartan algebra of $\mf{sp}(N)$.  We do this by writing 
\begin{equation} 
	\C^{2N} =\C^4 \oplus \C^{2 (N-2) }. 
\end{equation}
We take $X$ to act only on the $\C^4$ in this decomposition, and to be diagonal with eigenvalues $a,-a,b,-b$.    Clearly
\begin{equation} 
	\op{tr}_{\C^{16} \otimes \C^{2N}} (X^4) = 32 a^4 + 32 b^4. 
\end{equation}

Then, 
\begin{equation} 
	\wedge^2_0 \C^{2 N} = \C^4 \otimes \C^{2(N-2)} \oplus \wedge^2 \C^2 \oplus \wedge^2 \C^{2 (N-2)}  
\end{equation}
The eigenvalues of $X$ in this representation are $a,-a,b,-b$, with multiplicity $2 (N-2)$; and $\pm a \pm b$, with multiplicity $1$. Therefore
\begin{equation} 
	\op{tr}_{\wedge^2_0 \C^{2 N} } (X^4) = 4 (N-2) (a^4 + b^4) + 2(a+b)^4 + 2 (a-b)^4. 
\end{equation}
The matter contribution is therefore
\begin{equation} 
	 (32 + 4 (N-2)) (a^4 + b^4) + 2(a+b)^4 + 2 (a-b)^4. 
\end{equation}

The adjoint representation of $\mf{sp}(N)$ is the symmetric square $\Sym^2 \C^{2N}$ of the fundamental representation. The eigenvalues of $X$ in the adjoint representation
\begin{equation} 
	\Sym^2 \C^{2N} = \Sym^2 \C^4 \oplus \C^4 \otimes \C^{2(N-2)} \oplus \Sym^2 \C^{2(N-2)}  
\end{equation}
are $2a, -2a, 2b, -2b, \pm a \pm b$, each with multiplicity $1$; and $a,-a,b,-b$ with multiplicity $2 (N-2)$.  Therefore
\begin{equation} 
	\op{tr}_{\mf{sp}(N)} (X^4) = (32 + 4 (N-2) )  (a^4 + b^4) + 2(a+b)^4 + 2 (a-b)^4 
\end{equation}
matching the contribution from the matter representation.

\subsection{$\mf{sl}(N)$ example of anomaly free theory} 
This example was suggested by Roland Bittleston.  Here, matter is in the representation
\begin{equation} 
	R = \C^8 \otimes (F \oplus F^\vee) \oplus \wedge^2 F \oplus \wedge^2 F^\vee. 
\end{equation}
This is a real representation, as desired. To check that this cancels the anomaly, let us decompose the fundamental representation $F = \C^N$ as
\begin{equation} 
	F = \C^{2}_1 \oplus \C^2_2 \oplus \C^{N-4} 
\end{equation}
This decomposition selects a subalgebra $\mf{sl}_2 \oplus \mf{sl}_2 \oplus \mf{sl}_{N-4}$ of $\mf{sl}_N$.  The spaces $\C^2_1$, $\C^2_2$ are the fundamental representations of the two copies of $\mf{sl}_2$.  

There are two quartic invariant polynomials of $\mf{sl}_N$, and they are determined by their restriction to the subalgebra $\mf{sl}_2 \oplus \mf{sl}_2$.

As a representation of $\mf{sl}_2 \oplus \mf{sl}_2$, we have
\begin{equation} 
	R = (2N + 8) (\C^2_1 \oplus \C^2_2) \oplus 2 (\C^2_1 \otimes \C^2_2)  
\end{equation}
(ignoring trivial representations).  As a representation of $\mf{sl}_2 \oplus \mf{sl}_2$,  the adjoint representation of $\mf{sl}_N$ breaks up as  
\begin{equation} 
	\mf{sl}_N = \Sym^2 \C^2_1 \oplus \Sym^2 \C^2_2 \oplus 2 (\C^2_1 \otimes \C^2_2) \oplus (2N-8) (\C^2_1 \oplus \C^2_2)
\end{equation}
again ignoring trivial representations of $\mf{sl}_2$. 

The difference between $R$ and the adjoint is
\begin{equation} 
	R - \mf{sl}_N = 16(\C^2_1 \oplus \C^2_2) - S^2 \C^2_1 - S^2 \C^2_2 
\end{equation}
and we have already seen that trace of $X^4$ in the adjoint of $\mf{sl}_2$ cancels with that in $16$ copies of the fundamental.

We will mostly focus on this family of examples in what follows.

\subsection{Vanishing of one-loop all $+$ amplitude}
The anomaly cancellation condition we have been discussing guarantees that the self-dual gauge theory with matter has a lift to a local theory on twistor space \cite{Costello:2021bah}. The one-loop all $+$ amplitude is an amplitude of the self-dual theory itself, and not a form factor.

It is immediate that any local QFT on twistor space has vanishing amplitudes (when states have generic momenta).  Give twistor space coordinates $z,v_1,v_2$.  In terms of the spinors $\lambda_{\alpha}, \til{\lambda}_{\dot{\alpha}}$ used in the spinor-helicity description of amplitudes, we have $\lambda = (1,z)$,  and $\til{\lambda}_{\dot{\alpha}}$ can be thought of as momenta dual to $v^{\dot{\alpha}}$.  

Any massless on-shell state on $\R^4$ can be lifted to a $(0,1)$ form on twistor space of the form 
\begin{equation} 
	\delta_{z = z_0} e^{ \til{\lambda}_{\dot{\alpha}} v^{\dot \alpha} }. 
\end{equation}
(This expression is sometimes called the half-Fourier transform; it is basically the same as the Penrose transform). 

If we have a collection of states with $\lambda_{\alpha}^i = (1,z_i)$, then their twistor representatives do not touch.  Therefore the can't scatter off each other in any local field theory on twistor space. 

This tells us that for any matter so that equation \eqref{eqn:noaxion} holds, the one-loop all $+$ amplitude must vanish. We can see this in concrete terms too, however. Consider the four-point amplitude. It is (up to a multiplicative constant) 
\begin{equation} 
	\sum_{\sigma \in S_4} 	\frac{[\sigma_1 \sigma_2][\sigma_3 \sigma_4]}{\ip{\sigma_1 \sigma_2} \ip{\sigma_3 \sigma_4} } \op{Tr}_{\g \oplus \pi R} ( \t_{a_{\sigma_1} } \dots \t_{a_{\sigma_4}} ). 
\end{equation}
However, the expression
\begin{equation} 
	\frac{[12][34]} {\ip{12} \ip{34} } 
\end{equation}
is totally symmetric (which can be seen easily using conservation of momentum). 

It follows that the amplitude can be rewritten as 
\begin{equation}  
	\frac{[1 2][34]}{\ip{1 2} \ip{3 4} } \sum_{\sigma \in S_4} 	\op{Tr}_{\g \oplus \pi R} ( \t_{a_{\sigma_1} } \dots \t_{a_{\sigma_4}} ). 
\end{equation}
Here we see the symmetrized trace in $\g \oplus \pi R$.  However, we have chosen our representation precisely so that this is zero! 

The vanishing of the higher-point amplitudes is not quite so immediate. While it follows from the twistor space analysis, it would be useful to have an explicit derivation. 

\section{Computational methods: chiral algebra versus BCFW}
As we have already mentioned, local operators of self-dual gauge theory (plus axion or appropriate matter) are the same as conformal blocks of the chiral algebra. Form factors for a local operator are the same as correlators of the chiral algebra in that conformal block.

This leads to a bootstrap method for computing form factors.  The idea is very simple. If $\mc{O}$ is a local operator in self-dual gauge theory, we write
\begin{equation} 
	\ip{\mc{O} \mid J[\til{\lambda}_1](z_1) \dots \til{J}[\til{\lambda}_n] (z_n) } 
\end{equation}
be the correlation function of the chiral algebra in the corresponding conformal block, with some states of positive and negative helicity, and possibly some matter and/or axion state insertions too.  This correlator will be the same as the form factor of SDYM with the operator $\mc{O}$ placed at the origin. (If we instead place the operator at some point $x$ we introduce a factor of $e^{\i x \cdot P}$, where $P$ is the total momentum of the states we are scattering). 

We can expand this correlator as a sum of terms like
\begin{equation} 
	\ip{\mc{O} \mid J[k_1](z_1) \dots \til{J}[k_n] (z_n) }\label{eqn:correlator} 
\end{equation}
This amounts to expanding the external states in the form factor as a sum of soft modes.

If the operator is of scaling dimension $d$, then the dimension of the insertions must sum up to $-d$. For instance, the positive helicity state $J[k]$ contributes dimension $-k$, and $\til{J}[k]$ contributes dimension $-k-2$. 

Knowledge of the OPEs constrains the poles in equation \eqref{eqn:correlator}.  For instance, all poles involving $J[0]$ insertions are determined by tree-level OPEs, and poles involving $J[1]$ insertions are constrained by one-loop OPEs\footnote{We only presented a formula for the one-loop OPEs between $J[1]$ and $J[1]$ or $\til{J}[1]$, and not for more general one-loop OPEs between $J[1]$ and $J[k]$ or $\til{J}[k]$.  These more general OPEs can be derived by associativity.}. 

We can express the residue at the poles of \eqref{eqn:correlator} in terms of chiral algebra correlators which are either at lower loop number, or which involve fewer insertions.  In this way, we can inductively determine the chiral algebra correlator, and so the form factor.

Our current knowledge of the chiral algebra prevents this technique being carried out in full generality, as we do not have expressions for all OPEs. The OPEs we have determined are sufficient for two-loop computations of $\op{tr}(B^2)$, as we will see.

The reader may be reminded of the BCFW recursion \cite{Britto:2005fq} which is also used to determine amplitudes based on their poles.  The difficult part of the BCFW recursion at loop level is getting a precise understanding of the pole structure of the amplitude. For instance, one-loop splitting functions give a double pole, but the single pole underlying the double pole can be hard to pin down.

The advantage of the chiral algebra method is that, once we know the OPE, we have a precise expression for \emph{all} singularities in the form factor.  In this way, the chiral algebra method can be interpreted as a strong form of the BCFW recursion. The disadvantage is that the chiral algebra method has more limited applicability, since it requires working with a theory where the anomaly on twistor space is cancelled.

In fact, associativity of the chiral algebra is equivalent to saying that form factors like \eqref{eqn:correlator} have a perticularly nice analytic structure.   Associativity of the chiral algebra boils down to the following statements:
\begin{enumerate} 
	\item All form factors such as \eqref{eqn:correlator} in SDYM plus matter are rational functions, with poles only in the  $\ip{ij}$ variables. 
	\item If there is a pole in the form factor \eqref{eqn:correlator} at $z_i = z_j$, then the polar part is also a form factor where the states $J[k_i](z_i)$, $J[k_j](z_j)$ have been replaced by some normally ordered product of states at $z_j$, times some expressions in $[ij]$ and $\ip{ij}^{-1}$.  
	\item The normally ordered product of states appearing above are independent of the 4d local operator $\mc{O}$ defining the form factor.  
\end{enumerate}

\section{Formulae for two-loop amplitudes}
Now we can present the formula we derive for the two-loop single-trace all $+$ amplitudes. Let us first consider the case where we have $SU(N)$ gauge theory with matter in $8 F \oplus 8 F^\vee \oplus \wedge^2 F \oplus \wedge^2 F^\vee$.

First, we have the four-point amplitude \footnote{I am very grateful to Lance Dixon and Anthony Morales for pointing out several errors in the original version of this formula. These were caused by algebraic errors computing colour factors in equation \eqref{eqn:alg_relation} in the appendix, as well as a few dropped signs.}:
\begin{equation} 
	\begin{split} 
		\mc{A}^{TO}(1^+,2^+,3^+,4^+) 		= & 
		\left(   6 N - 4 - 8 N^{-1}   \right) 	\left( \frac{1}{ (4 \pi)^4 }	\frac{[12][34]} { \ip{12} \ip{34} } + 	\frac{1}{ (4 \pi)^4 }	\frac{[41][23]} { \ip{41} \ip{23} } \right)	\\ 
		&	- \left( 4 + 8 N^{-1}  \right)  \frac{1}{ (4 \pi)^4} \frac{[13][24]}{ \ip{13}\ip{24}} \\ 
		& - \frac{2}{  (4 \pi)^4   }  \frac{[12] [34]( \ip{13} \ip{24}  + \ip{14} \ip{23} )}{\ip{12}^2 \ip{34}^2} \\
		&	+ \frac{2}{  (4 \pi)^4   }  \frac{[14] [23]( \ip{13} \ip{42}  + \ip{12} \ip{43} )}{\ip{14}^2 \ip{23}^2}
\end{split}
\end{equation}
Then the $n$-point single trace amplitude is 
\begin{equation} 
	\mc{A}_2^{TO} (1^+, \dots, n^+ ) =   \sum_{1 \le i < j < k < l \le n}  \mc{A}_2^{TO}(i^+,j^+,k^+,l^+) \frac{ \ip{ij} \ip{jk} \ip{kl} \ip{li} } { \ip{12} \dots \ip{n1} } \label{eqn:npointfrom4point} 
\end{equation}
We note that if we expand this in powers of $N$, the leading term is of order $N$ and is (almost) proportional\footnote{Versions one and two of this paper had a typo going from equation \eqref{eqn:npointfrom4point} to equation \eqref{eqn:NfNcleading}.}   to the one-loop all $+$ amplitude of pure gauge theory:
\begin{equation}
	\begin{split} 	
	\mc{A}_2^{TO} (1^+, \dots, n^+ ) = \frac{6 N}{(4\pi)^4}      \sum_{1 \le i < j < k < l \le n} \frac{ [ij] \ip{jk} [kl] \ip{li} + \ip{ij} [jk] \ip{kl} [li]   } { \ip{12} \dots \ip{n1} }\\
		+ \text{ lower orders in } N.
	\end{split} \label{eqn:NfNcleading}
\end{equation}
The one-loop all $+$ amplitude only has the term $\ip{ij} [jk] \ip{kl} [li]$, not the second term. The expression we find is called the parity-even part of the one-loop all $+$ amplitude, because in the numerator we have added the amplitude to its parity conjugate. 
 
\subsection{Amplitudes in the case $N_F = N_c$}
Next, let us consider the case when $N_F = N_c$, and the anomaly is cancelled by an axion. Then, the four-point two-loop single-trace all $+$ amplitude is 
\begin{equation} 
	\mc{A}_2^{TO}(1^+,2^+,3^+,4^+) =  
				\left( N^2 - 3    \right)	\frac{1}{ (4 \pi)^4 }	\frac{[12][34]} { \ip{12} \ip{34} } + 
		 		\frac{2}{ (4 \pi)^4 }	\frac{[13][24]} { \ip{13}\ip{24} } 	
\end{equation}
The $n$-point amplitude is again given by equation \eqref{eqn:npointfrom4point}.  Once again, the leading terms in $N$ are
\begin{equation}
	\begin{split} 	
	\mc{A}_2^{TO} (1^+, \dots, n^+ ) = \frac{1}{2 (4\pi)^4}   N^2    \sum_{1 \le i < j < k < l \le n}  \frac{ [ij] \ip{jk} [kl] \ip{li}+\ip{ij} [jk] \ip{kl} [li]} { \ip{12} \dots \ip{n1} } \\
		+ \text{ lower orders in } N. \label{eqn:NfNcleading2}
	\end{split}
\end{equation}

It is interesting to note that processes involving an exchange of an axion field are subleading in $N$. Therefore the expression in \eqref{eqn:NfNcleading2} gives the leading order term in $N_f = N_c$ QCD \emph{even without} the axion. 

The rest of the paper, and the appendix, is devoted to computing these amplitudes. To implement our inductive technique using collinear singularities, we first need to compute certain tree-level and one loop form factors. This is achieved in section \ref{sec:treelevel_oneloop}.  In section \ref{sec:oneloopcomparison}, the one-loop form factors are compared to expressions in the literature, and found to match at four points as long as the anomaly cancellation criterion holds.

In section \ref{sec:twoloop},  the two loop form factors are computed using tree-level and one-loop form factors as input.  A crucial calculation of colour factors is placed in the appendix \ref{sec:computingalgebraicfactors}.

\section{Computation of tree-level and one-loop form factors}
\label{sec:treelevel_oneloop}
Now let us turn to actually computing form factors of SDYM with matter.  In \cite{Costello:2022upu} we computed one-loop form factors of the operator $\half \op{tr}(B^2)$ by using the one-loop corrections to the OPE. Here we will review that calculation and generalize it (very slightly) to include matter.  Along the way we do some ancillary form-factor computations that will feed into the two loop computation. 

We are computing the form factor of the operator placed at the origin.  The form factor of the operator placed at $x$ is obtained from that of the operator placed at the origin by multiplying by $e^{\i x \cdot P}$, where $P$ is the total momentum of the external states.  The amplitude is the integral of this expression over $x$. Clearly, the amplitude only differs from the form factor of the operator at the origin by a momentum-conserving $\delta$-function $\delta_{P = 0}$.  In what follows we will ignore the momentum-conserving $\delta$-function. 

Our starting point is that
\begin{equation} 
	\ip{\half \op{tr}(B^2) \mid \til{J}^a(z_1) \til{J}^b(z_2) } = -K^{ab} \ip{12}^2 
\end{equation}
(We use the shorthand that $J, \til{J}$ without any $[i,j]$ indices means $J[0,0], \til{J}[0,0]$).  This is essentially fixing our convention for what we mean by $\half \op{tr}(B^2)$.  

\subsection{Tree level form factors}

The tree level form factor involving on $J, \til{J}$ insertions were computed in \cite{Costello:2022wso}, and found to reproduce the Parke-Taylor formula:
\begin{equation}
	\begin{split} 
	\ip{\half \op{tr}(B^2) \mid \til{J}_{a_1}(z_1) \til{J}_{a_2}(z_2) J_{a_3}(z_3) \dots J_{a_n}(z_n) } \\
		= \frac{1}{n} \sum_{\sigma \in S_n} \op{tr}(\t_{a_{\sigma(1)}} \dots \t_{a_{\sigma(n)}} ) \frac{ \ip{12}^4} { \ip{\sigma(1) \sigma(2)} \dots \ip{\sigma(n) \sigma(1) } }. 
	\end{split}	
\end{equation}

With the inclusion of matter, we need to use the following OPEs:
\begin{equation} 
	\begin{split}
		M_i(z_1) J_a(z_2) \sim \frac{1}{\ip{12}} M_j(z_1) g_{ia}^j \\ 
		\til{M}^i(z_1) J_a(z_2) \sim -\frac{1}{\ip{12}} \til{M}^j(z_1) g_{ja}^i \\
		\til{M}^i (z_1) M_j (z_2) \sim -\frac{1}{\ip{12} } \til{J}^a(z_1) g_{ja}^i .
		 \end{split}
\end{equation}
From these OPEs, we find that we find the following form factors.
\begin{equation}
	\begin{split} 	
		\ip{ \half\op{tr}(B^2) \mid  \til{M}^j(z_1)  \til{J}_a(z_2) M_i(z_3) } &= g_{ia}^j \frac{ \ip{12}^2 } { \ip{13} } \\
	&=  -\ip{e^j, \t_a e_i } \frac{\ip{12}^2} {\ip{13}}. \end{split}
\end{equation}

Indeed, $M(z_2)$ has a non-singular OPE with $\til{J}(z_1)$, and has a first order zero at $z_2 = \infty$. The OPE of $M(z_2)$ with $\til{M}(z_3)$ determines this three-point form factor.

More generally, one can see by induction that
\begin{equation} 
	\begin{split} 
		\ip{\half\op{tr}(B^2) \mid \til{M}^j(z_1)  \til{J}_{a_2} (z_2) J_{a_3}(z_3) \dots J_{a_{n-1}} (z_{n-1}) M_i(z_{n} ) } \\
		= \sum_{\sigma \in S_{n-2} } \frac{ \ip{12}^3 \ip{2n}     }  { \ip{1 \sigma(2) } \ip{\sigma(2) \sigma(3)} \dots \ip{\sigma(n-2) \sigma(n-1)} \ip{\sigma(n-1) n } \ip{n1}      }  \ip{ e^j, \t_{a_{\sigma(2)}} \dots \t_{a_{\sigma(n-1)}} e_i }.    
	\end{split}
\end{equation}
The sum is over permutations of $2,\dots,n-1$.

In particular, taking $n = 4$, we have
\begin{equation}
	\begin{split} 	
		\ip{\half\op{tr}(B^2) \mid \til{M}^j(z_1)  \til{J}_{a_2} (z_2) J_{a_3}(z_3) M_i(z_{4} ) } &  \\ 
		&= \ip{ e^j, \t_{a_{2}}  \t_{a_{3}} e_i } \frac{ \ip{12}^3 \ip{24}    } { \ip{12} \ip{23} \ip{34} \ip{41} }  + \ip{ e^j, \t_{a_{3}}  \t_{a_{2}} e_i } \frac{ \ip{12}^3 \ip{24}    } { \ip{13} \ip{32} \ip{24} \ip{41} }   \\
		&=		g^j_{a_2 k} g^k_{a_3 i}    \frac{ \ip{12}^3 \ip{24}    } { \ip{12} \ip{23} \ip{34} \ip{41} }  + g^j_{a_3 k} g^k_{a_2 i}   \frac{ \ip{12}^3 \ip{24}    } { \ip{13} \ip{32} \ip{24} \ip{41} }   
	\end{split} \label{eqn:twofermionstree}
\end{equation}

We will also be interested in form factors with four matter field insertions. In this case only the four-point amplitude is relevant for us.  We find that
\begin{equation} 
	\begin{split} 
		\ip{ \half \op{tr}(B^2) \mid   \til{M}^{i_1} (z_1) M_{i_2}(z_2) \til{M}^{i_3}(z_3) M_{i_4}(z_4)   } \\
		= -\frac{ \ip{13}^2 } { \ip{12} \ip{34} } g^{i_1}_{i_2 a} g^{i_3}_{i_4 a} - \frac{ \ip{13}^2 } { \ip{14} \ip{23} } g^{i_1}_{i_4 a} g^{i_3}_{i_2 a}  
	\end{split} \label{eqn:treelevelfourfermion}
\end{equation}
Again this is determined by the OPEs.

\subsection{Tree-level form factors involving normally-ordered products}
\label{section:vanishing_oneNO}
Before we consider the three-point form factors, we need to consider tree-level two-point form factors one of whose insertions is a normally ordered product.

The form factor 
\begin{equation} 
	\ip{ \half \op{tr}(B^2) \mid \til{J}^a(z_1) \NO{\til{J}^b  J_c}  (z_2) }  
\end{equation}
vanishes. This is because it is obtained as a contour integral 
\begin{equation} 
	\oint_{\abs{z_3-z_2} = 1}  \d z_3 z_{23}^{-1}  	\ip{ \half \op{tr}(B^2) \mid \til{J}^a(z_1) \til{J}^b(z_2)  J_c  (z_3) }  
\end{equation}
The Parke-Taylor formula tells us that the form factor in the integrand is $\frac{z_{12}^3}{z_{13}z_{23}}$, so that the integral vanishes.  

The  form factor 
\begin{equation} 
	\ip{ \half \op{tr}(B^2) \mid \til{J}_a(z_1) \left( \NO{\til{M}_iM_j } +  \NO{\til{M}_j M_i } \right) (z_2) }  \label{eqn:normallyorderedcorrelator} 
\end{equation}
vanishes for symmetry reasons. Here, we have lowered the index in $\til{M}^i$ using the fact that the matter is in a real representation, and then symmetrized in the $i,j$ indices. 

To see this, we note that the three-point function 
\begin{equation} 
	\ip{ \half \op{tr}(B^2) \mid \til{J}_a(z_1) \til{M}^i(z_2)M_j  (z_3) }  
\end{equation}
is anti-symmetric in the $i-j$ indices, because the representation is a real representation so that the structure constants $g_{ija}$ of the representation are anti-symmetric. 

Therefore, \eqref{eqn:normallyorderedcorrelator} vanishes.  This particular expression is relevant because, in the OPE, the normally-ordered product $:\til{M}_i M_j:$ appears with the $i,j$ indices symmetrized. 

\subsection{Tree level correlators with two normally ordered products}

For the two-loop computation which is our ultimate goal, we also need to determine certain tree-level correlators with two normally ordered product insertions. We will find that
\begin{align}  
	\ip{\half \op{tr}(B^2) \mid \NO{ \til{J}_{(a} J_{b)} } (z)  \NO{ \til{J}_{(c} J_{d)}} (w)  } &=   -2\op{tr}(\t_{(a} \t_{b)} \t_{(c} \t_{d)} ) + \op{tr}(\t_{a} \t_{c} \t_{b} \t_{d}) + \op{tr}(\t_{a} \t_{d} \t_{b} \t_{c} ) \label{eqn:twonormalorder1} 	\\
		\ip{\half \op{tr}(B^2) \mid \NO{ \til{J}_{(a} J_{b)}}  (z)  \NO{ \til{M}_{(i} M_{j)} } (w)}  &= \half\left( g^j_{ka} g^k_{ib} + g^j_{kb} g^k_{ia}    \label{eqn:twonormalorder2}\right) 	\\ 
		\ip{\half \op{tr}(B^2) \mid \NO{\til{M}_{(i} M_{j)}} (z) \NO{ \til{M}_{(k} M_{l)}} (w)} &=  - \half \left(  g_{i l a} g_{k j a} +    g_{j l a} g_{k i a} 	 \right).	\label{eqn:twonormalorder3}
\end{align}
Here we are symmetrizing with respect to the bracketed indices, in every normally ordered product.  The symmetrization of two indices includes a symmetry factor of $\half$.  

First, we have
\begin{equation}
	\begin{split} 
		\ip{\half \op{tr}(B^2) \mid \til{J}_{(a}(z_1) J_{b)}(z_2) \til{J}_{(c} (z_3) J_{d)} (z_4) } =& 2\op{tr}(\t_{(a} \t_{b)} \t_{(c} \t_{d)} ) \frac{\ip{13}^4  } {\ip{12} \ip{23} \ip{34} \ip{41}   }   \\ 
		&+ ( \op{tr}(\t_{a} \t_{c} \t_{b} \t_{d}) + \op{tr}(\t_{a} \t_{d} \t_{b} \t_{c} ) )  \frac{\ip{13}^4  } {\ip{14} \ip{42} \ip{23} \ip{31}   }   \\ 
		&+ 2\op{tr}(\t_{(a} \t_{b)} \t_{(c} \t_{d)} ) \frac{\ip{13}^4  } {\ip{13} \ip{34} \ip{42} \ip{21}   }    
	\end{split}	
\end{equation}
Adding the first and last lines together this becomes
\begin{equation} 
	\begin{split} 
		2\op{tr}(\t_{(a} \t_{b)} \t_{(c} \t_{d)} ) \frac{\ip{13}^4  \left( \ip{13} \ip{42} - \ip{23} \ip{41} \right)   } {\ip{12} \ip{23} \ip{34} \ip{41} \ip{13} \ip{42}    }  \\
		+ ( \op{tr}(\t_{a} \t_{c} \t_{b} \t_{d}) + \op{tr}(\t_{a} \t_{d} \t_{b} \t_{c} ) )  \frac{\ip{13}^4  } {\ip{14} \ip{42} \ip{23} \ip{31}   }   . 
	\end{split}
\end{equation}
Using the Schouten identity this is
\begin{equation} 
	\begin{split} 
		-2\op{tr}(\t_{(a} \t_{b)} \t_{(c} \t_{d)} ) \frac{\ip{13}^4   \ip{12} \ip{34}   } {\ip{12} \ip{23} \ip{34} \ip{41} \ip{13} \ip{42}    }  \\
		+ ( \op{tr}(\t_{a} \t_{c} \t_{b} \t_{d}) + \op{tr}(\t_{a} \t_{d} \t_{b} \t_{c} ) )  \frac{\ip{13}^4  } {\ip{14} \ip{42} \ip{23} \ip{31}   }   . 
	\end{split}
\end{equation}
The expression is now in a form in which there is no pole as $z_2 \to z_1$ and $z_4 \to z_3$, in which case we find
\begin{equation}
	\begin{split} 	
	 		\ip{\half \op{tr}(B^2) \mid \til{J}_{(a}(z_1) J_{b)}(z_1) \til{J}_{(c} (z_3) J_{d)} (z_3) }\\
			=  -2\op{tr}(\t_{(a} \t_{b)} \t_{(c} \t_{d)} ) + \op{tr}(\t_{a} \t_{c} \t_{b} \t_{d}) + \op{tr}(\t_{a} \t_{d} \t_{b} \t_{c} )
	\end{split}
\end{equation}

For \eqref{eqn:twonormalorder2}, we first recall from \eqref{eqn:twofermionstree} that 
\begin{equation}
	\begin{split} 
		\ip{\half\op{tr}(B^2) \mid \til{M}^i(z_1)  \til{J}_{(a_2} (z_2) J_{a_3)}(z_3) M_j(z_{4} ) } \\=   
		 \ip{ e^i, \t_{(a_{2}}  \t_{a_{3})} e_j }  \ip{12}^3 \ip{24}  \left( \frac{1} { \ip{12} \ip{23} \ip{34} \ip{41} }  +  \frac{1 } { \ip{13} \ip{32} \ip{24} \ip{41} } \right). 
		\end{split}
\end{equation}
Using the Schouten identity as we did earlier, we have
\begin{equation} 
	\frac{1} { \ip{12} \ip{23} \ip{34} \ip{41} }  +  \frac{1 } { \ip{13} \ip{32} \ip{24} \ip{41} } = -\frac{1}{\ip{12} \ip{24} \ip{43} \ip{31} }. 
\end{equation}
Therefore
\begin{equation}
	\begin{split} 
		\ip{\half\op{tr}(B^2) \mid \til{M}^i(z_1)  \til{J}_{(a_2} (z_2) J_{a_3)}(z_3) M_j(z_{4} ) } \\=   
	-	\ip{ e^i, \t_{(a_{2}}  \t_{a_{3})} e_j }  \ip{12}^3 \ip{24}  \frac{1} { \ip{12} \ip{24} \ip{43} \ip{31} }   
	\end{split}
\end{equation}
This is in a form where we can send $z_3 \to z_2$ and $z_4 \to z_1$, giving
\begin{equation}
	\begin{split} 	
		\ip{\half \op{tr}(B^2) \mid \NO{\til{J}_{(a} J_{b)}} (z) \NO{\til{M}^{i} M_{j}} (w)}  &= \ip{ e^i, \t_{(a}  \t_{b)} e_j }   \\
		&=\half \left( g^i_{ka} g^k_{jb} + g^i_{kb} g^k_{ja} \right) 
	\end{split}
\end{equation}

Now let us discuss the correlator \eqref{eqn:twonormalorder3}. We know from \eqref{eqn:treelevelfourfermion} that  
\begin{equation} 	
	\ip{ \half \op{tr}(B^2) \mid \til{M}^{i_1} (z_1) M_{i_2}(z_2) \til{M}^{i_3}(z_3) M_{i_4}(z_4)   } 
		= -\frac{ \ip{13}^2 } { \ip{12} \ip{34} } g^{i_1}_{i_2 a} g^{i_3}_{i_4 a} - \frac{ \ip{13}^2 } { \ip{14} \ip{23} } g^{i_1}_{i_4 a} g^{i_3}_{i_2 a}  \label{eqn:4pt4m}	
\end{equation}
We are symmetrizing with respect to $i_1, i_2$ and $i_3,i_4$. Since matter is in a real representation, when we lower the index $g_{i_1 i_2 a}$ is anti-symmetric in $i_1,i_2$.  Because we are interested in the normally ordered product, we need to take the non-singular part in the limit as $z_2 \to z_1$ and $z_4 \to z_3$.   Only the second term in \eqref{eqn:4pt4m}  contributes, and we find that
\begin{equation} 
	\ip{\half \op{tr}(B^2) \mid \NO{\til{M}_{(i} M_{j)}}  (z) \NO{\til{M}_{(k} M_{l)}} (w)} = - \half \left(  g_{i l a} g_{k j a} +    g_{j l a} g_{k i a} 	 \right).	
\end{equation}
(Here we define $g_{ija} = -\ip{e_i, \t_a e_j}$).

\subsection{One loop form factors}

\subsubsection{Two point one loop form factors}
\label{section_2pt1l} 
First we show one loop two point form factors vanish. There are two possible such form factors, 
\begin{equation} 
	\begin{split} 
		 	\ip{\half \op{tr}(B^2) \mid J_a[1](z_1) \til{J}^b[1](z_2) } \\
	\ip{\half \op{tr}(B^2) \mid M_i[1](z_1) \til{M}^j[1](z_2) } 		 
	\end{split}
\end{equation}
(A form factor with an insertion of $J[2]$, $\til{J}[2]$, etc.\ must vanish by $SU(2)$ invariance). 

Since $J_a[1](z_1)$ has a first order zero at $z_1 = \infty$, the first form factor is determined by the poles at $z_1 = z_2$.   The pole at $z_1 = z_2$ is given by the OPE between $J_a[1]$ and $\til{J}_b[1]$, and is proportional to
\begin{equation} 
	\ip{\half \op{tr}(B^2) \mid :\til{J}^c[1](z_1) \til{J}^d[1](z_1): }. 
\end{equation}
This vanishes, as the normally ordered product is defined to be the contour integral 
\begin{equation} 
	\oint_{\abs{w - z_1}  = \eps} \frac{1}{z_{1} - w}	\ip{\half \op{tr}(B^2) \mid \til{J}^c[1](z_1) \til{J}^d[1](w) }.  
\end{equation}
which vanishes because the correlation function has a second order zero at $z_1 = w$.

Similarly, the operator $M_i[1](z_1)$ is regular at $z_1 = \infty$.  Therefore the form factor 
\begin{equation} 
	\ip{\half \op{tr}(B^2) \mid M_i[1](z_1) \til{M}^j[1](z_2) }\label{eqn:2ptmm} 
\end{equation}
is determined by the pole at $z_1 = z_2$, and the regular term at $z_1 = z_2$.  The one-loop OPE of $M[1]$ with $\til{M}[1]$ is the normally ordered product $:\til{J} \til{J}:$. Therefore, the pole at $z_1 = z_2$ in \eqref{eqn:2ptmm} vanishes, by the previous argument.  There remains the possibility that
\begin{equation} 
	\ip{\half \op{tr}(B^2) \mid M_i[1](z_1) \til{M}^j[1](z_2) } = C \delta_i^j  
\end{equation}
for some constant $C$.  This, however, can be excluded by a symmetry argument: it is not invariant under the $SU(2)$ which rotates the $z$-plane. This is because $M_i[1]$ has spin $0$, and so is a scalar, but $\til{M}^j[1]$ has spin $-1$.

\subsection{Three point one loop form factors}
\label{section_3pt1l} 
We will calculate three-point 
\begin{align} 	
	\ip{\half \op{tr}(B^2) \mid \til{J}^a[1](z_1) J_b[1](z_2) J_c(z_3) } \label{eqn:3pt1} \\
		\ip{\half \op{tr}(B^2) \mid \til{M}^i[1](z_1) M_j[1](z_2) J_a(z_3) } \label{eqn:3pt2} 		\\
\ip{\half \op{tr}(B^2) \mid \til{M}^i[1](z_1) M_j(z_2) J_a[1](z_3) } 		 \label{eqn:3pt3}\\	
	\ip{\half \op{tr}(B^2) \mid \til{J}^a(z_1) J_b[1](z_2) J_c[1](z_3) } \label{eqn:3pt4} \\	
\end{align}

The form factor \eqref{eqn:3pt1} vanishes. To see this, we study the behaviour in the variable $z_3$. It vanishes to second order at $z_3 = \infty$, and has first order poles at $z_3 = z_1$ and $z_3 = z_2$.  The residue at these poles vanishes, because the residue is a two-point form factor discussed above. Therefore the whole form factor vanishes.

The same argument shows that the form factor \eqref{eqn:3pt2} vanishes.  Similarly, for \eqref{eqn:3pt3}, the form factor vanishes by applying this argument to the variable $z_2$. As a function of $z_2$, there is a first-order zero at $\infty$ and first-order poles at $z_1,z_3$, whose residue vanishes because they are two-point functions we have already considered. 

\subsubsection{The form factor \eqref{eqn:3pt4}}
\label{section_3pt1lnonzero}

Let us first compute the form factor \eqref{eqn:3pt4}. As a function of $z_3$, it has a first order zero at $z_3 = \infty$, and a potential first order poles at $z_1$ and second order pole at $z_2$.   The residue of the first order pole at $z_1$ vanishes, because it is a two point function discussed already.  

Recall that we have the OPE,  slightly rewritten from what we had before,
\begin{equation} 
\begin{split} 
		 	J_a[1](z_1) J_b[1](z_2)
				\sim & \frac{[12]}{\ip{12} }   \frac{1}{ (2\pi \i)^2 8  }	   (f_{ea}^c f_{db}^e  + f_{da}^e f_{eb}^c)   :J_c[0,0] \til{J}^d[0,0] :(z_1) \\
				& \frac{[12]}{\ip{12} }   \frac{1}{ (2\pi \i)^2 8  }	   (g_{ka}^i g_{jb}^k  + g_{ja}^k g_{kb}^i)   :M_i[0,0] \til{M}^j[0,0] :(z_1) \\
		&	- \frac{[12]}{\ip{12} }    \frac{1}{ (2 \pi \i)^3 48  }   \op{Tr}_{\g \oplus \pi R}([\t_a,\t_b]\t_c )  \partial_z \til{J}^c (z_1)
			+ \frac{[12] }{\ip{12}^2}  \frac{1}{ (2 \pi \i)^2 24  } \op{Tr}_{\g \oplus \pi R}([\t_a,\t_b] \t_c )  \til{J}^c(z_1) 	
\end{split}
\end{equation}

We can rewrite  $\op{Tr}_{\g \oplus \pi R} ([\t_b,\t_c] \t_d)$ as $f_{bcd} (D_{\g} - D_R)$, where $D_{\g}$, $D_R$ are Dynkin indices for the adjoint representation and the representation $R$. (The dual Coxeter number satisfies $2 \sh^\vee = D_{\g}$). Further, the terms in this OPE which contain matter states $M^i$, $\til{M}^i$ can be ignored, as they are not present in the tree-level form factor. We have already checked that the tree-level correlator of $\til{J}$ with the normally ordered product $:J \til{J}:$ vanishes, so we can ignore this term too. 

We conclude that we have
\begin{equation}
	\begin{split} 	
		\ip{\half \op{tr}(B^2) \mid \til{J}^a(z_1) J_b[1](z_2) J_c[1](z_3) } \sim& f_{bcd} (D_{\g} - D_R)  \frac{[23]}{\ip{23}^2 24 (2 \pi \i)^2  } \ip{\half \op{tr}(B^2) \mid \til{J}^a(z_1)   \til{J}^d ( z_2)  }\\
		&- f_{bcd} (D_{\g} - D_R)  \frac{[23]}{ \ip{23} (2  \pi \i)^3 48  } \ip{\half \op{tr}(B^2) \mid \til{J}^a(z_1) \partial  \til{J}^d ( z_2)  }
	\end{split}
\end{equation}
where $\sim$ means we are taking the form factor up to expressions regular at $z_2 = z_3$.

Inserting the Parke-Taylor amplitude on the right hand side, we get
\begin{equation}
	\begin{split} 	
		\ip{\half \op{tr}(B^2) \mid \til{J}^a(z_1) J_b[1](z_2) J_c[1](z_3) } \sim& -f_{abc} (D_{\g} - D_R) \left( \frac{[23]\ip{12}^2 }{\ip{23}^2 (2\pi \i)^2 24 }  -\frac{[23]}{\ip{23} (2 \pi \i)^3 48 }\partial_{z_2} \ip{12}^2   \right) \\
		&\sim  - f_{abc} (D_{\g} - D_R) \left( \frac{[23]\ip{12}^2 }{\ip{23}^2 (2 \pi \i)^2 24 }  + \frac{[23] \ip{12} }{\ip{23} (2 \pi \i)^2 24  }   \right) \\
		&\sim   f_{abc} \frac{D_{\g} - D_R} { 96 \pi^2   }   \frac{[23](\ip{12}^2 +\ip{12}\ip{23}  )  }{\ip{23}^2 } \\
		&\sim   f_{abc} \frac{D_{\g} - D_R}{ 96 \pi^2 }  \frac{[23] \ip{12} \ip{13} }{\ip{23}^2}.
	\end{split}
\end{equation}
We have written the amplitude up to regular terms in $\ip{23}$, but in fact this is the full expression:
\begin{equation} 
	\ip{\half \op{tr}(B^2) \mid \til{J}^a(z_1) J_b[1](z_2) J_c[1](z_3) }  =  f_{abc} \frac{D_{\g} - D_R}{ 96 \pi^2 }  \frac{[23] \ip{12} \ip{13} }{\ip{23}^2} . \label{eqn:oneloopthreepoint}
\end{equation}
This is the only expression with the correct pole at $z_2 = z_3$, the allowed second order pole at $z_1 = \infty$, and the required first order zero at $z_2,z_3 = \infty$.

\subsection{Vanishing of certain one-loop form factors with normally ordered products}

We will show that
\begin{equation} 
	\ip{\half \op{tr}(B^2) \mid :\til{J}^a J_b[1]: J_c[1](z_3) }  = 0. \label{eqn:vanishing_normalorder2}  
\end{equation}
Indeed the correlator with the insertion of the normal ordered product is obtained as the limit as $z_2 \to z_1$ of \eqref{eqn:oneloopthreepoint}, which is clearly zero.

Next we will show that  
\begin{equation} 
	\ip{\half \op{tr}(B^2) \mid \NO{ \til{M}^i M_j [1] } (z_1) J_a[1](z_3) } = 0. \label{eqn:vanishing_normalorder} 
\end{equation}
We will first compute
\begin{equation} 
	\ip{\half \op{tr}(B^2) \mid \til{M}^i (z_1) M_j [1]  (z_2) J_a[1](z_3) } \label{eqn:3pt5} 
\end{equation}
and then take the limit as $z_2 \to z_1$. 

The form factor \eqref{eqn:3pt5} does not vanish.  The precise coefficient of this form factor will not be important for us.  However, it is not hard to show that
\begin{equation} 
	\ip{\half \op{tr}(B^2) \mid \til{M}^i(z_1) M_j[1](z_2) J_a[1](z_3) } \propto g^i_{ja} \frac{[23] \ip{12}}{\ip{23}} .\label{eqn:oneloopope}	
\end{equation}
To see this, we consider the poles in $z_2$. There is no pole at $z_2 = z_1$, as we have already determined that $\ip{\half \op{tr}(B^2) \mid \til{J}[1] J[1]}$ vanishes.   A similar argument tells us that there is no pole at $z_1 = z_3$, as the residue would be the correlator of $\til{M}[1] M[1]$, which vanishes.  
The only pole is at $z_2 = z_3$.  The OPE of $M[1]$ with $J[1]$ is schematically
\begin{equation} 
	M[1](z_2) J[1](z_3) \sim \frac{1}{\ip{23}} M[2] + \frac{[23]}{\ip{23}} : M \til{J}:  
\end{equation}
The correlator of $M[2]$ and $\til{M}$ vanishes, by $SU(2)$ invariance.  We have seen that 
\begin{equation} 
	 \ip{\half \op{tr}(B^2) \mid \til{M}^i (z_1)  :M_j\til{J}:(z_2) } = g^i_{ja}\ip{12}  
\end{equation}
Therefore
\begin{equation} 
	\ip{\half \op{tr}(B^2) \mid \til{M}^i(z_1) M_j[1](z_2) J_a[1](z_3) } \sim  g^i_{ja} \frac{[23] \ip{12}}{\ip{23}}
\end{equation}
up to an overall constant which is not important. 

At $z_3 = \infty$, the correlator vanishes; at $z_2 = \infty$, it is regular; and at $z_1 = \infty$, it can have a first order pole. Further, as we have seen, the only pole in the correlator is a first order pole at $z_2 = z_3$.  This tells us that \eqref{eqn:oneloopope} is the only global expression with the correct behaviour at $\infty$ and the allowed poles. 

This implies equation \eqref{eqn:vanishing_normalorder}.  

\subsection{$n$-point one loop gluon form factors}
The full correlator is 
\begin{multline} 
	\ip{\half \op{tr}(B^2) \middle| \til{J}^{a_1}(z_1) J_{a_2}[1](z_2) J_{a_3}[1](z_3) J_{a_4}(z_4) \dots J_{a_n}(z_n) } \\= -\frac{1}{n 96 \pi^2 } \sum_{\sigma \in S_n}  \frac{   [23]\ip{12}^2  \ip{13}^2     }{\ip{23}  \ip{\sigma_1 \sigma_2} \ip{\sigma_2 \sigma_3} \dots \ip{\sigma_n \sigma_1} } \op{Tr}_{\g \oplus \Pi R} \left(\t_{a_{\sigma_1}} \dots \t_{a_{\sigma_n}} \right). \label{eqn:fullcorrelator1} 
\end{multline}
	We will check that this is correct by induction, starting with  $n=3$.  Because we are dividing by $n$, we can take the sum over $\sigma$ to be over those permutations $\sigma$ which send $1$ to $1$.  There are two such, which contribute the two different trace orders that contribute to equation \eqref{eqn:oneloopthreepoint}.

To prove the result by induction, we assume it holds for $n-1$. Then we consider the expression as a function of $z_n$. There must be a second order zero at $z_n = \infty$, which the expression \eqref{eqn:fullcorrelator1} has because $z_n$ appears twice in the denominator, in expressions like $\ip{in}\ip{nj}$.   To prove that the expression \eqref{eqn:fullcorrelator1} is the correct form factor, we show that its poles when $z_n = z_i$ are those determined by the OPE of $J_{a_n}(z_n)$ with other insertions.  

When the theory does not have the axion field, the only poles in the OPE of $J_{a_n}(z_n)$ with any other operator are first-order poles, giving the Kac-Moody type OPE. Consider the pole in the expression \eqref{eqn:fullcorrelator1} at $z_n = z_i$.  There are two ways such a pole can arise: if we have a an $\ip{in}$ in the Parke-Taylor denominator and a $\t_{a_n} \t_{a_i}$ in the trace, or if we have an $\ip{ni}$ in the Parke-Taylor denominator at a $\t_{a_i} \t_{a_n}$ in the trace. Clearly these expressions contribute with opposite signs, so that the coefficient of the  pole at $z_n = z_i$ is given by the $n-1$ point correlator with $\t_{a_i}$ replaced by $[\t_{a_{n}}, \t_{a_i}]$.  Therefore, by induction, the expression in equation \eqref{eqn:fullcorrelator1} is the full correlator.

To write this expression in a more symmetric form, we introduce the generating functions $J[\mu](z)$. We have
\begin{multline} 
	\ip{\half \op{tr}(B^2) \middle| \til{J}^{a_1}(z_1) J_{a_2}[\mu](z_2) \dots J_{a_n}[\mu] (z_n) } \\= -\frac{1}{n 96 \pi^2 } \sum_{\sigma \in S_n} \sum_{2 \le i < j \le n}   \frac{   [ij] \ip{1i}^2  \ip{1j}^2     }{\ip{ij}  \ip{\sigma_1 \sigma_2} \ip{\sigma_2 \sigma_3} \dots \ip{\sigma_n \sigma_1} } \op{Tr}_{\g \oplus \Pi R} \left(\t_{a_{\sigma_n}} \dots \t_{a_{\sigma_1}} \right). 
\end{multline}

\subsection{Correction of an error in \cite{Costello:2022upu}.}
In \cite{Costello:2022upu} we stated that this formula holds in the case when the anomaly is cancelled with an axion as well as when it is cancelled with matter.  This is not correct; the formula as stated is only valid in the case when there is no axion. The issue is in the inductive step, which goes from the $3$-point form factor to the $n$-point form factor by considering poles with $J[0]$.  For this to work, we need the OPE of $J[0]$ with $J[k]$ or $\til{J}[k]$ to be given by the classical Kac-Moody formula.  When there is an axion field, this is not correct, and there is an extra possible OPE whereby $J[0]$ and $J[1]$ can produce $E[1]$ \eqref{eqn:EFope}.    This leads to a more complicated expression, and I do not currently know the complete one-loop form factor in this situation.

There are some situations, however, where \eqref{eqn:fullcorrelator1} remains correct even with the axion. Suppose we have $SU(N_c)$ gauge group, and we take matter so $N_f = N_c$.  Then, the twistor space anomaly can be cancelled by an axion.  If we look for the \emph{single-trace} one-loop  form factor, the argument above applies, because the OPE between $J[0]$ and $J[1]$ which produces $E[1]$ only contributes to the  double-trace form factor.   

In this case, the formula \eqref{eqn:fullcorrelator1} yields zero.  This is because when we drop double trace terms, the trace in the adjoint and the trace in the matter representation
\begin{equation} 
	R = N_c F \oplus N_c F^\vee 
\end{equation}
($F$ being the fundamental) cancel.  (This is perhaps easiest to see at the beginning of the induction, where there is an explicit factor of $D_{\g} - D_R$, which vanishes in this case).  

It is easy to see, by considering Feynman diagrams, that an axion exchange can not contribute to the one-loop single-trace computation.  This argument therefore shows that, even without the axion, the one-loop form factor for $\op{tr}(B^2)$ vanishes when $N_f = N_c$.  This is already known, at least when the form factor is integrated to give the one minus amplitude.  It follows algebraically from the fact that this amplitude vanishes for $N=1$ supersymmetric gauge theory. 

\section{Comparison to amplitudes in the literature}
\label{sec:oneloopcomparison}
We will use the conventions in \cite{Bern:2005ji}. In that paper, a factor of $(4 \pi)^2$ is absorbed into the coupling constant, which we restore here.

The known four-point colour-ordered one-loop amplitude for pure gauge theory is \cite{Bern:1991aq, bern2005shell}
\begin{equation} 
	A_4(1^-, 2^+, 3^+, 4^+ )  = \i \frac{ 1 }{96 \pi^2} \frac{ \ip{24} [24]^3 }{ [12] \ip{23} \ip{34} [41] }. 
\end{equation}
Colour-ordered in the conventions of \cite{bern2005shell} means that this expression accompanies $ N_c \op{tr}(\t_1 \t_2 \t_3 \t_4)$, where the trace is taken in the fundamental representation. This amplitude is also multiplied by $N_p$, the number of particles travelling in the loop.  Since we are interested in pure gauge theory, we take $N_p =  2$.  

We would like to compare this to our amplitude, which accompanies the trace in the adjoint representation. For $\mf{sl}_N$, the single-trace part of the trace in the adjoint representation is $2 N_c$ times the trace in the fundamental.  This introduces an additional factor of $2$, cancelling the factor of two coming from $N_p$.  

This implies that the full amplitude, including all colour indices, can be written in terms similar to what we wrote above as
\begin{equation} 
	\sum_{\sigma \in S_3}  A_4(1^-, \sigma_2, \sigma_3, \sigma_4) \op{Tr}_{\g} (\t_1 \t_{\sigma_2} \t_{\sigma_3} \t_{\sigma_4} ) \label{eqn:amplitude4pt} 
\end{equation}
For $\g = \mf{sl}_{N_c}$, trace in the adjoint can be expanded as single and double trace in the fundamental. 

The structure of the Feynman diagrams tells us that matter in a representation $R$ contributes an amplitude whose colour factors are contracted by the trace in $R$.   For supersymmetric gauge theory, with matter in the adjoint representation, the amplitude vanishes. This tells us that the matter contribution must be the same as \eqref{eqn:amplitude4pt}, with the opposite sign and trace taken in $R$.  

This implies that the full amplitude, including all colour indices, can be written in terms similar to what we wrote above as
\begin{equation} 
	\sum_{\sigma \in S_3}  A_4(1^-, \sigma_2, \sigma_3, \sigma_4) \op{Tr}_{\g\oplus \Pi R} (\t_1 \t_{\sigma_2} \t_{\sigma_3} \t_{\sigma_4} ) \label{eqn:amplitude_matter} 
\end{equation}
Set
\begin{equation}
	\begin{split} 	
		\what{A}_4 &=  	\what{A}_4(1^{-1}, 2^+, 3^+, 4^+ )=\\ & -	\frac{1}{96 \pi^2} 	\frac{1}{ \ip{12} \ip{23} \ip{34} \ip{41} } \left(  \frac{ [23] \ip{12}^2 \ip{13}^2   }{\ip{23} } +  \frac{   [34] \ip{13}^2 \ip{14}^2   }{\ip{34} } +  \frac{ [24] \ip{12}^2 \ip{14}^2   }{\ip{24} }       \right) \label{eqn:whata}
\end{split}
\end{equation}

Our amplitude can be written, in the same way, as the symmetrization under the permutation of the $2,3,4$ indices of  
\begin{multline} 
	\sum_{\sigma \in S_3}  \what{A}_4(1^-, \sigma_2, \sigma_3, \sigma_4) \op{Tr}_{\g\oplus \Pi R} (\t_1 \t_{\sigma_2} \t_{\sigma_3} \t_{\sigma_4} ) 
\end{multline}

We have carefully chosen our matter so that the symmetrized trace $\op{Tr}_{\g \oplus \pi R} (t_{(a} \dots \t_{d)} )$ vanishes.  Since we are taking the trace in a real representation, it is automatically dihedrally symmetric.  This means the trace relation is
\begin{equation} 
	\op{Tr}_{\g \oplus \pi R}(\t_1 \t_2 \t_3 \t_4    + \t_1 \t_3 \t_4 \t_2 + \t_1 \t_4 \t_2 \t_3) = 0. \label{eqn:tracerelationagain} 
\end{equation}

Therefore, any  amplitude which is cyclically invariant in the $2,3,4$ indices should be treated as zero.  To prove that our amplitude matches the standard amplitude we need to show that
\begin{equation} 
	\what{A}_4 =   A_4 + \text{ terms cyclically symmetric in } 2,3,4. 
\end{equation}

To see that this is true, we will use conservation of momentum to rewrite $A_4$. We have
\begin{equation} 
	   A_4 =  i \frac{1}{96 \pi^2} \frac{ - [13] [24] s_{24}^2     }{ [12] [13][ 14] \ip{23} \ip{24} \ip{34} }
\end{equation}
where
\begin{equation} 
	  s_{24 } = [ 24 ] \ip{24}. 
\end{equation}

Conservation of momentum implies
\begin{equation} 
	s_{24} = s_{13} = -s_{14} - s_{12}. 
\end{equation}
Conservation of momentum and the Schouten identity implies that 
\begin{equation} 
	\frac{[13][24] }{\ip{13}\ip{24}} 
\end{equation}
is totally symmetric.  So the amplitude is
\begin{equation} 
A_4 =	 i \frac{1}{96 \pi^2} \frac{  [13] [24] s_{13} (s_{12} + s_{14})      }{ [12] [13][ 14] \ip{23} \ip{24} \ip{34} } 
\end{equation}
This can be rewritten as
\begin{equation} 
	 i \frac{1}{96 \pi^2} \frac{  [14] [23] \ip{13} s_{13} s_{12}       }{ [12] [13][ 14] \ip{14} \ip{23}^2  \ip{34} } +   i \frac{1}{96 \pi^2} \frac{  [12] [34] s_{13} s_{14}  \ip{13}      }{ [12] [13][ 14]\ip{12}  \ip{23}  \ip{34}^2 } 
\end{equation}
Cancelling out terms in the numerator and denominator this is
\begin{equation} 
	 i \frac{1}{96 \pi^2} \frac{   [23] \ip{13}^2 \ip{12}      }{\ip{14} \ip{23}^2  \ip{34} } +   i \frac{1}{96 \pi^2} \frac{   [34]  \ip{13}^2 \ip{14} }     {\ip{12}  \ip{23}  \ip{34}^2 } 
\end{equation}
which can be rewritten as
\begin{equation} 
A_4 =	- i \frac{1}{96 \pi^2} \frac{1}{ \ip{12}\ip{23} \ip{34} \ip{41} } \left(  \frac{ [23] \ip{13}^2 \ip{12}^2 } { \ip{23}    } +    \frac{   [34] \ip{13}^2 \ip{14}^2 }     {   \ip{34} }\right) 
\end{equation}
This is very close to $\what{A}_4$ \eqref{eqn:whata}. 
There are only two differences.  First, there is a multiplicative factor of $\i$. This simply  comes different normalization of the Parke-Taylor amplitude, and ultimately, from a factor of $\i$ in the action.

There is also an additive factor of 
\begin{equation} 
	\frac{ [24]\ip{12}^2\ip{14}^2 } { \ip{24} \ip{12} \ip{23} \ip{34} \ip{41} }. 
\end{equation}
It turns out that this expression is cyclically symmetric under the $234$ indices, and so does not contribute.

Let us check this. We have 
\begin{equation} 
	\frac{ [24]\ip{12}^2\ip{14}^2 } { \ip{12} \ip{23} \ip{34} \ip{41} } = -\frac{ [24]  \ip{12} \ip{14} } { \ip{24} \ip{23} \ip{34} } 
\end{equation}
The denominator of the right hand side is antisymmetric under the $S_3$ action. It suffices to show the numerator is also antisymmetric, i.e.  changes sign under any transposition.  

This follows from conservation of momentum. Conservation of momentum gives us the equations
\begin{equation} 
	\sum_{j}	[ij]\ip{jk}   = 0. 
\end{equation}
Applying this to the numerator, we find 
\begin{equation} 
	-	[24] \ip{14} \ip{12} = [23]\ip{13} \ip{12}  
\end{equation}
which tells us the numerator is anti-symmetric under the $34$ transposition. The other transpositions are similar. 

This completes the proof that the $4$-point amplitude found in \cite{Costello:2022upu} matches the standard $4$-point amplitude.  It is rather satisfying that we had to use the trace identity \eqref{eqn:tracerelationagain}, which was introduced to cancel the twistor space anomaly.    

It would be very interesting to show directly that the amplitudes computed by the chiral algebra method match the $5$-point and higher amplitudes, but we do not attempt to do this.

It is worth noting that the match of the overall coefficient of $\frac{1}{96 \pi^2}$ in the amplitudes is not meaningful, at least the way the analysis is performed here\footnote{It is possible in principle to derive this by carefully following how chiral algebra states are matched to states in self-dual gauge theory by the Penrose transform. We attempted this in \cite{Costello:2022upu} but we were off by a factor of two.  For this paper, it is simpler just to use known one-loop QCD results to fix the normalization.}.  If one changes the overall coefficient of the one-loop correction to the chiral algebra OPE, one finds an isomorphic chiral algebra, where the isomorphism rescales $J[n]$ and $\til{J}[n]$ by $c^n$ for a constant $c$. We have normalized the one-loop correction so as to match the coefficient of the one-loop amplitude computed here.

For the two loop amplitude, however, there are no more tunable parameters, so all coefficients are meaningful.

\section{Two-loop form-factors}
\label{sec:twoloop}
In this section we will use the same methods to compute the leading-trace two-loop amplitude. We will focus on the $\mf{sl}_N$ gauge theory with matter in $8 F \oplus 8 F^\vee \oplus \wedge^2 F \oplus \wedge^2 F^\vee$.  

The amplitude we will compute is the integral of the form factor of the operator $\half \op{tr}(B^2)$ in self-dual gauge theory.  Since we are computing the amplitude, and not the form factor, we are free to add to this operator any operator which is a total derivative.  This will only change the form factor by an expression which vanishes when conservation of momentum is imposed.    

What we will compute is the form factor of the operator 
\begin{equation} 
	\half \op{tr}(B^2) + \hbar^2 C \op{tr}(F^2) 
\end{equation}
for some constant $C$.  That is, we have corrected our original operator $\half \op{tr}(B^2)$ by a two-loop counter-term $\hbar^2 C \op{tr}(F^2)$ (the factor of $\hbar^2$ is simply to remind us that this term is added at two loops). 

Since the operator $\op{tr}(F^2)$ is a total derivative, the form factor of this operator vanishes when momentum conservation is imposed.  Indeed, the form factor of $\op{tr}(F^2)$ is  
\begin{equation} 
	\ip{ \half \op{tr}(F^2) \mid J[\til{\lambda}_1](z_1) \dots J[\til{\lambda}_n](z_n) } = \frac{ \left( \sum_{1 \le i < j \le n}  [ij]\ip{ij} \right)^2   }{ \ip{12} \dots \ip{n1} } \label{eqn:F2formfactor} 
\end{equation}
which is manifestly zero when conservation of momentum is imposed. This identity was first proved in \cite{Dixon:2004za}; we will give a simple derivation based on the tree-level chiral algebra. 

To prove this identity, we proceed by induction.  For $n = 2$, 
\begin{equation} 
	\ip{ \half \op{tr}(F^2) \mid J[\til{\lambda}_1](z_1) J[\til{\lambda_2}] (z_2) } = -[12]^2 
\end{equation}
holds by symmetry, up to the normalization of the operator $\op{tr}(F^2)$. The normalization is chosen so that this expression is the parity conjugate of the two-point form factor of $\half \op{tr}(B^2)$.

The expression \eqref{eqn:F2formfactor} is then characterized inductively by the fact that it has the poles at $z_i = z_j$ given by the tree-level OPEs in the chiral algebra.

The two-loop form factor of $\op{tr}(B^2)$ has a term of the form \eqref{eqn:F2formfactor} but with a coefficient I do not know how to compute. The coefficient of $\op{tr}(F^2)$ is chosen simply to remove this term.

\subsection{The two-point form factor}
The computation proceeds in a similar way to the one-loop computation discussed above.  We start by looking at the two-point form factor.

At two loops, all insertions are positive helicity states (we will study only insertions of gluons).  There are two possibilities for the two-loop two-point form factor:
\begin{equation} 
	\begin{split} 
		\ip{\half \op{tr}(B^2) \mid J[1] (z_1) J[3](z_2) } \\
	\ip{\half \op{tr}(B^2) \mid J[2] (z_1) J[2](z_2) }  
	\end{split}
\end{equation}
The first expression vanishes, as the operator $\op{tr}(B^2)$ is $SU(2)_+$ invariant, and the insertions $J[1]$ and $J[3]$ are in the spin $\tfrac{1}{2}$ and spin $\tfrac{3}{2}$ representations of $SU(2)_+$. 

The second expression may be non-zero.  However, we can determine it up to an overall constant by symmetry considerations.  The field $J[2]$ is of spin $0$.  Therefore, under the global conformal transformations of $\CP^1$, the correlation functions transform as scalars.  This immediately implies that that the correlation function is a constant, independent of $z_1,z_2$. 

Symmetry under $SU(2)_+$ then dictates that 
\begin{equation} 
	\ip{\half \op{tr}(B^2) \mid J[2] (z_1) J[2](z_2) }  = 2 C [12]^2 
\end{equation}
for some constant $C$ (the factor of $2$ is included for later convenience). 

We also have the tree-level form factor
\begin{equation} 
	\ip{ \half \op{tr}(F^2) \mid J[2](z_1) J[2](z_2)} = -  [12]^2.   
\end{equation}
We conclude that
\begin{equation} 
	\ip{ \half \op{tr}(B^2) + C  \op{tr}(F^2) \mid J[2] (z_1) J[2](z_2) } = 0.  
\end{equation}
For the rest of the computation, we will compute the form factor of the renormalized expression
\begin{equation} 
	  \half \op{tr}(B^2) + \hbar^2 C \op{tr}(F^2). 
\end{equation}

\subsection{Three point form factors}
\label{sec:threepoint2loop}
In the three point form factor, any expression with an insertion of a $J[3]$ or $J[4]$ necessarily vanishes, because it can not be symmetric under $SU(2)_+$.  

The remaining possibilities are
\begin{equation} 
	\begin{split}
\ip{ \half \op{tr}(B^2) + \hbar^2 C \op{tr}(F^2) \mid J[2](z_1)  J[2](z_2) J[0](z_3) } \\
	\ip{ \half \op{tr}(B^2) + \hbar^2 C \op{tr}(F^2) \mid J[2](z_1)  J[1](z_2) J[1](z_3) }	 
	\end{split}
\end{equation}
The first expression is easily seen to vanish. Indeed, as a function of $z_3$, it has a second order zero at $\infty$, and no poles because we already know that the two-point form factor vanishes.  It must therefore be zero.

Let us show that the second expression also vanishes. We first note that symmetry under $SU(2)_+$ implies the expression is divisible by $[12][13]$.  Further, symmetry under $SU(2)_-$ implies the expression is proportional to
\begin{equation} 
	\ip{ \half \op{tr}(B^2) + \hbar^2 C \op{tr}(F^2) \mid J[2](z_1)  J[1](z_2) J[1](z_3) }	\propto \frac{[12][13]} {\ip{23} } F(z_1,z_2,z_3) \label{eqn:twoloop3pt} 
\end{equation}
where $F(z_1,z_2,z_3)$ is a meromorphic function on $\CP^1 \times \CP^1 \times \CP^1$ which is invariant under the global conformal transformations $SL_2(\C)$.  Since the conformal group acts triply transitively on $\CP^1$, all such functions $F$ are constant.  

Finally, we note that the form factor \eqref{eqn:twoloop3pt} can have no poles at $z_2 = z_3$.  Therefore it must be zero.  To see this, the poles at $z_2 = z_3$ must be from the OPEs between $J[1]$ and $J[1]$. The tree-level OPE produces $J[2]$, and we know that the correlation function of $J[2] J[2]$ vanishes. We also know that the one-loop form factor with an insertion of $J[2]$ vanishes, so that the result of the one-loop OPE in \eqref{eqn:twoloop3pt} vanishes. 

\subsection{The four point form factor}
For the four-point form factor, expressions involving an insertion of $J[0](z_k)$ automatically vanish. Since the three point form factors vanish, such an expression has no poles in $z_k$, and also has a second-order zero at $z_k = \infty$.

The only four-point form factor not of this form is
\begin{equation} 
	\ip{\half \op{tr}(B^2) + C \hbar^2 \op{tr}(F^2) \mid J_{a_1}[1](z_1) J_{a_2}[1](z_2) J_{a_3}[1](z_3) J_{a_4}[1](z_4) }. 
\end{equation}
Let us consider the poles that can arise as a function of $z_2$.  Because we are considering the trace-ordered single trace form factor, poles when $z_2 = z_4$ do not contribute. 

Let us first consider the pole at $z_2 = z_1$.  We have the one-loop OPE
\begin{equation}
\begin{split} 
		 	J_a[1](z_1) J_b[1](z_2)
				\sim & \frac{[12]}{\ip{12} }   \frac{1}{ (2\pi \i)^2 8  }	   (f_{ea}^c f_{db}^e  + f_{da}^e f_{eb}^c)   :J_c \til{J}^d :(z_1)  \\
				& \frac{[12]}{\ip{12} }   \frac{1}{ (2\pi \i)^2 8  }	   (g_{ka}^i g_{jb}^k  + g_{ja}^k g_{kb}^i)   :M_i[0,0] \til{M}^j[0,0] :(z_1) \\
		&	- \frac{[12]}{\ip{12} }    \frac{1}{ (2 \pi \i)^2 48  }(D_{\g} - D_R)  f_{abc}  \frac{1}{2 \pi \i} \partial_z \til{J}^c (z_1)  \\
	&		+ \frac{[12] }{\ip{12}^2}  \frac{1}{ (2 \pi \i)^2 24  }  (D_{\g} - D_R)  f_{abc}  \til{J}^c(z_1) 
\end{split}\label{eqn:ope4}
\end{equation}

Let consider the correlator of these with $J[1](z_3) J[1](z_4)$.  The first thing we have to show is that 
\begin{proposition}
	For any of the terms on the right hand side  of \eqref{eqn:ope4}, the correlator with $J[1](z_3) J[1](z_4)$ is regular at $z_3 = z_1$, $z_3 = z_2$, $z_4 = z_1$, $z_4 = z_2$. 
\end{proposition}
\begin{proof}
	We first note that for these expressions, we can compute the form factor of the operator $\half \op{tr}(B^2)$, and not the renormalized operator where we add $ \hbar^2 C \op{tr}(F^2)$.  This is because the form factors of $\op{tr}(F^2)$ do not involve any $\til{J}$ insertions.   

	To check the statement,  we note that a pole at $z_3 = z_1$ or $z_3 = z_2$ would come only from the tree-level OPE and would therefore give a correlator which is one of the following: 
\begin{equation} 
	\begin{split} 
		\ip{ \half \op{tr}(B^2) \mid : J[1] \til{J} : (z_1)J[1] (z_3) J[1](z_4) } \\
		\ip{ \half \op{tr}(B^2)\mid : J\til{J}[1] : (z_1) J[1] (z_3) J[1](z_4) } \\	
		\ip{ \half \op{tr}(B^2) \mid : M\til{M}[1] : (z_1)J[1] (z_3)  J[1](z_4) } \\
		\ip{ \half \op{tr}(B^2) \mid : M[1]\til{M} : (z_1) J[1] (z_3) J[1](z_4) } \\
		\ip{ \half \op{tr}(B^2) \mid  \partial \til{J}[1]  (z_1) J[1] (z_3) J[1](z_4) } \\
		\ip{ \half \op{tr}(B^2) \mid  \til{J}[1]  (z_1) J[1](z_3) J[1](z_4) } 
	\end{split}
\end{equation}
	All of these are zero, by the calculations in section \ref{section_3pt1l}, section \ref{section_2pt1l} and equations \eqref{eqn:vanishing_normalorder2}, \eqref{eqn:vanishing_normalorder}. 

	We conclude that the terms in the four-point function which have a pole at $z_1 = z_2$, have additional poles only at   $z_3 = z_4$.  This pole must be given by the one-loop OPE of $J[1](z_3)$ with $J[1](z_4)$, as the tree-level OPE yields $J[2]$ and  one-loop correlators with a $J[2]$ insertion vanish.  

\end{proof}

\subsection{Vanishing of certain cross terms}
This tells us that to compute the four-point function of $J[1](z_1)$, $J[1](z_2)$, $J[1](z_3)$, $J[1](z_4)$, we need to compute the two-point function of the OPE of $J[1](z_1) J[1](z_2)$ with the OPE of $J[1](z_3) J[1](z_4)$.  Only the one-loop OPE will be relevant in each case, because one and two-loop correlators with a $J[2]$ insertion vanish.  

We are therefore left with computing the the correlator between one of the terms in the expansion \eqref{eqn:ope4} placed at $z_1$, with another term in the expansion placed at (say) $z_3$.  
However, several terms vanish.  We have seen in section \ref{section:vanishing_oneNO} that 
\begin{equation} 
	\begin{split} 
		\ip{ \half \op{tr}(B^2) \mid \til{J}^a(z_1) :\til{J}^b  J_c: (z_2) } &= 0\\
		\ip{ \half \op{tr}(B^2) \mid \til{J}_a(z_1) :\til{M}_iM_j + \til{M}_j M_i : (z_2) } &=0  
	\end{split}
\end{equation}
From this we see that, in the correlator
\begin{equation} 
	\ip{ \half \op{tr}(B^2)  + \hbar^2 C \op{tr}(F^2)   \mid J[1] (z_1) J[1](z_2) J[1](z_3) J[1](z_4) } 
\end{equation}
where are considering poles at $z_2 = z_1$ and $z_4 = z_3$, the only terms that contribute are those tree-level correlators which have either two normally ordered products, or no normally ordered products.

\subsection{Terms with no normally ordered products}
The terms with no normally ordered products are, from the OPE \eqref{eqn:ope4}, 
\begin{equation}
	\begin{split}  
		\frac{[12][34] }{\ip{12}^2 \ip{34}^2 }(D_{\g} - D_R)^2   f_{a_1 a_2 c}  f_{a_3 a_4 d}  \frac{1}{(4 \pi)^4 \cdot (12)^2 } \\	\ip{\half \op{tr}(B^2) \mid 	\left( - \ip{12} \frac{1}{2 \pi \i} \partial_z \til{J}^c (z_1)  + 2      \til{J}^c(z_1)\right)  \left( - \ip{34} \frac{1}{2 \pi \i} \partial_z \til{J}^d (z_3)  + 2    \til{J}^d(z_3)\right) }
	\end{split}
\end{equation}
Using the fact that the two-point Parke-Taylor amplitude is $-\ip{12}^2 \delta^{cd}$ this becomes\footnote{There was an overall sign error in this formula in version 1, pointed out by Anthony Morales.}
\begin{equation} 
	-\frac{[12][34] }{\ip{12}^2 \ip{34}^2 }(D_{\g} - D_R)^2   f_{a_1 a_2 c}  f_{a_3 a_4 c}  \frac{1}{(4 \pi)^4 \cdot (12)^2 }  \left( - \ip{12} \frac{1}{2 \pi \i} \partial_{z_1}  + 2    \right) \left( - \ip{34} \frac{1}{2 \pi \i} \partial_{z_3} + 2\right) \ip{13}^2   
\end{equation}
which, when we expand out, becomes
\begin{equation} 
	\begin{split} 	
		-\frac{[12][34] }{\ip{12}^2 \ip{34}^2 }(D_{\g} - D_R)^2   f_{a_1 a_2 c}  f_{a_3 a_4 c}  \frac{1}{(4 \pi)^4 \cdot (12)^2 }  \left( 4 \ip{13}^2 -4 \ip{12} \ip{13} + 4 \ip{13} \ip{34} - 2 \ip{12} \ip{34} \right).
	\end{split}
\end{equation}
Since
\begin{equation} 
	\ip{13} - \ip{12} + \ip{34} = \ip{24} 
\end{equation}
this becomes
\begin{equation} 	
	-\frac{[12][34] }{\ip{12}^2 \ip{34}^2 }(D_{\g} - D_R)^2   f_{a_1 a_2 c}  f_{a_3 a_4 c}  \frac{1}{(4 \pi)^4 \cdot (12)^2 }  ( 4 \ip{13} \ip{24} - 2 \ip{12} \ip{34} ). 	
\end{equation}
It is convenient to use the Schouten identity to rewrite this slightly as 
\begin{equation} 
	- \frac{[12][34] }{\ip{12}^2 \ip{34}^2 }(D_{\g} - D_R)^2   f_{a_1 a_2 c}  f_{a_3 a_4 c}  \frac{1}{(4 \pi)^4 \cdot (12)^2 }  ( 2 \ip{13} \ip{24} + 2 \ip{14} \ip{23} ). 	
\end{equation}

There is one more simplification that we can make.  Recall that our Lie algebra is $\g = \mf{sl}(N)$ and the representation is $R = \C^{8} \otimes F \oplus \C^8 \otimes F^\vee \oplus \wedge^2 F \oplus \wedge^2 F^\vee$. 

The Dynkin index of the adjoint is $D_{\g} = 2N$. 
The Dynkin index of the fundamental is $1$, and that of $\wedge^2 F$ is $N-2$.  Therefore
\begin{equation} 
	D_{\g} - D_{R} =  -12  
\end{equation}
This gives a factor of $ (12)^2$, cancelling all powers of $12$ and leaving us with 
\begin{equation} 
	-	f_{a_1 a_2 c}    f_{ca_3 a_4 } \frac{2}{  (4 \pi)^4   }  \frac{[12] [34]( \ip{13} \ip{24}  + \ip{14} \ip{23} )}{\ip{12}^2 \ip{34}^2}. 
\end{equation}
The factor of $1/(4 \pi)^4$ appears in any two-loop amplitude and is often absorbed into the coupling constant. 

We also need to take account of the terms which arise from poles at $z_1 = z_3$ and $z_1 = z_4$. These are related by symmetry, and are
\begin{equation}
	\begin{split} 	
	-	f_{a_1 a_2 c}    f_{ca_3 a_4 } \frac{2}{  (4 \pi)^4   }  \frac{[12] [34]( \ip{13} \ip{24}  + \ip{14} \ip{23} )}{\ip{12}^2 \ip{34}^2} \\
- f_{a_1 a_3 c}    f_{ca_2 a_4 } \frac{2}{  (4 \pi)^4   }  \frac{[13] [24]( \ip{12} \ip{34}  + \ip{14} \ip{32} )}{\ip{13}^2 \ip{24}^2} \\
- f_{a_1 a_4 c}    f_{ca_2 a_3 } \frac{2}{  (4 \pi)^4   }  \frac{[14] [23]( \ip{13} \ip{42}  + \ip{12} \ip{43} )}{\ip{14}^2 \ip{23}^2}
	\end{split}
\end{equation}
We note that this expression has a zero at $z_i = \infty$ for $i = 1,2,3,4$.  This means it has the correct behaviour to be an amplitude. 

As a final step, we will extract the trace-ordered part of this, i.e.\ the coefficient of $\op{tr}(\t_{a_1} \t_{a_2} \t_{a_3} \t_{a_4} )$.  This is \footnote{There was a relative sign error in this equation in version 1, corrected by Anthony Morales.} 
\begin{equation}
	\begin{split} 
	-	 \frac{2}{  (4 \pi)^4   }  \frac{[12] [34]( \ip{13} \ip{24}  + \ip{14} \ip{23} )}{\ip{12}^2 \ip{34}^2} \\
		+ \frac{2}{  (4 \pi)^4   }  \frac{[14] [23]( \ip{13} \ip{42}  + \ip{12} \ip{43} )}{\ip{14}^2 \ip{23}^2}
\end{split}\label{eqn:amplitude_nono_final}
\end{equation} 
\subsection{The contribution of terms with two normally ordered products}\label{sec:2no}
Recall we are computing the four-point two-loop chiral algebra correlator 
\begin{equation} 
	\ip{\half\op{tr}(B^2)  + \hbar^2 C \op{tr}(F^2)   \mid  J_{a_1}[1](z_1) J_{a_2}[1] (z_2) J_{a_3}[1](z_3) J_{a_4}[1](z_4) }. 
\end{equation}
The remaining step is to compute the first-order pole at $z_2 = z_1$ and at $z_3 = z_4$ from the OPEs whose output is a normally-ordered product $:\til{J} J:$ or $:\til{M} M:$.  This is a rather cumbersome calculation.   Let us introduce the notation 
\begin{equation} 
	\begin{split} 
		T_{ab}^{cd} &=  f_{ea}^c f_{be}^d  + f_{ae}^d f_{eb}^c \\
	S_{abj}^i &= g_{ka}^i g_{jb}^k  + g_{ja}^k g_{kb}^i
	\end{split}
\end{equation}
The polar part of the correlator we are computing has an overall factor of
\begin{equation} 
	 	\frac{[12][34]}{\ip{12}\ip{34} }   \frac{1}{ (2\pi \i)^4 2^6  }	 
\end{equation}
This is accompanied by a rather complicated contraction of the gauge theory indices, given by 
\begin{equation} 
	\begin{split} 	
		R_{a_1 a_2 a_3 a_4} = \ip{ \half \op{tr}(B^2) \mid  T_{a_1 a_2}^{b_1 b_2}  :J_{b_1} \til{J}_{b_2} :(z_1) T_{a_3 a_4}^{b_3 b_4}  :J_{b_3} \til{J}_{b_4} :(z_3)   } \\
+ \ip{ \half \op{tr}(B^2) \mid  T_{a_1 a_2}^{b_1 b_2}  :J_{b_1} \til{J}_{b_2} :(z_1) S_{a_3 a_4 i_4}^{i_3}    :M_{i_3}\til{M}^{i_4}   :(z_3)    } \\
+ \ip{ \half \op{tr}(B^2) \mid  S_{a_1 a_2 i_2}^{i_1}    :M_{i_1}\til{M}^{i_2} :(z_1) T_{a_3 a_4}^{b_3 b_4}  :J_{b_3} \til{J}_{b_4} :(z_3)  }\\ 
+ \ip{ \half \op{tr}(B^2) \mid S_{a_1 a_2 i_2}^{i_1}    :M_{i_1}\til{M}^{i_2} :(z_1) S_{a_3 a_4 i_4}^{i_3}    :M_{i_3}\til{M}^{i_4}   :(z_3)   } 	
	\end{split}\label{eqn:R_definition}
\end{equation}

We have already computed all these correlators in equations \eqref{eqn:twonormalorder1}, \eqref{eqn:twonormalorder2}, \eqref{eqn:twonormalorder3} and found that they are independent of the position of the operators and given by the following expressions:
\begin{equation} 
	\begin{split} 
		 \ip{\half \op{tr}(B^2) \mid :\til{J}_{(a} J_{b)}(z) :\til{J}_{(c} J_{d)} (w)} &=   -2\op{tr}(\t_{(a} \t_{b)} \t_{(c} \t_{d)} ) + \op{tr}(\t_{a} \t_{c} \t_{b} \t_{d}) + \op{tr}(\t_{a} \t_{d} \t_{b} \t_{c} )  	\\
		\ip{\half \op{tr}(B^2) \mid :\til{J}_{(a} J_{b)}(z) :\til{M}_{(i} M_{j)} (w)}  &= \half\left( g^j_{ka} g^k_{ib} + g^j_{kb} g^k_{ia}  \right)  	\\ 
		\ip{\half \op{tr}(B^2) \mid :\til{M}_{(i} M_{j)}(z) :\til{M}_{(k} M_{l)} (w)} &= -  \half \left(  g_{i l a} g_{k j a} +    g_{j l a} g_{k i a} 	 \right).
	\end{split}
\end{equation}
This gives us an explicit, but complicated, expression for the tensor $R_{a_1 a_2 a_3 a_4}$.  The contribution of two normally ordered products in the OPE to the pole at $z_1 = z_2$ and $z_3 = z_4$ is then  
\begin{equation} 
	\frac{[12][34]}{\ip{12}\ip{34} }   \frac{1}{ (2\pi \i)^4 2^6  }	  R_{a_1a_2 a_3 a_4} 
\end{equation}
for $R_{a_1 a_2 a_3 a_4}$ as in equation \eqref{eqn:R_definition}. This tensor is $\mf{sl}(N)$ invariant and symmetric under the permutations $(12)$, $(34)$,  $(13)(24)$, and $(14)(23)$.   

It is important to also include the contributions of the poles at $z_1 = z_3$, $z_2 = z_4$ and $z_1 = z_4$, $z_2 = z_3$, giving the expression
\begin{equation} 
	\begin{split} 
	\frac{[12][34]}{\ip{12}\ip{34} }   \frac{1}{ (2\pi \i)^4 2^6  }	  R_{a_1a_2 a_3 a_4} 	\\
+ \frac{[13][24]}{\ip{13}\ip{24} }   \frac{1}{ (2\pi \i)^4 2^6  }	  R_{a_1a_3 a_2 a_4} 	\\
+\frac{[14][23]}{\ip{14}\ip{23} }   \frac{1}{ (2\pi \i)^4 2^6  }	  R_{a_1a_4 a_2 a_3} 	
	\end{split}\label{eqn:normallyordered_amplitude}	\end{equation}

In appendix \ref{sec:computingalgebraicfactors}, the single-trace components of the tensor $R_{a_1 a_2 a_3 a_4}$ are computed. We find that the contributions can be divided into those from the gauge sector and the various matter representations.   The contributions are the following.
\begin{enumerate} 
	\item The pure gauge sector contributes
		\begin{equation} 
			R^{gauge}_{a_1 a_2 a_3 a_4} = 		 -2N^2 \op{tr} (\t_{a_1} \t_{a_2} \t_{a_3} \t_{a_4} )  +  8\op{tr}( \t_{a_1} \t_{a_3} \t_{a_2} \t_{a_4} ) 		- 8  \op{tr}( \t_{a_1} \t_{a_2} \t_{a_3} \t_{a_4} ) 		+ \text{ permutations }
		\end{equation}
	\item $N_F$ copies of the fundamental representation contributes
		\begin{equation} 
			R^{F}_{a_1 a_2 a_3 a_4} =	( 4 N_F  N  + 2 N_F N^{-1}  )	  \op{tr}_F(\t_{a_1} \dots \t_{a_4} ) + \text{ permutations } 
		\end{equation}
	\item Matter in $\wedge^2 F \oplus \wedge^2 F^\vee$ contributes 
		\begin{equation}
			\begin{split} 	
				R^{\wedge^2 F}_{a_1 a_2 a_3 a_4} =	 	\left( 2 N^2 -8 N - 8 -  48 N^{-1}     \right)   \op{tr}_F(\t_{a_1} \t_{a_2} \t_{a_3} \t_{a_4} ) \\ 
			-   ( 16 + 16 N^{-1}     ) \op{tr}_F( \t_{a_1} \t_{a_3} \t_{a_2} \t_{a_4} ) + \text{ permutations }
\end{split}
		\end{equation}
	\item Adjoint matter contributes
		\begin{equation}
			R_{a_1 a_2 a_3 a_4}^{a} = 		2  N^2 \op{tr}_F( \t_{a_1} \t_{a_2} \t_{a_3}\t_{a_4} ) + 8N_a  \op{tr}_F(\t_{a_1} \t_{a_2} \t_{a_3} \t_{a_4} ) -  8N_a  \op{tr}_F (\t_{a_1} \t_{a_3} \t_{a_2} \t_{a_4} ) + \text{ permutations }.
		\end{equation}
\end{enumerate}
In each case, we add to the expression displayed the sum over the non-trivial permutations in the subgroup $\Z/2 \times \Z/2$ of $S_4$ generated by $(12)$ and $(34)$.

As a cross check, we note that the contribution from adjoint matter cancels with the contribution from pure gauge.  This means that form factor vanishes for $N=1$ supersymmetric gauge theory.  This is to be expected, because the one-loop all $+$ form factor vanishes in this case, and the collinear singularities of the two loop amplitude are expressed in terms of the one-loop form factor.

 We can compute the term in the correlator coming from normally ordered products using the expressions for $R_{a_1 a_2 a_3 a_4}$ and  equation \eqref{eqn:normallyordered_amplitude}.  We are interested in the trace-ordered term, that is, the coefficient of $\op{tr}_F(\t_{a_1} \t_{a_2} \t_{a_3} \t_{a_4})$. 

To compute the trace-ordered term, we note that any term like 
	\begin{equation} 
		f(N) \op{tr}(\t_{a_1} \t_{a_2} \t_{a_3} \t_{a_4})  + \text{ permutations } 
	\end{equation}
	in $R_{a_1 a_2 a_3 a_4}$  will yield
	\begin{equation} 
		\frac{f(N)}{2^2 (4 \pi)^4 } \left( \frac{[12] [34]  }{\ip{12} \ip{34} } +  \frac{[41] [23]  }{\ip{41} \ip{23} } \right)  
	\end{equation}
	in the trace-ordered amplitude. (Here $f$ is a function of $N$).

Any term like 
\begin{equation} 
	g(N) \op{tr}(\t_{a_1} \t_{a_3} \t_{a_2} \t_{a_4})  + \text{ permutations } 
	\end{equation}
	in $R_{a_1 a_2 a_3 a_4}$  will yield
	\begin{equation} 
		\frac{2 g(N)}{2^2 (4 \pi)^4} \frac{[13][24] }{\ip{13} \ip{24}}. \label{eqn:1324}  
	\end{equation}
	The extra factor of two in the numerator of \eqref{eqn:1324} arises for the following reason.  Only  $R_{a_1 a_3 a_2 a_4}$ can contribute a trace-ordered expression $\op{tr}(\t_{a_1} \t_{a_2} \t_{a_3} \t_{a_4})$. $R_{a_1 a_3 a_2 a_4}$ is obtained by applying the permutation $(23)$ to $R_{a_1 a_2 a_3 a_4}$, which has four terms, given by $\op{tr}(\t_{a_1} \t_{a_3} \t_{a_2} \t_{a_4})$ and its permutations under the $\Z/2 \times \Z/2$ generated by $(12)$ and $(34)$.   Two of these four terms become trace-ordered when the permutation $(23)$ is applied. 

We conclude that
 \begin{enumerate} 	
	\item The pure gauge sector contributes the trace-ordered amplitude
		\begin{equation}
			\begin{split} 	
				- (2 N^2 + 8)	\left( \frac{1}{2^2 (4 \pi)^4 }	\frac{[12][34]} { \ip{12} \ip{34} } + 	\frac{1}{2^2 (4 \pi)^4 }	\frac{[41][23]} { \ip{41} \ip{23} } \right)	  + 
			16	\frac{1}{2^2 (4 \pi)^4 }	\frac{[13][24]} { \ip{13}\ip{24} } 	
\end{split}
		\end{equation}
	\item $N_F$ copies of the fundamental and anti-fundamental representation contribute the trace-ordered amplitude
		\begin{equation} 
			( 4 N_F  N  + 2 N_F N^{-1}  ) \left(	\frac{1}{2^2 (4 \pi)^4 }	\frac{[12][34]} { \ip{12} \ip{34} }   + 	\frac{1}{2^2 (4 \pi)^4 }	\frac{[41][23]} { \ip{41} \ip{23} } \right) 
		\end{equation}
	\item Matter in $\wedge^2 F \oplus \wedge^2 F^\vee$ contribute the trace-ordered amplitude
		\begin{equation} 
			 	\left( 2 N^2 -8 N - 8 -  48 N^{-1}     \right)   \left(	\frac{1}{2^2 (4 \pi)^4 }	\frac{[12][34]} { \ip{12} \ip{34} }   + 	\frac{1}{2^2 (4 \pi)^4 }	\frac{[41][23]} { \ip{41} \ip{23} } \right)
			-   ( 32 + 32 N^{-1}     )  \frac{1}{2^2 (4 \pi)^4} \frac{[13][24]}{ \ip{13}{24}} 
		\end{equation}
 \end{enumerate}
	We are computing the amplitude for the anomaly-free theory with matter in $8 F \oplus 8 F^\vee \oplus \wedge^2 F \oplus \wedge^2 F^\vee$. This gives 
\begin{equation}
	\begin{split} 	
		\left(   24 N - 16 - 32 N^{-1}   \right) 	\left( \frac{1}{2^2 (4 \pi)^4 }	\frac{[12][34]} { \ip{12} \ip{34} } + 	\frac{1}{2^2 (4 \pi)^4 }	\frac{[41][23]} { \ip{41} \ip{23} } \right)	\\ 
		- \left(    16 + 32 N^{-1}  \right)  \frac{1}{2^2 (4 \pi)^4} \frac{[13][24]}{ \ip{13}{24}}
\end{split} \label{eqn:amplitude_no_final} 
\end{equation}

\subsection{The complete four-point single-trace trace-ordered amplitude}
So far we have separately computed the contributions from the OPEs which yield normally ordered products, in \eqref{eqn:amplitude_no_final}, and those which do not, in equation \eqref{eqn:amplitude_nono_final}. Adding these together (and also cancelling some factors of $4$) yields 
\begin{equation} 
	\begin{split} 
		\mc{A}^{TO}(1^+,2^+,3^+,4^+) 		= & 
		\left(   6 N - 4 - 8 N^{-1}   \right) 	\left( \frac{1}{ (4 \pi)^4 }	\frac{[12][34]} { \ip{12} \ip{34} } + 	\frac{1}{ (4 \pi)^4 }	\frac{[41][23]} { \ip{41} \ip{23} } \right)	\\ 
		&	- \left( 4 + 8 N^{-1}  \right)  \frac{1}{ (4 \pi)^4} \frac{[13][24]}{ \ip{13}{24}} \\ 
		& - \frac{2}{  (4 \pi)^4   }  \frac{[12] [34]( \ip{13} \ip{24}  + \ip{14} \ip{23} )}{\ip{12}^2 \ip{34}^2} \\
		&	+ \frac{2}{  (4 \pi)^4   }  \frac{[14] [23]( \ip{13} \ip{42}  + \ip{12} \ip{43} )}{\ip{14}^2 \ip{23}^2}
\end{split}
\end{equation}

\subsection{$n$-point single trace two loop amplitudes}
Now we can compute the $n$-point single trace amplitude inductively from the four-point amplitude.  

We first study the chiral algebra correlator
\begin{equation} 
	\ip{\half \op{tr}(B^2) + \hbar^2 C \op{tr}(F^2) \mid J_{a_1}[1](z_1) J_{a_2}[1](z_2) J_{a_3}[1](z_3) J_{a_4}[1](z_4) J_{z_5}(z_5) \dots J_{a_n}(z_n)}. \label{eqn:2loopcorrelator} 
\end{equation}

To describe this, we use the following notation. An element $\sigma \in S_n$ induces a total order $\sigma(1),\sigma(2) \dots$ on the set $\{1,\dots,n\}$.    It therefore induces a cyclic order on any subset of $\{1,\dots,n\}$, and in particular on the subset $1,2,3,4$.   We denote the elements $1,2,3,4$ as they occur in this order by $1_{\sigma}, 2_{\sigma}, 3_{\sigma}, 4_{\sigma}$. 

For example, if $n = 5$ and $\sigma = (13542)$, then the ordering induced by $\sigma$ is $31524$, so that  then $1_{\sigma} = 3$, $2_{\sigma} = 1$, $3_{\sigma} = 2$, $4_{\sigma} = 4$, as this is the order in which these elements appear in the list $31524$. 
Then, 
\begin{proposition}
	The correlator \eqref{eqn:2loopcorrelator} is 
	\begin{equation} 
		\frac{1}{n} \sum_{\sigma \in S_n} \mc{A}^{TO}_{2}(1_{\sigma}^+,2_{\sigma}^+,3_{\sigma}^+,4_{\sigma}^+) \frac{ \ip{ 1_{\sigma}, 2_{\sigma} } \ip{ 2_{\sigma}, 3_{\sigma} } \ip{ 3_{\sigma}, 4_{\sigma} } \ip{ 4_{\sigma}, 1_{\sigma} }} { \ip{\sigma(1) \sigma(2)}, \dots, \ip{\sigma(n) \sigma(1)}} \op{tr}(\t_{\sigma(1)} \dots \t_{\sigma(n)} ) 
\label{eqn:completeamplitude1}  
	\end{equation}
\end{proposition}
\begin{proof}
	When $n = 4$,  this becomes the statement that $\mc{A}_{2}^{TO}$ is the trace-ordered four-point amplitude. Indeed, in this case, $\sigma (1) = 1_{\sigma}$, $\sigma(2) = 2_{\sigma}$, and so on.  Thus the terms with angle brackets in the numerator and denominator of \eqref{eqn:completeamplitude1} cancel.   Further, $\mc{A}^{TO}_2$ is invariant under cyclic permutations of $1,\dots,4$.  Thus, we can rewrite \eqref{eqn:completeamplitude1} in the case $n=4$ as  
	\begin{equation} 
		\frac{1}{4} \sum_{\sigma \in S_4} \mc{A}^{TO}_{2}(\sigma(1)^+, \sigma(2)^+, \sigma(3)^+,\sigma(4)^+ )\op{tr}(\ \t_{\sigma(1)} \dots \t_{\sigma(4)} ).
	\end{equation}
This is the standard expression for the full amplitude in terms of the trace-ordered partial amplitude.

To prove the formula for $n > 4$, let us proceed by induction by looking at the poles in the variable  $z_n$.  We need to check that this expression has the poles dictated by the OPE of $J_{a_n}(z_n)$ with the other insertions. These OPEs do not receive quantum corrections. The OPE of $J_{a_n}(z_n)$ with the operator $J_{a_k}(z_k)$ (for $k \not \in \{1,\dots,4\}$) is 
	\begin{equation} 
		f_{a_k a_n}^c J_c(z_k) \frac{1}{\ip{kn} }. 
	\end{equation}
	The OPE with $J_{a_k}(z_k)[1]$ (for $k \in \{1,\dots,4\})$ is
	\begin{equation} 
		f_{a_k a_n}^c J_c(z_k)[1] \frac{1}{\ip{kn} }. 
	\end{equation}
	The poles at $n = k$ in the expression \eqref{eqn:completeamplitude1} arise when the indices $k,n$ are adjacent in the trace.  There are two ways this can happen, with opposite signs, leading to the commutator $f_{a_k a_n}^c$.  The residue at the pole is precisely the same formula with $n-1$ insertions.  By induction, we find that the expressions \eqref{eqn:2loopcorrelator} and \eqref{eqn:completeamplitude1} are the same. 
\end{proof}

Now let us rewrite this expression where the symmetry between the $n$ incoming states has been restored. 
\begin{proposition} 
	The complete two-loop, single-trace, all $+$ amplitude for $SU(N)$ gauge theory with matter in $8 F \oplus 8 F^\vee \oplus \wedge^2 F \oplus \wedge^2 F^\vee$  is 
	\begin{equation} 
		\mc{A}_2 (1^+, \dots, n^+ ) = \frac{1}{n} \sum_{\sigma \in S_n}  \sum_{1 \le i < j < k < l \le n}  \mc{A}_2^{TO}(\sigma(i)^+,\sigma(j)^+,\sigma(k)^+,\sigma(l)^+) \frac{ \ip{\sigma(i)\sigma(j)} \ip{\sigma(j) \sigma(k)} \ip{\sigma(k)\sigma(l)} \ip{\sigma(l)\sigma(i)} } { \ip{\sigma(1)\sigma(2)} \dots \ip{\sigma(n) \sigma(1)} } \label{eqn:completeamplitude_symmetric} 
	\end{equation}
\end{proposition}
\begin{proof}
	The full amplitude is the unique expression which is totally symmetric, and which, when we drop any terms involving $[ij]$ where one of $i,j$ is greater than $4$, reduces to the expression \eqref{eqn:completeamplitude1}. 

	Clearly equation \eqref{eqn:completeamplitude_symmetric} is symmetric. We need to check that if we specialize to the terms only including square brackets $[ij]$ when $1 \le i,j \le 4$, we find the previous expression \eqref{eqn:completeamplitude1}.   To see this, note that (because of the factor of $1/n$ in the front) we can view this expression as being a sum over cyclic orders on the set $\{1,\dots,n\}$, where $\sigma$ gives a cyclic ordering by $\sigma(1),\sigma(2),\dots,\sigma(n)$.  In the sum in equation \eqref{eqn:completeamplitude_symmetric} over $i,j,k,l$, only one value of $i,j,k,l$ will give a non-zero answer.  This is the term where $1 \le \sigma(i), \sigma(j), \sigma(k), \sigma(l) \le 4$.    Retaining only this term,   $\mc{A}_2^{TO}(\sigma(i)^+,\sigma(j)^+,\sigma(k)^+,\sigma(l)^+)$ becomes $\mc{A}_2^{TO}(1_{\sigma}, 2_{\sigma}, 3_{\sigma}, 4_{\sigma)} )$ where $1_{\sigma}$, $2_{\sigma}$ \ldots indicates the first (second \ldots) element of the set $\{1,2,3,4\}$ that appears in the ordered set $\sigma(1),\sigma(2),\dots,\sigma(n)$.  This reduces \eqref{eqn:completeamplitude_symmetric} to \eqref{eqn:completeamplitude1}.   
\end{proof}
Finally, we can study the single trace expression.  This is 
\begin{equation} 
	\mc{A}_2^{TO} (1^+, \dots, n^+ ) = \sum_{\sigma \in S_n}  \sum_{1 \le i < j < k < l \le n}  \mc{A}_2^{TO}(i^+,j^+,k^+,l^+) \frac{ \ip{ij} \ip{j k} \ip{kl} \ip{li} } { \ip{12} \dots \ip{n 1} } \label{eqn:completeamplitude2} 
\end{equation}

This is a rather complicated expression, but the terms which are leading order in $N$ simplify.  We can write the two-loop single-trace amplitude as a series in $N$:
\begin{equation} 
	\mc{A}_{2}^{TO} = \sum_{k = -1}^{1} N^k \mc{A}^{TO}_{2,(k)} . 
\end{equation}
(The order $N^2$ term vanishes).

Retaining only the order $N$ term, we find
\begin{equation} 
	\begin{split} 
		\mc{A}^{TO}_{2,(1)} (1^+,2^+,3^+,4^+) =  6N	 \frac{1}{ (4 \pi)^4 }\left(	\frac{[12][34]} { \ip{12} \ip{34} } + 	\frac{[41][23]} { \ip{41} \ip{23} }   \right)  
	\end{split}
\end{equation}
Therefore the $n$-point amplitude is 
\begin{equation}
	\begin{split} 	
		\mc{A}^{TO}_{2,(\ge 1)} (1^+, \dots, n^+ ) &= \frac{6 N }{ (4 \pi)^4}   \sum_{1 \le i < j < k < l \le n} \left( \frac{[ij][kl]} { \ip{ij} \ip{kl} } 	+  \frac{[jk][li]} { \ip{jk} \ip{li} } 	  	 \right)    \frac{ \ip{ij} \ip{jk} \ip{kl} \ip{li} } { \ip{12} \dots \ip{n1} }\\
		&=   \frac{6 N}{ (4 \pi)^4}   \sum_{1 \le i < j < k < l \le n} 	   	     \frac{ [ij] \ip{jk} [kl] \ip{li}  + \ip{ij} [jk] \ip{kl} [li]   } { \ip{12} \dots \ip{n1} }
\end{split}
\end{equation}
This is a sum of two terms, one of which is the one-loop all $+$ amplitude and the other is the same amplitude with the order of insertions reversed.  This sum is called the parity-even part of the one-loop all $+$ amplitude, because the numerator is even under a parity transformation of all the external amplitudes. 


\section{Two loop form-factors involving the axion}
Here we will compute the formula for two-loop all $+$ amplitudes for $SU(N)$ gauge theory with $N_f = N_c$, but with an axion to cancel the anomaly.  

The calculation closely parallels the previous result.  Firstly, the form factor 
\begin{equation} 
	\ip{\half \op{tr}(B^2) \mid J[2] J[2] } 
\end{equation}
can be set to zero, by adding on a counter-term $\hbar C \op{tr}(F^2)$ to $\half \op{tr}(B^2)$. (The other two point form factors vanish for $SU(2)$ symmetry reasons).

Next, we can check that the three point form factors vanishes. The argument closely parallels that in section \ref{sec:threepoint2loop}.  The only three-point form factor which is not obviously zero is 
\begin{equation} 
	\begin{split} 
		\ip{\half \op{tr}(B^2) + \hbar C \op{tr}(F^2) \mid J[2](z_1) J[1](z_2) J[1](z_3) } .
	\end{split}
\end{equation}
For symmetry reasons, as before, this must be proportional to $[12][13]/\ip{23}$. Any pole at $z_2 = z_3$  would come from an OPE of $J[1]$ with $J[1]$.  We will show that there is no such pole. 

The tree-level OPE of $J[1]$ with $J[1]$ produces $J[2]$, and the correlator with two insertions of $J[2]$ vanishes by construction. The one-loop OPE produces $:J \til{J}:$, $E[2]$ or $F[0]$.  Let us first verify that the one-loop correlator $\ip{\half \op{tr}(B^2) \mid J[2](z_1 ) : J \til{J} :(z_2) }$ vanishes.    Any one-loop three-point correlation function of $\op{tr}(B^2)$ is single trace, and the single-trace one-loop form factor with $N_f = N_c$ vanishes. 

Next, let us consider the correlator between $J[2]$ and $E[2]$ or $F[0]$. It  is not possible to have a non-zero correlator involving insertions $J[2]$ and $E[2]$ or $J[2]$ and $F[0]$, because $E,F$ are neutral under constant gauge transformations whereas $J[2]$ lives in the adjoint representation.

Finally, one might be concerned with the possibility of a two-loop correction to the OPE of $J[1]$ with $J[1]$. It is convenient to phrase the absence of two-loop corrections as a lemma:
\begin{lemma} 
	The correlator of $J[2]$ with the two-loop correction to the OPE of $J[1]$ with $J[1]$ vanishes.\label{lemma:2loopope} 
\end{lemma}
\begin{proof} 
The two-loop OPE would produce an expression with a total of two axion fields or $\til{J}$ or $\til{M}$ fields. These expressions must be of dimension $-2$, and of spin $\ge 0$. and (to have a non-zero correlator with $J[2]$) must be in the adjoint representation of the gauge group.  The only quantities  of this form are expressions like $:J \dots J  E[1] E[1]:$. However, a simple analysis of the Feynman diagrams that describe the quantum chiral algebra \cite{Costello:2022wso} shows that no OPE of $J[k]$ with $J[l]$ can produce a bilinear in the axion fields. 
\end{proof}
This tells us that there are no poles at $z_2 = z_3$, so we conclude that the three point function \eqref{eqn:twoloop3pt} vanishes. 

\subsection{Four point single trace correlators}
We will follow the argument in the case where the anomaly was cancelled only with matter.  As before, the only non-zero four-point functions are
\begin{equation} 
	\ip{\half \op{tr}(B^2) \mid J_{a_1}[1](z_1) J_{a_2}[1](z_2) J_{a_3}[1](z_3) J_{a_4}[1](z_4)} . 
\end{equation}
When we consider single trace correlators, the OPEs that produce an axion do not contribute. This is because, as we saw in lemma \ref{lemma:2loopope}, there are no two-loop corrections to the OPE of $J_{a_1}[1](z_1)$ with $J_{a_2}[1](z_2)$.   One-loop OPEs involving the axion produce terms like $K_{a_1 a_2} F[0]$ and $K_{a_1 a_2} E[2]$.  In both of these terms the colour indices are contracted by the Killing form, and so they can only contribute to the double-trace correlator.

The single-trace correlators only involve the OPEs which do not involve the axion, and so are the same as the ones in the case of only matter.  Therefore the computation is entirely parallel.  

There are a few changes. Firstly, the terms in the four-point function like
\begin{equation} 
	\frac{[12][34] }{\ip{12}^2 \ip{34}^2 } \frac{1}{(4 \pi)^4 \cdot (12)^2 }  ( 2 \ip{13} \ip{24} + 2 \ip{14} \ip{23} ). 	 
\end{equation}
is accompanied by $ (D_{\g} - D_R)^2 $.  In the case of $\mf{sl}_N$ gauge group with $N_f = N_c$, the matter is $N F \oplus N F^\vee$, where $F$, $F^\vee$ are the fundamental and anti-fundamental.      The Dynkin index of the adjoint is $2N$, and that of the fundamental or anti-fundamental is $1$. Therefore $D_{\g} - D_R$ vanishes, and this term drops out.

We have discussed the contribution of the remaining terms already in section \ref{sec:2no}.  The pure gauge sector contributes
		\begin{equation}
			\begin{split} 	
				- (2 N^2 + 8)	\left( \frac{1}{2^2 (4 \pi)^4 }	\frac{[12][34]} { \ip{12} \ip{34} } + 	\frac{1}{2^2 (4 \pi)^4 }	\frac{[41][23]} { \ip{41} \ip{23} } \right)	  + 
				\frac{8}{2^2 (4 \pi)^4 }	\frac{[13][24]} { \ip{13}\ip{24} } 	
\end{split}
		\end{equation}
And, $N_F$ copies of the fundamental and anti-fundamental representation contributes
		\begin{equation} 
			( 4 N_F  N  + 2 N_F N^{-1}  ) \left(	\frac{1}{2^2 (4 \pi)^4 }	\frac{[12][34]} { \ip{12} \ip{34} }   + 	\frac{1}{2^2 (4 \pi)^4 }	\frac{[41][23]} { \ip{41} \ip{23} } \right) 
		\end{equation}
Taking $N_F = N$ and adding these two terms together, we find the complete single-trace four-point amplitude is
\begin{equation} 
	\mc{A}_2^{TO}(1^+,2^+,3^+,4^+) =  
				\left(2 N^2 - 6    \right)	\left( \frac{1}{2^2 (4 \pi)^4 }	\frac{[12][34]} { \ip{12} \ip{34} } + 	\frac{1}{2^2 (4 \pi)^4 }	\frac{[41][23]} { \ip{41} \ip{23} } \right)	  + 
		 		\frac{2}{ (4 \pi)^4 }	\frac{[13][24]} { \ip{13}\ip{24} } 	
\end{equation}
\subsection{$n$-point single trace form factors}
The formula given before for the $n$-point single trace form factor in terms of the $4$-point form factor is valid in this situation:
\begin{equation} 
	 \mc{A}^{TO}_2 (1^+, \dots, n^+ ) =   \sum_{1 \le i < j < k < l \le n}  \mc{A}_2^{TO}(i^+,j^+,k^+,l^+) \frac{ \ip{ij} \ip{jk} \ip{kl} \ip{li} } { \ip{12} \dots \ip{n1} } . 
\end{equation}
The proof has a minor subtlety, however.  When we have cancelled the anomaly by including the axion, it is possible to have a one-loop OPE involving $J[0]$, as the OPE between $J[0]$ and $J[1]$ can produce $E[1]$.  In general this causes problems in the inductive argument (as we have already mentioned).  However, OPEs which produce an axion can not contribute to single-trace form factors, so the problem does not arise in the single-trace sector.

\subsection{Extracting amplitudes without the axion}
As before, we can write the two-loop single-trace amplitude as a series in $N$:
\begin{equation} 
	\mc{A}_{2}^{TO} = \sum_{k = -1}^{2} N^k \mc{A}^{TO}_{2,(k)} . 
\end{equation}
As one can see by considering the structure of Feynman diagrams, processes involving exchanges of the axion field do not contribute to the $\mc{A}_{2,(2)}^{TO}$ term, which is the leading order term.  

Therefore, $\mc{A}_{2,(2)}^{TO}$ computes the leading order in $N$ amplitude of ordinary QCD, with $N_F = N_c$ and no axion. 

The leading-order  $n$-point amplitude is 
\begin{equation}
	\begin{split} 	
		\mc{A}^{TO}_{2,(2)} (1^+, \dots, n^+ ) 
		&=   \frac{N^2}{ 2 (4 \pi)^4}   \sum_{1 \le i < j < k < l \le n} 	   	     \frac{ [ij] \ip{jk} [kl] \ip{li} +  \ip{ij} [jk] \ip{kl} [li]} { \ip{12} \dots \ip{n1} }
\end{split}
\end{equation}
This is again two copies of the familiar one-loop all $+$ amplitude of pure gauge theory, one with the order of particles reversed.

\section{Outlook}
We have used the chiral algebra structure to bootstrap two-loop amplitudes in QCD, in two cases: one where the anomaly is cancelled by pure matter, and one where the anomaly is cancelled by a mixture of matter with $N_F = N_c$, and an axion field.

One can ask, what other amplitudes and form factors can one compute using this method? 

There are some amplitudes it would be rather easy to compute. When the anomaly is cancelled just by matter, the multi-trace two-loop amplitudes can be computed by the same technique. (It is not hard to show that the three-trace amplitude vanishes, and the two-trace amplitude can be computed using the chiral algebra  methods explained here).

To go to higher loop number,  one would need a better understanding of the two-loop and higher quantum corrections to the chiral algebra.  Some, but not all, two-loop and higher corrections to the OPE have been computed explicitly in \cite{Zeng:2023qqp}.  I am optimistic that the remaining OPEs are largely fixed by associativity, but so far this has not been verified.  

If one did have a complete understanding of the chiral algebra, one could compute multi-loop form factors of local operators of self-dual Yang-Mills theory as correlators of the chiral algebra. Form factors of self-dual Yang-Mills theory are equal to certain form factors of full Yang-Mills theory, at particular orders in the loop expansion and with particular helicity configurations. For instance, the all $+$ form factor of the operator $\op{tr}(B^n)$ at $n$ loops is a quantity  that is the same in self-dual Yang-Mills and in full Yang-Mills. 

A full understanding of the chiral algebra would (in principle) allow a computation of these form factors.  This would be a major step forward, as there are almost no form factor computations in QCD beyond low loop numbers.  As we see from the analysis of this paper, however, there is a lot of tedious algebra to be done to generate form factors from chiral algebra OPEs. 

In a different direction, one might ask if the methods discussed here give any insight into the non-rational terms in the two-loop amplitudes and form factors.  Let us discuss this for the two-loop all $+$ amplitudes of QCD with $N_f = N_c$.  The chiral algebra method requires one to cancel the twistor space anomaly by an axion field.  With the axion, all two-loop all $+$  amplitudes are rational.  Axion exchange does not contribute at leading order in $N$, however, it does contribute at subleading order.

This suggests that we could calculate the transcendental terms in the two-loop amplitude with $N_f = N_c$, by studying those Feynman diagrams in which an axion is exchanged. Because the axion is introduced as part of a Green-Schwarz mechanism, an axion exchange is effectively a one-loop process.  This means that the transcendental parts of the amplitude with $N_f = N_c$ can in principle be computed using one-loop and tree-level Feynman diagrams which contain either one or two axion exchanges.   

\subsection*{Acknowledgements}
I would like to thank Roland Bittleston, Lance Dixon,  Lionel Mason and Natalie Paquette for helpful comments and conversations.  I am particularly grateful to Anthony Morales for pointing out a number of computational errors  in the first version of this paper, and for kindly cross-checking the colour algebra computations in the appendix.  Any remaining errors are, of course, my responsibility.

I am grateful to the Krembil Foundation, the NSERC Discovery Grant program and the Perimeter Institute for Theoretical Physics for support. Research at Perimeter Institute is supported in part by the Government of Canada through the Department of Innovation, Science and Economic Development Canada and by the Province of Ontario through the Ministry of Colleges and Universities. 

\appendix 

\section{Associativity of the OPE with matter included} 
The calculation in \cite{Costello:2022upu} goes \emph{mutatis mutandis} up to the point where we find that the identity
\begin{multline}
	C  \left(\op{Tr}( \t_a \t_b \t_c \t_d ) + \op{Tr}(\t_b \t_a \t_c \t_d) \right)  -\half D \tfrac{1}{2\sh^\vee- D_{R} }  \op{Tr}\left(  [\t_a,\t_b][\t_c, \t_{\d}]\right)   - D \tfrac{1}{2\sh^\vee - D_{R}}	\op{Tr}\left([\t_b,\t_c] [\t_a, \t_d] \right)     \\
	=  \what{\lambda}_{\g,R}^2 \left( K_{ac} K_{bd} + K_{ab} K_{cd} + K_{ad} K_{bc} \right).\label{eqn:Lie_algebra} 
\end{multline}
needs to hold. Here $D_{R}$ is the Dynkin index of the representation $R$.   

To verify that this holds, we need to show that the terms on the left hand side of equation \eqref{eqn:Lie_algebra} that are not totally symmetric cancel. In what follows we write $\op{Tr}$ for trace in the adjoint minus trace in the representation $R$. We have 
\begin{equation} 
	\begin{split} 
		3   \left(\op{Tr}( \t_a \t_b \t_c \t_d ) + \op{Tr}(\t_b \t_a \t_c \t_d) \right) & - 2 \left( \op{Tr}(\t_a \t_b \t_c \t_d) +  \op{Tr}(\t_a \t_c \t_d \t_b) +  \op{Tr}(\t_a \t_d \t_b \t_c)  \right)  \\
		& =\op{Tr}( \t_a \t_b \t_c \t_d) + \op{Tr}(\t_a \t_c  \t_d \t_b) - 2 \op{Tr}(\t_a \t_d \t_b \t_c) \\
		&= \op{Tr}([\t_d,\t_a]  \t_b \t_c) + \op{Tr}([\t_a,\t_c] \t_d \t_b ) .	
	\end{split}
\end{equation}
By dihedral symmetry we have
\begin{equation} 
	\op{Tr}([\t_d,\t_a]  \t_b \t_c) =  \tfrac{1}{2} \op{Tr}([\t_d,\t_a]  [\t_b, \t_c]) 
\end{equation}
so that
\begin{equation} 
	\begin{split} 
		 3   \left(\op{Tr}( \t_a \t_b \t_c \t_d ) + \op{Tr}(\t_b \t_a \t_c \t_d) \right) & - 2 \left( \op{Tr}(\t_a \t_b \t_c \t_d) +  \op{Tr}(\t_a \t_c \t_d \t_b) +  \op{Tr}(\t_a \t_d \t_b \t_c)  \right)  \\
		&= \tfrac{1}{2} \op{Tr}([\t_d,\t_a]  [\t_b, \t_c]) +  \half \op{Tr}([\t_a,\t_c] [\t_d ,\t_b]) \\
		&= -\tfrac{1}{2} \op{Tr}([\t_a,\t_d]  [\t_b, \t_c]) +  \half \op{Tr}([\t_b, [\t_a,\t_c]] \t_d ) \\
		&= -\tfrac{1}{2} \op{Tr}([\t_a,\t_d]  [\t_b, \t_c]) +  \half \op{Tr}( [[\t_b,\t_a],\t_c] \t_d )  +   \half \op{Tr}( [\t_a,[\t_b,\t_c]] \t_d )    \\
		&= - \op{Tr}([\t_a,\t_d]  [\t_b, \t_c]) +  \half \op{Tr}( [[\t_b,\t_a],\t_c] \t_d ) \\
		&= - \op{Tr}([\t_a,\t_d]  [\t_b, \t_c]) -  \half \op{Tr}( [\t_b,\t_a][\t_c, \t_d] ) .
	\end{split}
\end{equation}

Therefore we have
\begin{equation} 
	\begin{split} 
		\tfrac{3}{2}   \left( \op{Tr}( \t_a \t_b \t_c \t_d ) + \op{Tr}(\t_b \t_a \t_c \t_d) \right)& - \lambda_{\g,R}^2 ( K_{ac} K_{bd} + K_{ab} K_{cd} + K_{ad} K_{bc} )\\
		&	= - \frac{1}{2 } \op{Tr}([\t_a,\t_d]  [\t_b, \t_c]) -  \frac{1}{4 }  \op{Tr}( [\t_b,\t_a][\t_c, \t_d] ) .
	\end{split}
\end{equation}
Therefore, the Lie algebra identity \eqref{eqn:Lie_algebra} holds if we take
\begin{equation} 
	\begin{split} 
		C & =  \frac{3}{2 (2\pi \i)^2 12  }	\\
		D &= -\frac{2\sh^\vee- D_{R}}{ (2 \pi \i)^3 24  } 
	\end{split}
\end{equation}

\section{Computing algebraic factors}
\label{sec:computingalgebraicfactors}
We defined 
\begin{equation} 
	\begin{split} 
T_{ab}^{cd} &=  f_{ea}^c f_{be}^d  + f_{ae}^d f_{eb}^c \\
	S_{abj}^i &= g_{ka}^i g_{jb}^k  + g_{ja}^k g_{kb}^i
	\end{split}
\end{equation}
where $g_{ka}^i$ is the matrix defining the action of a Lie algebra generator $\t_a$ on the matter representation $R$. Thus, $S$ is the matrix $\t_a \t_b + \t_b \t_a$, acting in the matter representation.  

We note that
\begin{equation} 
	T_{a_1 a_2}^{b_1 b_2} \t_{b_1} \t_{b_2} = - [\t_{a_1}, \t_e] [\t_{a_2}, \t_e]  -  [\t_{a_2}, \t_e] [\t_{a_1}, \t_e]  
\end{equation}

The pole in the two-loop $4$-point form factor at $z_1 = z_2$, $z_3 = z_4$  coming from the one loop corrected OPEs is 
\begin{equation} 
	 	\frac{[12][34]}{\ip{12}\ip{34} }   \frac{1}{ (2\pi \i)^4 2^6  }	 
\end{equation}
multiplied by the rather complicated algebraic expression
\begin{equation} 
	\begin{split} 	
\ip{ \half \op{tr}(B^2) \mid  T_{a_1 a_2}^{b_1 b_2}  :J_{b_1} \til{J}_{b_2} :(z_1) T_{a_3 a_4}^{b_3 b_4}  :J_{b_3} \til{J}_{b_4} :(z_3)   } \\
+ \ip{ \half \op{tr}(B^2) \mid  T_{a_1 a_2}^{b_1 b_2}  :J_{b_1} \til{J}_{b_2} :(z_1) S_{a_3 a_4 i_4}^{i_3}    :M_{i_3}\til{M}^{i_4}   :(z_3)    } \\
+ \ip{ \half \op{tr}(B^2) \mid  S_{a_1 a_2 i_2}^{i_1}    :M_{i_1}\til{M}^{i_2} :(z_1) T_{a_3 a_4}^{b_3 b_4}  :J_{b_3} \til{J}_{b_4} :(z_3)  }\\ 
+ \ip{ \half \op{tr}(B^2) \mid S_{a_1 a_2 i_2}^{i_1}    :M_{i_1}\til{M}^{i_2} :(z_1) S_{a_3 a_4 i_4}^{i_3}    :M_{i_3}\til{M}^{i_4}   :(z_3)   } 	
			\end{split}
\end{equation}
where
\begin{equation} 
	\begin{split} 
		 \ip{\half \op{tr}(B^2) \mid :\til{J}_{(a} J_{b)}:(z) :\til{J}_{(c} J_{d)}: (w)} &=   -2\op{tr}(\t_{(a} \t_{b)} \t_{(c} \t_{d)} ) + \op{tr}(\t_{a} \t_{c} \t_{b} \t_{d}) + \op{tr}(\t_{a} \t_{d} \t_{b} \t_{c} )  	\\
		\ip{\half \op{tr}(B^2) \mid :\til{J}_{(a} J_{b)}: (z)  :\til{M}_{(i} M_{j)}: (w)}  &=\half ( g^j_{ka} g^k_{ib} + g^j_{kb} g^k_{ia}  )  	\\ 
		\ip{\half \op{tr}(B^2) \mid :\til{M}_{(i} M_{j)}:(z) :\til{M}_{(k} M_{l)}: (w)} &=   -\half \left(  g_{i l a} g_{k j a} +    g_{j l a} g_{k i a} 	 \right).
	\end{split}
\end{equation}
We thus need to compute the four terms
\begin{equation} 
	\begin{split} 
		T_{a_1 a_2}^{b_1 b_2} T_{a_3 a_4}^{b_3 b_4}  \left(  -2\op{tr}(\t_{(b_1} \t_{b_2)} \t_{(b_3} \t_{b_4)} ) + \op{tr}(\t_{b_1} \t_{b_3} \t_{b_2} \t_{b_4}) + \op{tr}(\t_{b_1} \t_{b_4} \t_{b_2} \t_{b_3} )  \right) 	\\
	\half	 T_{a_1 a_2}^{b_1 b_2} S_{a_3 a_4 j}^{i}  \left( g^{j}_{kb_1} g^k_{ib_2} + g^{i}_{kb_2} g^k_{jb_1}   \right) 	\\
  	\half	 T_{a_3 a_4}^{b_1 b_2} S_{a_1 a_2 j}^{i}  \left( g^{j}_{kb_1} g^k_{ib_2} + g^{i}_{kb_2} g^k_{jb_1}   \right) 	\\  	
   - \half   S_{a_1 a_2 i_2}^{i_1}  S_{a_3 a_4 i_4}^{i_3}\left(  g_{i_1 i_4 a} g_{i_3 i_2 a} +    g_{i_2 i_4 a} g_{i_3 i_1 a} 	 \right).
	\end{split}
\end{equation}
Our goal is to simplify these algebraic expressions.

We will do this for gauge group $\mf{sl}_N$ with various choices of matter. We will take $N_F$ copies of the fundamental plus the anti-fundamental, $N_{\wedge^2 F}$ copies of $\wedge^2 F \oplus \wedge^2 F^\vee$, and $N_{a}$ copies of the adjoint.

We find that the total contribution is
\begin{equation} 
	\begin{split} 
		  -2N^2 \op{tr}_F (\t_{a_1} \t_{a_2} \t_{a_3} \t_{a_4} )  +  8\op{tr}_F( \t_{a_1} \t_{a_3} \t_{a_2} \t_{a_4} ) 		- 8  \op{tr}_F( \t_{a_1} \t_{a_2} \t_{a_3} \t_{a_4} ) \\\
	+ ( 4 N_F  N  + 2 N_F N^{-1}  )	  \op{tr}_F(\t_{a_1} \dots \t_{a_4} ) \\
		+ N_{\wedge^2 F} 	\left( 2 N^2 -8 N - 8 -  48 N^{-1}     \right)   \op{tr}_F(\t_{a_1} \t_{a_2} \t_{a_3} \t_{a_4} )  
		+  N_{\wedge^2 F} ( -16 - 16 N^{-1}     ) \op{tr}_F( \t_{a_1} \t_{a_3} \t_{a_2} \t_{a_4} ) \\
+ 2N_a  N^2 \op{tr}_F( \t_{a_1} \t_{a_2} \t_{a_3}\t_{a_4} ) + 8N_a  \op{tr}_F(\t_{a_1} \t_{a_2} \t_{a_3} \t_{a_4} ) -  8N_a  \op{tr}_F (\t_{a_1} \t_{a_3} \t_{a_2} \t_{a_4} )
\\
		+ \text { permutations }
	\end{split} 
\end{equation}
where permutations are under the $\Z/2 \times \Z/2$ inside $S_4$ permuting $(12)$ and $(34)$. 

As a consistency check,  we note that, with $N=1$ supersymmetry, where there is only  adjoint matter, everything cancels. This is to be expected as the two loop all $+$ amplitude vanishes in this case.  

In the theory with fundamental matter and $N_F = N$, the planar (order $N^2$)  term is non-zero at two loops, although at one loop the planar term vanishes with only external gluons. This is not a contradiction: at one loop, the planar term does not vanish if two of the external fields are fermions. These one-loop diagrams with two fermions contribute to the two-loop all $+$ amplitude.

\subsection{First term}

The first term is the only term which is independent of the matter content. We should view this term as the contribution of pure glue. It is
\begin{equation} 
	2 T_{a_1 a_2}^{b_1 b_2} T_{a_3 a_4}^{b_3 b_4} \left(   -\op{tr}_F(\t_{b_1} \t_{b_2} \t_{b_3} \t_{b_4} ) + \op{tr}_F(\t_{b_1} \t_{b_3} \t_{b_2} \t_{b_4} )  \right).  
\end{equation}
(Here we have used the fact that $T_{a_1 a_2}^{b_1 b_2}$ is symmetric under permutation of $b_1,b_2$). 

Using the definition of $T_{a_1 a_2}^{b_1 b_2}$, and working in an orthonormal basis for the Lie algebra, this is
\begin{equation}
	\begin{split} 
		- 2 \op{tr}_F \left(  [\t_{a_1}, \t_{e} ] [ \t_{a_2},\t_{e} ] [\t_{a_3}, \t_{f} ][ \t_{a_4}, \t_{f} ]	\right)\\ 
		+ 2   \op{tr}_F \left(  [\t_{a_1}, \t_{e} ] [ \t_{a_3},\t_{f} ] [\t_{a_2}, \t_{e} ][ \t_{a_4}, \t_{f} ]	\right) \\
		+ \text{ permutations } 
	\end{split}	
\end{equation}
where there are $8$ terms in total, which are the permutations of the two terms written down under the $\Z/2 \times \Z/2$ subgroup of $S_4$ generated by $(12)$ and $(34)$. 

We are only interested in the single-trace form factor.  The quadratic Casimir for $\mf{sl}_N$ can be written as
\begin{equation} 
	\sum \t_e \otimes \t_e = \sum \t_{i}^j \otimes \t_{j}^i - \frac{1}{N} \op{Id} \otimes \op{Id}. 
\end{equation}
Since the quadratic Casimir only appears in commutators, we can use the Casimir for $\mf{gl}_N$ instead, dropping the term $N^{-1} \op{Id} \otimes \op{Id}$. 

To extract the single trace terms, we use the following rules. 
First, inside any trace, we can replace an occurrence of $\t_e \t_e$ by $N$.   Next, an expression like 
\begin{equation} 
	\op{tr}_F(  \dots \t_e \t_{a_1} \dots \t_{a_k} \t_e \dots ) 
\end{equation}
where none  the indices $a_1,\dots,a_k$ nested between the two $t_e$'s are contracted,  is contracted, and they are not adjacent, is double trace. 

Finally, an expression like
\begin{equation} 
	\op{tr}_F ( \t_a \t_e \t_b \t_f \t_c \t_e \t_d \t_f ) 
\end{equation}
where two pairs of adjoint indices are contracted, but with a $\t_f$ occuring between the  two $\t_e$'s, is single trace. This has the permuted trace ordering $\op{tr}_F ( \t_a \t_d \t_c \t_b)$. 

Using these rules, the single-trace part of $	- 2 \op{tr}_F \left(  [\t_{a_1}, \t_{e} ] [ \t_{a_2},\t_{e} ] [\t_{a_3}, \t_{f} ][ \t_{a_4}, \t_{f} ]	\right)$ is 
\begin{equation} -2N^2 \op{tr}_F (\t_{a_1} \t_{a_2} \t_{a_3} \t_{a_4} )   
\end{equation}

The expression $2   \op{tr}_F \left(  [\t_{a_1}, \t_{e} ] [ \t_{a_3},\t_{f} ] [\t_{a_2}, \t_{e} ][ \t_{a_4}, \t_{f} ]	\right)$ is entirely single trace.  It has $16$ terms, each of which is a permutation of $\op{tr}_F(\t_{a_1} \dots \t_{a_4})$ by some element in $S_4$, with some sign.   If one writes out all the $16$ terms, one finds that $8$ terms cancel and we are left with
\begin{equation}
\begin{split} 
	2   \op{tr}_F \left(  [\t_{a_1}, \t_{e} ] [ \t_{a_3},\t_{f} ] [\t_{a_2}, \t_{e} ][ \t_{a_4}, \t_{f} ]	\right) + \text{ permutations }= \\
		8\op{tr}_F( \t_{a_1} \t_{a_3} \t_{a_2} \t_{a_4} ) 		- 8  \op{tr}_F( \t_{a_1} \t_{a_2} \t_{a_3} \t_{a_4} ) 		+ \text{ permutations }
	\end{split}	
\end{equation}
The total contribution of the first term is
\begin{equation} 
	  -2N^2 \op{tr}_F (\t_{a_1} \t_{a_2} \t_{a_3} \t_{a_4} )  +  8\op{tr}_F( \t_{a_1} \t_{a_3} \t_{a_2} \t_{a_4} ) 		- 8  \op{tr}_F( \t_{a_1} \t_{a_2} \t_{a_3} \t_{a_4} ) 		+ \text{ permutations }
\end{equation}

\subsection{Second and third terms}
The second term is 
\begin{equation} 
\half	T_{a_1 a_2}^{b_1 b_2} S_{a_3 a_4 i}^j \left(   g^j_{kb_1 } g^k_{ib_2} + g^j_{k b_2 } g^k_{i b_1}   \right). 	\label{eqn:secondterm} 
\end{equation}
Here we are raising and lowering matter indices using the fact that the matter representation has a symmetric invariant pairing. 

Since $S_{a_3 a_4 i}^j$ is the matrix element of $\t_{a_3} \t_{a_4} + \t_{a_4} \t_{a_3}$, we can write this as 
\begin{equation} 
\half	 T_{a_1 a_2}^{b_1 b_2} \op{tr}_R \left( (\t_{b_1} \t_{b_2} + \t_{b_2} \t_{b_1} )  (\t_{a_3} \t_{a_4} + \t_{a_4} \t_{a_3} ) \right) 	 
\end{equation}
where $R$ is the matter representation.
Since
\begin{equation} 
	T_{a_1 a_2}^{b_1 b_2} \t_{b_1} \t_{b_2} = - [\t_{a_1}, \t_e] [\t_{a_2}, \t_e]  -  [\t_{a_2}, \t_e] [\t_{a_1}, \t_e]  
\end{equation}
we find that \eqref{eqn:secondterm} is
\begin{equation} 
	- \op{tr}_R \left( [\t_{a_1},\t_e] [\t_{a_2},\t_e] \t_{a_3} \t_{a_4}  \right) 	 + \text{ permutations } 
\end{equation}
where as before the permutations are under $\Z/2 \times \Z/2$ swapping $(12)$ and $(34)$.  

The sum of the second and third terms is then
\begin{equation}
	\begin{split} 
		- \op{tr}_R \left( [\t_{a_1},\t_e] [\t_{a_2},\t_e] \t_{a_3} \t_{a_4}  \right) 	\\
	-	\op{tr}_R \left( \t_{a_1} \t_{a_2}[ \t_{a_3}, \t_e] [\t_{a_4},\t_e]  \right) 	\\
		+ \text{ permutations }   
	\end{split}	
\end{equation} 
\subsection{The fourth term}  
The fourth term is
\begin{equation} 
	-\half  S^{i_1}_{a_1 a_2 i_2} S^{i_3}_{a_3 a_4 i_4}  \left(  g_{i_1  i_4 a} g_{i_3 i_2  a} +    g_{i_2 i_4  a} g_{i_3 i_1 i a} 	 \right).
\end{equation}
This is
\begin{equation} 
 	-\op{tr}_R \left( \t_e (\t_{a_1} \t_{a_2} + \t_{a_2} \t_{a_1} )  \t_e  (\t_{a_3} \t_{a_4} + \t_{a_4} \t_{a_3} )   \right)  
\end{equation}
which is
\begin{equation} 
	- 	\op{tr}_R \left( \t_e \t_{a_1} \t_{a_2}   \t_e  \t_{a_3} \t_{a_4}    \right)   + \text{ permutations } 
\end{equation}
Putting this together with the second and third terms we get
\begin{equation}
	\begin{split} 	
	 	 	 \op{tr}_R \left( \t_e \t_{a_1} \t_{a_2}   \t_e  \t_{a_3} \t_{a_4}    \right)   -   \op{tr}_R (\t_e \t_{a_1} \t_e \t_{a_2} \t_{a_3} \t_{a_4} ) -   \op{tr}_R ( \t_{a_1} \t_e \t_{a_2} \t_e \t_{a_3} \t_{a_4} )\\
		-  \op{tr}_R (\t_{a_1} \t_{a_2} \t_e \t_{a_3} \t_e \t_{a_4} )  -\op{tr}_R (\t_{a_1} \t_{a_2}  \t_{a_3} \t_e \t_{a_4} \t_e )   + \op{tr}_R (\t_{a_1} \t_e \t_e \t_{a_2} \t_{a_3} \t_{a_4} )+   \op{tr}_R (\t_{a_1} \t_{a_2} \t_{a_3}  \t_e \t_e  \t_{a_4} ) \\
		+ \text{ permutations } \label{eqn:app_matter}
	\end{split}
\end{equation}
This is the complete contribution of the matter terms.

\subsection{Contribution of matter in the fundamental and anti-fundamental}
Let us now take $R$ to be $N_F$ copies of the fundamental plus anti-fundamental, and determine the contribution of $R$. 
\begin{equation} 
	\sum \t_e \otimes \t_e = \sum t^i_j \otimes \t^j_i - \frac{1}{N} \op{Id} \otimes \op{Id}. 
\end{equation}
Each term in equation \eqref{eqn:app_matter} gives a single-trace expression coming from the $-N^{-1} \op{Id} \otimes \op{Id}$.  Together, these give
\begin{equation}
	2 N_F  N^{-1} \op{tr}_F(\t_{a_1} \t_{a_2} \t_{a_3} \t_{a_4} ) + \text{ permutations }. 
\end{equation}
(The factor of $2$ is because we have $N_F$ fundamental and $N_F$ anti-fundamental.    Fundamental and anti-fundamental contribute the same, because the contributions are related by a permutation in $\Z/2 \times \Z/2$).

There are also contributions from the $\mf{gl}_N$ quadratic Casimir $\sum \t_i^j \otimes \t_j^i$.  These only occur when the two copies of $\t_e$ are adjacent. We can replace $\t_e \t_e$ by $N $.  These terms contribute 
\begin{equation} 
	4  N_F  N \op{tr}(\t_{a_1} \t_{a_2} \t_{a_3} \t_{a_4} ) + \text{ permutations }. 
\end{equation}
The total contribution of matter in the fundamental and anti-fundamental is
\begin{equation} 
	( 4 N_F  N  + 2 N_F N^{-1}  )	  \op{tr}(\t_{a_1} \dots \t_{a_4} ) + \text{ permutations }.
\end{equation}

\subsection{Contribution of adjoint matter}
Next we will calculate what happens with adjoint matter. This will give us the two-loop amplitudes for $N=1$ supersymmetric gauge theory. These we know to be zero, so by showing that the matter contribution cancels with the pure gauge theory contribution we will find a non-trivial cross-check.  

We need to compute the single trace part of 
\begin{equation}
	\begin{split}
		 \op{tr}_{F^\vee \otimes F}  \left( \t_e \t_{a_1} \t_{a_2}   \t_e  \t_{a_3} \t_{a_4}    \right)   -   \op{tr}_{F^\vee \otimes F} (\t_e \t_{a_1} \t_e \t_{a_2} \t_{a_3} \t_{a_4} ) -    \op{tr}_{F^\vee \otimes F} ( \t_{a_1} \t_e \t_{a_2} \t_e \t_{a_3} \t_{a_4} )\\
		-  \op{tr}_{F^\vee \otimes F} (\t_{a_1} \t_{a_2} \t_e \t_{a_3} \t_e \t_{a_4} )  -  \op{tr}_{F^\vee \otimes F} (\t_{a_1} \t_{a_2}  \t_{a_3} \t_e \t_{a_4} \t_e )
		+  \op{tr}_{F^\vee \otimes F} (\t_{a_1} \t_e \t_e \t_{a_2} \t_{a_3} \t_{a_4} )+   \op{tr}_{F^\vee \otimes F} (\t_{a_1} \t_{a_2} \t_{a_3}  \t_e \t_e  \t_{a_4} ) \\
		+ \text{ permutations }. 
	\end{split} \label{eqn:adjoint_initial}
\end{equation}
We will use the fact that
\begin{equation} 
	\op{tr}_{F^\vee \otimes F} (\t_{a_1} \t_{a_2} \t_{a_3} \t_{a_4} ) = \sum_{I \subset \{1,2,3,4\} } (-1)^{\abs{I}}  \op{tr}_{F} ( \t_{\br{I}}) \op{tr}_{F} ( \t_{I^c} ) 
\end{equation}
where $\br{I}$ is the set $I$ with opposite ordering.  

Also, the identity in $\mf{gl}_N$ acts by $0$ in the adjoint representation, so that we can use the $\mf{gl}_N$ Casimir instead of the $\mf{sl}_N$ Casimir.  

We find that
\begin{equation} 
	\begin{split} 
		\op{tr}_{F^\vee \otimes F}  \left( \t_j^i \t_{a_1} \t_{a_2}   \t_i^j  \t_{a_3} \t_{a_4}    \right) &= (2 N^2 + 8)  \op{tr}_F(\t_{a_1} \t_{a_2} \t_{a_3} \t_{a_4} ) -8 \op{tr}_F (\t_{a_1} \t_{a_3} \t_{a_2} \t_{a_4} ) \\
		+ \text{ irrelevant terms } \\
		\op{tr}_{F^\vee \otimes F} (\t_j^i \t_{a_1} \t_i^j \t_{a_2} \t_{a_3} \t_{a_4} ) &=  2 N^2 \op{tr}_F(\t_{a_1} \t_{a_2} \t_{a_3} \t_{a_4} ) \\ 
		+ \text{ irrelevant terms } \\
		\op{tr}_{F^\vee \otimes F} (\t_{a_1} \t_e \t_e \t_{a_2} \t_{a_3} \t_{a_4}) &= 4 N^2 \op{tr}_F( \t_{a_1} \t_{a_2} \t_{a_3} \t_{a_4}) \\
+ \text{ irrelevant terms } 
	\end{split}
\end{equation}
where irrelevant terms are those that are either double trace, or average to zero under the $\Z/2\times \Z/2$ action.

These equations tell us that \eqref{eqn:adjoint_initial} is
\begin{equation} 
	\begin{split} 	
		2 N^2 \op{tr}_F( \t_{a_1} \t_{a_2} \t_{a_3}\t_{a_4} ) + 8 \op{tr}_F(\t_{a_1} \t_{a_2} \t_{a_3} \t_{a_4} ) -  8 \op{tr}_F (\t_{a_1} \t_{a_3} \t_{a_2} \t_{a_4} ) \\
		+ \text { permutations }
	\end{split}
\end{equation}

\subsection{Contribution of matter in $\wedge^2 F$} 
Matter in $\wedge^2 F$ and $\wedge^2 F^\vee$ is more challenging to deal with.   Let $P : F \otimes F \to F \otimes F$ be the permutation operator.  Then, for any elements $\t_{a_i} \in \mf{sl}_N$, we have
\begin{equation} 
	\op{tr}_{\wedge^2 F} (\t_{a_1} \dots \t_{a_n} ) = \half \op{tr}_{F \otimes F} ( \t_{a_1} \dots \t_{a_n}  (1 - P )) . 
\end{equation}
This is because $\wedge^2 F$ is the image of the projector $\half (1-P)$.

We can write traces in $F \otimes F$ as follows.  For any subset $I \subset \{1,\dots,n\}$, let $\t_I$ be the ordered product of the elements $\t_{a_i}$ for $i \in I$.  Let $I^c$ be the complement of $I$; similarly we have $\t_{I^c}$.  Then,
\begin{equation} 
	\op{tr}_{F \otimes F} (\t_{a_1} \dots \t_{a_n} ) = \sum_{I \subset \{1,\dots,n\}} \op{tr}_F (\t_{I}) \op{tr}_{F}(\t_{I^c}). 
\end{equation}
This follows from the fact that the Lie algebra element $\t_a$ acts as $\t_a \otimes 1 + 1 \otimes \t_a$ in the representation $F \otimes F$. 

Similarly,
\begin{equation} 
	\op{tr}_{F \otimes F} (\t_{a_1} \dots \t_{a_n} P ) = \sum_{I \subset \{1,\dots,n\}} \op{tr}_F (\t_{I}\t_{I^c}). 
\end{equation}
Therefore
\begin{equation} 
	\op{tr}_{\wedge^2 F}  (\t_{a_1} \dots \t_{a_n}  ) = \half \sum_{I \subset \{1,\dots,n\} } \left( \op{tr}_{F} (\t_I) \op{tr}_{F} (\t_{I^c}) - \op{tr}_{F} (\t_I \t_{I^c})  \right) \label{eqn:wedge2expansion}
\end{equation}

Focusing on the contribution from $\wedge^2 F$, we would like to extract the single-trace part of 
\begin{equation}
	\begin{split} 	
		 \op{tr}_{\wedge^2 F} \left( \t_e \t_{a_1} \t_{a_2}   \t_e  \t_{a_3} \t_{a_4}    \right)   -   \op{tr}_{\wedge^2 F} (\t_e \t_{a_1} \t_e \t_{a_2} \t_{a_3} \t_{a_4} ) -   \op{tr}_{\wedge^2 F} ( \t_{a_1} \t_e \t_{a_2} \t_e \t_{a_3} \t_{a_4} )\\
		-  \op{tr}_{\wedge^2 F} (\t_{a_1} \t_{a_2} \t_e \t_{a_3} \t_e \t_{a_4} )  - \op{tr}_{\wedge^2 F} (\t_{a_1} \t_{a_2}  \t_{a_3} \t_e \t_{a_4} \t_e )
		+ \op{tr}_{\wedge^2 F} (\t_{a_1} \t_e \t_e \t_{a_2} \t_{a_3} \t_{a_4} )+   \op{tr}_{\wedge^2 F} (\t_{a_1} \t_{a_2} \t_{a_3}  \t_e \t_e  \t_{a_4} ) \\
		+ \text{ permutations }. 
	\end{split} \label{eqn:wedge2initialexpression}
\end{equation}
The first thing to note is that the quadratic Casimir $\t_e \t_e$ acts as
\begin{equation} 
	2 N - 2 - 4 N^{-1}  
\end{equation}
in $\wedge^2 F$, so we can rewrite the previous expression as
\begin{equation}
	\begin{split} 	
 \op{tr}_{\wedge^2 F} \left( \t_e \t_{a_1} \t_{a_2}   \t_e  \t_{a_3} \t_{a_4}    \right)   -   \op{tr}_{\wedge^2 F} (\t_e \t_{a_1} \t_e \t_{a_2} \t_{a_3} \t_{a_4} ) -   \op{tr}_{\wedge^2 F} ( \t_{a_1} \t_e \t_{a_2} \t_e \t_{a_3} \t_{a_4} )\\
		-  \op{tr}_{\wedge^2 F} (\t_{a_1} \t_{a_2} \t_e \t_{a_3} \t_e \t_{a_4} )  - \op{tr}_{\wedge^2 F} (\t_{a_1} \t_{a_2}  \t_{a_3} \t_e \t_{a_4} \t_e )
		+ (4 N - 4 - 8 N^{-1} )\op{tr}_{\wedge^2 F} (\t_{a_1}  \t_{a_2} \t_{a_3} \t_{a_4} ) \\
		+ \text{ permutations }. \label{eqn:wedge2trace} 
	\end{split}
\end{equation}

The next step will be to expand the quadratic Casimir into the $\mf{gl}_N$ part $\t_i^j \otimes \t_j^i$ and the identity contribution $-N^{-1} \t^i_i \otimes \t^j_j$. 

The identity matrix $\t^i_i$  in $\mf{gl}_N$ acts, in the representation $\wedge^2 F$, by $2$ (because by the Leibniz rule we must add the actions on each tensor factor).  The identity term in the Casimir is $-N^{-1} \t^i_i \otimes \t^j_j$, with two copies of the identity matrix. This acts by $-4 N^{-1}$ in $\wedge^2 F$.  Therefore, each of the first five terms in equation \eqref{eqn:wedge2trace3} contributes $-4 N^{-1}$.  The total contribution is us $12 N^{-1} \op{tr}_{\wedge^2 F} (\t_{a_1} \dots \t_{a_4})$.  This brings equation \eqref{eqn:wedge2initialexpression} to the form  
\begin{equation}
	\begin{split} 	
 \op{tr}_{\wedge^2 F} \left( \t_e \t_{a_1} \t_{a_2}   \t_e  \t_{a_3} \t_{a_4}    \right)   -   \op{tr}_{\wedge^2 F} (\t_e \t_{a_1} \t_e \t_{a_2} \t_{a_3} \t_{a_4} ) -   \op{tr}_{\wedge^2 F} ( \t_{a_1} \t_e \t_{a_2} \t_e \t_{a_3} \t_{a_4} )\\
		-  \op{tr}_{\wedge^2 F} (\t_{a_1} \t_{a_2} \t_e \t_{a_3} \t_e \t_{a_4} )  - \op{tr}_{\wedge^2 F} (\t_{a_1} \t_{a_2}  \t_{a_3} \t_e \t_{a_4} \t_e )
		+ (4 N - 4 + 4 N^{-1} )\op{tr}_{\wedge^2 F} (\t_{a_1}  \t_{a_2} \t_{a_3} \t_{a_4} ) \\
		+ \text{ permutations }. \label{eqn:wedge2trace1.5} 
	\end{split}
\end{equation}
where we are now using the $\mf{gl}_N$ Casimir.

We need to write this in terms of single traces.  We focus on the last term first. We have
\begin{equation}
	\begin{split} 
		\op{tr}_{\wedge^2 F} (\t_{a_1} \t_{a_2} \t_{a_3} \t_{a_4} ) = (N- 6) \op{tr}_F  (\t_{a_1} \t_{a_2} \t_{a_3} \t_{a_4} ) \\
		- 2 \op{tr}_F (\t_{a_1} \t_{a_3} \t_{a_2} \t_{a_4} ) + \text{ irrelevant terms }  \label{eqn:wedge2trace2}
	\end{split}	
\end{equation}
where irrelevant terms include double traces and expressions that vanish when we average over the $\Z/2 \times \Z/2$.
This means we can rewrite \eqref{eqn:wedge2trace1.5} as
\begin{equation}
	\begin{split} 	
 \op{tr}_{\wedge^2 F} \left( \t_e \t_{a_1} \t_{a_2}   \t_e  \t_{a_3} \t_{a_4}    \right)   -   \op{tr}_{\wedge^2 F} (\t_e \t_{a_1} \t_e \t_{a_2} \t_{a_3} \t_{a_4} ) -   \op{tr}_{\wedge^2 F} ( \t_{a_1} \t_e \t_{a_2} \t_e \t_{a_3} \t_{a_4} )\\
		-  \op{tr}_{\wedge^2 F} (\t_{a_1} \t_{a_2} \t_e \t_{a_3} \t_e \t_{a_4} )  - \op{tr}_{\wedge^2 F} (\t_{a_1} \t_{a_2}  \t_{a_3} \t_e \t_{a_4} \t_e ) \\
		+ (4 N - 4 + 4 N^{-1} )  
		\left(  (N- 6) \op{tr}_F  (\t_{a_1} \t_{a_2} \t_{a_3} \t_{a_4} ) 
		- 2 \op{tr}_F (\t_{a_1} \t_{a_3} \t_{a_2} \t_{a_4} ) \right)  
		+ \text{ permutations }. \label{eqn:wedge2trace3} 
	\end{split}
\end{equation}

Now we need to determine the contribution of the five terms
\begin{equation}
	\begin{split} 	
 \op{tr}_{\wedge^2 F} \left( \t_e \t_{a_1} \t_{a_2}   \t_e  \t_{a_3} \t_{a_4}    \right)   -   \op{tr}_{\wedge^2 F} (\t_e \t_{a_1} \t_e \t_{a_2} \t_{a_3} \t_{a_4} ) -   \op{tr}_{\wedge^2 F} ( \t_{a_1} \t_e \t_{a_2} \t_e \t_{a_3} \t_{a_4} )\\
		-  \op{tr}_{\wedge^2 F} (\t_{a_1} \t_{a_2} \t_e \t_{a_3} \t_e \t_{a_4} )  - \op{tr}_{\wedge^2 F} (\t_{a_1} \t_{a_2}  \t_{a_3} \t_e \t_{a_4} \t_e ) 
\label{eqn:app3terms}
\end{split}
\end{equation}
where, as above, we are using the $\mf{gl}_N$ Casimir. We will do this using the expansion of trace in $\wedge^2 F$ in terms of single and double traces, given by \eqref{eqn:wedge2expansion} which I reproduce here for convenience:
\begin{equation} 
	\op{tr}_{\wedge^2 F}  (\t^i_j \t_{a_1} \t_{a_2} \t^j_i \t_{a_3} \t_{a_4} ) = \half \sum_{I \subset \{ \circ,1,2,\bullet,3,4   \} } \left( \op{tr}_{F} (\t_I) \op{tr}_{F} (\t_{I^c}) - \op{tr}_{F} (\t_I \t_{I^c})  \right) \label{eqn:wedge2expansion2}
\end{equation}
Here the element $\circ$ in the set $\{\circ,1,2,\bullet,3,4\}$ indicates $\t^i_j$ and $\bullet$ indicates $\t^j_i$.

The following terms will be ultimately single trace:
\begin{enumerate} 
	\item Double trace expressions where $t^i_j$ and $t_j^i$ are in different traces.
	\item Single trace expressions where $t^i_j$ and $t_j^i$ are adjacent.
\end{enumerate}
There are terms of order $N^2$, order $N$, and order $1$\footnote{In the first version of this article, I missed the order $N^2$ terms in this equation, and made some further elementary errors in the colour algebra computation.  This led to an error in the formula for the two-loop amplitude which was pointed out by Lance Dixon and Anthony Morales. I am very grateful to them for pointing out this error, and to Anthony Morales for performing an independent verification of the colour algebra in this section.}.   Let us discuss them in turn, first studying the contribution of the expression
\begin{equation} 
	 	\op{tr}_{\wedge^2 F} \left( \t_i^j \t_{a_1} \t_{a_2}   \t_j^i  \t_{a_3} \t_{a_4}    \right) . 
\end{equation}
The order $N^2$ term arises from those terms in \eqref{eqn:wedge2expansion2} which are double traces of the form 
\begin{equation}
	\op{tr}_F (\t^i_j \t_i^j ) \op{tr}_F(\t_{a_1} \dots \t_{a_4}) .
\end{equation}
This can happen in two ways: where the set $I$ in equation \eqref{eqn:wedge2expansion2} consists of $\{\circ \bullet\}$,  or where the complement of $I$ consists of $\{\circ \bullet\}$. But taking into account the factor of $\half$, we find the order $N^2$ term is $N^2 \op{tr}(\t_{a_1} \dots \t_{a_4} )$.  

The order $N$ term arises from those terms in \eqref{eqn:wedge2expansion2} of the form $-\half \op{tr}_F (\t_I \t_{I^c})$  where $\t^i_j \t^j_i$ are adjacent.  There several ways this can happen. First, $I$ can contain both $\circ$ and $\bullet$. For $\t^i_j$ and $\t^j_i$ to be adjacent in the trace, $I$ can not contain $1$ or $2$. This gives $4$ terms,  all of which are related by a permutation in $\Z/2 \times \Z/2$ to $-N\op{tr}_F (\t_{a_1} \dots \t_{a_4})$.   There are similar $4$ possibilities from the case when $I^c$ contains both $\circ$ and $\bullet$.  Together, taking account the factor of $\half$, this contributes $-4 N \op{tr}_F(\t_{a_1} \dots \t_{a_4})$.

Or, $I$ contains exactly one of $\circ$ or $\bullet$. The two contributions are symmetric, so we can assume that $I$ contains $\circ$; this absorbs the factor of $\half$.  Then, for $\t^i_j$ and $\t^j_i$ to be adjacent in the trace, the final element of the set $I^c$ must be $\bullet$. This happens if and only if $I$ also contains the elements $3$ and $4$, together with an arbitrary subset of $\{1,2\}$. This gives $4$ terms, all of which are related by a permutation in $\Z/2 \times \Z/2$ to $-N\op{tr}_F (\t_{a_1} \dots \t_{a_4})$. Therefore the complete order $N$ term is  
\begin{equation}
	-8 N \op{tr}(\t_{a_1} \dots \t_{a_4}) + \text{ irrelevant terms } 
\end{equation}
where irrelevant terms are those which vanish when we average over the $\Z/2 \times \Z/2$ subgroup of $S_4$ permuting $(12)$ and $(34)$.

Finally, there are $32$ order $1$ terms.  These come from the terms $\half \op{tr}(\t_I) \op{tr}(\t_{I^c})$ in the expansion \eqref{eqn:wedge2expansion2} where $\t^i_j$ and $\t^j_i$ are in different traces. Without loss of generality, we can assume that $\circ \in I$ and $\bullet \in I^c$; this absorbs the factor of $\half$.  Writing $I = \{\circ\} \cup J$ for $J \subset \{1,\dots,4\}$, we find the order $1$ terms are
\begin{equation} 
	\sum_{J \subset \{1,\dots,4\}  }  \op{tr}(\t_J \t_{\til{J}^c} )    
\end{equation}
where $\til{J}^c$ means the complement of the subset $J$, but ordered according to the ordering as a subset of $\{3,4,1,2\}$.  

Each of the $16$ order $1$ terms can be permuted, using an element of the $\Z/2 \times \Z/2$ inside $S_4$, to $\op{tr}_F(\t_{a_1} \t_{a_2} \t_{a_3} \t_{a_4})$.   Thus, we can write   the order $1$ terms as $16 \op{tr}_F(\t_{a_1} \t_{a_2} \t_{a_3} \t_{a_4})$ plus irrelevant terms.  

In sum, we have
\begin{equation}
	\begin{split} 	
	\op{tr}_{\wedge^2 F} \left( \t_i^j \t_{a_1} \t_{a_2}   \t_j^i  \t_{a_3} \t_{a_4}    \right)  &= N^2 \op{tr}_F(\t_{a_1} \t_{a_2} \t_{a_3} \t_{a_4} )  -  8N\op{tr}_F(\t_{a_1} \t_{a_2} \t_{a_3} \t_{a_4} ) + 16   \op{tr}_F(\t_{a_1} \t_{a_2} \t_{a_3} \t_{a_4})     \\  &+ \text{ irrelevant terms } .
	\end{split} \label{eqn:alg_relation}
\end{equation}	
		
In a similar way, we can compute
\begin{equation} 
	 	\op{tr}_{\wedge^2 F} \left( \t_i^j \t_{a_1}\t^j_i  \t_{a_2}    \t_{a_3} \t_{a_4}    \right) . 
\end{equation}
We follow the previous analysis. We use the formula
\begin{equation} 
	\op{tr}_{\wedge^2 F}  (\t^i_j \t_{a_1}  \t^j_i \t_{a_2} \t_{a_3} \t_{a_4} ) = \half \sum_{I \subset \{ \circ,1,\bullet,2,3,4   \} } \left( \op{tr}_{F} (\t_I) \op{tr}_{F} (\t_{I^c}) - \op{tr}_{F} (\t_I \t_{I^c})  \right) \label{eqn:wedge2expansion3}
\end{equation}
where $\circ$ corresponds to $\t^i_j$ and $\bullet$ corresponds to $\t^j_i$.

The order $N^2$ term is $N^2 \op{tr}_F(\t_{a_1} \t_{a_2} \t_{a_3} \t_{a_4})$, exactly as before.   

To compute the order $N$ term, we argue as before. The order $N$ term comes from the $-\half \op{tr}(\t_I \t_{I^c})$ terms where $\t^i_j$ and $\t^j_i$ are adjacent in the trace. 

Suppose the the set $I$ contains exactly one of $\circ$ or $\bullet$.  If $I$ contains $\circ$, then the final element of $I^c$ must be $\bullet$, in order for $\t^i_j$ and $\t^j_i$ to be adjacent in the trace.  There are two ways this can happen, if  $I^c = \{\bullet\}$ or $I^c = \{1,\bullet\}$. Similarly, if $I$ contains $\bullet$, then this must be the final element of the set $I$, so that either $I = \{1,\bullet\}$ or $I = \{\bullet\}$.   These $4$ terms (taking account the factor of $\half$) contribute
\begin{equation} 
	-2 N \op{tr}(\t_{a_1} \t_{a_2} \t_{a_3} \t_{a_4} ).  
\end{equation}

If $I$ contains both $\circ$ and $\bullet$, then it can not contain $1$, leaving $8$ possibilities, and a similar $8$ possibilities from the case where $I$ contains neither $\circ$ or $\bullet$.  Taking into account the factor of $\half$, these contribute
\begin{equation} 
		-6 N \op{tr}(\t_{a_1} \t_{a_2} \t_{a_3} \t_{a_4} )  	-2 N \op{tr}(\t_{a_1} \t_{a_2} \t_{a_3} \t_{a_4} ).   
\end{equation}
so that the total order $N$ term is
\begin{equation} 
		-8 N \op{tr}(\t_{a_1} \t_{a_2} \t_{a_3} \t_{a_4} )  	-2 N \op{tr}(\t_{a_1} \t_{a_2} \t_{a_3} \t_{a_4} ).   
\end{equation}

There are $32$ order $1$ terms as before, coming from $\half \op{tr}(\t_I) \op{tr}(\t_{I^c})$ where $\t^i_j$ and $\t^j_i$ are in different traces. Taking account the factor of $\half$, and considering traces related by the $\Z/2 \times \Z/2$ permutations to be equivalent, we find the order $1$ term is 
\begin{equation} 
	12 \op{tr}_F (\t_{a_1} \t_{a_2} \t_{a_3} \t_{a_4}) + 4 \op{tr}_F (\t_{a_1} \t_{a_3} \t_{a_2} \t_{a_4} ). 
\end{equation}

Putting this together gives
\begin{equation} 
	\begin{split} 	
		\op{tr}_{\wedge^2 F} \left( \t_i^j \t_{a_1}\t^j_i  \t_{a_2}    \t_{a_3} \t_{a_4}    \right)  &= N^2 \op{tr}_F (\t_{a_1} \t_{a_2} \t_{a_3} \t_{a_4} )  -8N \op{tr}(\t_{a_1} \t_{a_2} \t_{a_3} \t_{a_4} )  - 2 N \op{tr}( \t_{a_1} \t_{a_3} \t_{a_2} \t_{a_4} )\\
		&	+ 12 \op{tr}_F (\t_{a_1} \t_{a_2} \t_{a_3} \t_{a_4}) + 4 \op{tr}_F (\t_{a_1} \t_{a_3} \t_{a_2} \t_{a_4} ) \\ 
		&+ \text{ irrelevant terms } . \label{eqn:wedge2singletrace} 
	\end{split}
\end{equation} 
Irrelevant terms are either double trace, or expressions that sum to zero when we sum over permutations in $\Z/2 \times \Z/2$.

Therefore the single trace contribution of the five terms in equation \eqref{eqn:app3terms}  is
\begin{equation}
	\begin{split} 	 
	  (-3 N^2 + 24 N - 32 )\op{tr}_F(\t_{a_1} \t_{a_2} \t_{a_3} \t_{a_4} )  
 +  ( 8 N - 16 ) \op{tr}( \t_{a_1} \t_{a_3} \t_{a_2} \t_{a_4} ) \\
		+ \text{ irrelevant terms }
	\end{split} 
\end{equation}
Finally, our initial expression \eqref{eqn:wedge2initialexpression} becomes
\begin{equation}
	\begin{split} 		
	  (-3 N^2 + 24 N - 32)\op{tr}_F(\t_{a_1} \t_{a_2} \t_{a_3} \t_{a_4} )  
 +  ( 8 N - 16 ) \op{tr}( \t_{a_1} \t_{a_3} \t_{a_2} \t_{a_4} ) \\
+ (4 N - 4 + 4 N^{-1} )  
		\left(  (N- 6) \op{tr}_F  (\t_{a_1} \t_{a_2} \t_{a_3} \t_{a_4} ) 
		- 2 \op{tr}_F (\t_{a_1} \t_{a_3} \t_{a_2} \t_{a_4} ) \right)  
\\
		+ \text{ irrelevant terms }. \label{eqn:wedge2trace_almost_final} 
	\end{split}
\end{equation}
This can be simplified to
\begin{equation}
	\begin{split} 	
		\left(  N^2 -4 N - 4 -  24 N^{-1}     \right)   \op{tr}_F(\t_{a_1} \t_{a_2} \t_{a_3} \t_{a_4} )  
		+  ( -8 - 8 N^{-1}     ) \op{tr}( \t_{a_1} \t_{a_3} \t_{a_2} \t_{a_4} ) 	\\	
+ \text{ irrelevant terms }. 
	\end{split}
\end{equation}
If we include the contribution from the $\wedge^2 F^\vee$ representation, we double this expression to find  
\begin{equation} 
\begin{split} 	
\left(  2 N^2 -8 N - 8 -  48 N^{-1}     \right)   \op{tr}_F(\t_{a_1} \t_{a_2} \t_{a_3} \t_{a_4} )  	+  ( -16 - 16 N^{-1}     ) \op{tr}( \t_{a_1} \t_{a_3} \t_{a_2} \t_{a_4} ) 	\\	
	+ \text{ irrelevant terms }. \label{eqn:wedge2trace_final} 
	\end{split}
\end{equation}
Restoring the sum over permutations in $\Z/2 \times \Z/2$, the final answer is
\begin{equation} 
\begin{split} 	
\left(  2 N^2 -8 N - 8 -  48 N^{-1}     \right)   \op{tr}_F(\t_{a_1} \t_{a_2} \t_{a_3} \t_{a_4} )  	+  ( -16 - 16 N^{-1}     ) \op{tr}( \t_{a_1} \t_{a_3} \t_{a_2} \t_{a_4} ) 	\\	
	+ \text{ permutations}. \label{eqn:wedge2trace_very_final} 
	\end{split}
\end{equation}

\providecommand{\href}[2]{#2}\begingroup\raggedright\endgroup


\begin{thebibliography}{10}

\bibitem{Costello:2022upu}
K.~Costello and N.~M. Paquette, \emph{{Associativity of One-Loop Corrections to
  the Celestial Operator Product Expansion}},
  \href{http://dx.doi.org/10.1103/PhysRevLett.129.231604}{\emph{Phys. Rev.
  Lett.} {\bf 129} (2022) 231604}, [\href{http://arxiv.org/abs/2204.05301}{{\tt
  2204.05301}}].

\bibitem{Arkani-Hamed:2022rwr}
N.~Arkani-Hamed, L.~J. Dixon, A.~J. McLeod, M.~Spradlin, J.~Trnka and
  A.~Volovich, \emph{{Solving Scattering in $N$ = 4 Super-Yang-Mills Theory}},
  \href{http://arxiv.org/abs/2207.10636}{{\tt 2207.10636}}.

\bibitem{Bern:2000dn}
Z.~Bern, L.~J. Dixon and D.~A. Kosower, \emph{{A Two loop four gluon helicity
  amplitude in QCD}},
  \href{http://dx.doi.org/10.1088/1126-6708/2000/01/027}{\emph{JHEP} {\bf 01}
  (2000) 027}, [\href{http://arxiv.org/abs/hep-ph/0001001}{{\tt
  hep-ph/0001001}}].

\bibitem{Badger:2013gxa}
S.~Badger, H.~Frellesvig and Y.~Zhang, \emph{{A Two-Loop Five-Gluon Helicity
  Amplitude in QCD}},
  \href{http://dx.doi.org/10.1007/JHEP12(2013)045}{\emph{JHEP} {\bf 12} (2013)
  045}, [\href{http://arxiv.org/abs/1310.1051}{{\tt 1310.1051}}].

\bibitem{Badger:2015lda}
S.~Badger, G.~Mogull, A.~Ochirov and D.~O'Connell, \emph{{A Complete Two-Loop,
  Five-Gluon Helicity Amplitude in Yang-Mills Theory}},
  \href{http://dx.doi.org/10.1007/JHEP10(2015)064}{\emph{JHEP} {\bf 10} (2015)
  064}, [\href{http://arxiv.org/abs/1507.08797}{{\tt 1507.08797}}].

\bibitem{Dunbar:2016aux}
D.~C. Dunbar and W.~B. Perkins, \emph{{Two-loop five-point all plus helicity
  Yang-Mills amplitude}},
  \href{http://dx.doi.org/10.1103/PhysRevD.93.085029}{\emph{Phys. Rev. D} {\bf
  93} (2016) 085029}, [\href{http://arxiv.org/abs/1603.07514}{{\tt
  1603.07514}}].

\bibitem{Dunbar:2016gjb}
D.~C. Dunbar, G.~R. Jehu and W.~B. Perkins, \emph{{Two-loop six gluon all plus
  helicity amplitude}},
  \href{http://dx.doi.org/10.1103/PhysRevLett.117.061602}{\emph{Phys. Rev.
  Lett.} {\bf 117} (2016) 061602}, [\href{http://arxiv.org/abs/1605.06351}{{\tt
  1605.06351}}].

\bibitem{Dunbar:2017nfy}
D.~C. Dunbar, J.~H. Godwin, G.~R. Jehu and W.~B. Perkins, \emph{{Analytic
  all-plus-helicity gluon amplitudes in QCD}},
  \href{http://dx.doi.org/10.1103/PhysRevD.96.116013}{\emph{Phys. Rev. D} {\bf
  96} (2017) 116013}, [\href{http://arxiv.org/abs/1710.10071}{{\tt
  1710.10071}}].

\bibitem{Dalgleish:2020mof}
A.~R. Dalgleish, D.~C. Dunbar, W.~B. Perkins and J.~M.~W. Strong, \emph{{Full
  color two-loop six-gluon all-plus helicity amplitude}},
  \href{http://dx.doi.org/10.1103/PhysRevD.101.076024}{\emph{Phys. Rev. D} {\bf
  101} (2020) 076024}, [\href{http://arxiv.org/abs/2003.00897}{{\tt
  2003.00897}}].

\bibitem{Dunbar:2016cxp}
D.~C. Dunbar, G.~R. Jehu and W.~B. Perkins, \emph{{The two-loop n-point
  all-plus helicity amplitude}},
  \href{http://dx.doi.org/10.1103/PhysRevD.93.125006}{\emph{Phys. Rev. D} {\bf
  93} (2016) 125006}, [\href{http://arxiv.org/abs/1604.06631}{{\tt
  1604.06631}}].

\bibitem{Kosower:2022bfv}
D.~A. Kosower and S.~P\"ogel, \emph{{A Unitarity Approach to Two-Loop All-Plus
  Rational Terms}},  \href{http://arxiv.org/abs/2206.14445}{{\tt 2206.14445}}.

\bibitem{Kosower:2022iju}
D.~A. Kosower and S.~P\"ogel, \emph{{Yang\textendash{}Mills All-Plus: Two Loops
  for the Price of One}},
  \href{http://dx.doi.org/10.22323/1.416.0031}{\emph{PoS} {\bf LL2022} (2022)
  031}, [\href{http://arxiv.org/abs/2208.06209}{{\tt 2208.06209}}].


\bibitem{Dixon:2024mzh}
L.~J.~Dixon and A.~Morales,
``On gauge amplitudes first appearing at two loops,''
JHEP \textbf{08}, 129 (2024)
doi:10.1007/JHEP08(2024)129
[\href{http://arxiv.org/abs/2407.13967} {{\tt 2407.13967}} ].


\bibitem{Alday:2010ku}
L.~F. Alday, D.~Gaiotto, J.~Maldacena, A.~Sever and P.~Vieira, \emph{{An
  Operator Product Expansion for Polygonal null Wilson Loops}},
  \href{http://dx.doi.org/10.1007/JHEP04(2011)088}{\emph{JHEP} {\bf 04} (2011)
  088}, [\href{http://arxiv.org/abs/1006.2788}{{\tt 1006.2788}}].

\bibitem{Basso:2013vsa}
B.~Basso, A.~Sever and P.~Vieira, \emph{{Spacetime and Flux Tube S-Matrices at
  Finite Coupling for N=4 Supersymmetric Yang-Mills Theory}},
  \href{http://dx.doi.org/10.1103/PhysRevLett.111.091602}{\emph{Phys. Rev.
  Lett.} {\bf 111} (2013) 091602}, [\href{http://arxiv.org/abs/1303.1396}{{\tt
  1303.1396}}].

\bibitem{Caron-Huot:2020bkp}
S.~Caron-Huot, L.~J. Dixon, J.~M. Drummond, F.~Dulat, J.~Foster,
  O.~G\"urdo\u{g}an et~al., \emph{{The Steinmann Cluster Bootstrap for $N$ =
  Super Yang-Mills Amplitudes}},
  \href{http://dx.doi.org/10.22323/1.376.0003}{\emph{PoS} {\bf CORFU2019}
  (2020) 003}, [\href{http://arxiv.org/abs/2005.06735}{{\tt 2005.06735}}].

\bibitem{Costello:2022wso}
K.~Costello and N.~M. Paquette, \emph{{Celestial holography meets twisted
  holography: 4d amplitudes from chiral correlators}},
  \href{http://arxiv.org/abs/2201.02595}{{\tt 2201.02595}}.

\bibitem{Guevara:2021abz}
A.~Guevara, E.~Himwich, M.~Pate and A.~Strominger, \emph{{Holographic Symmetry
  Algebras for Gauge Theory and Gravity}},
  \href{http://arxiv.org/abs/2103.03961}{{\tt 2103.03961}}.

\bibitem{Strominger:2017zoo}
A.~Strominger, \emph{{Lectures on the Infrared Structure of Gravity and Gauge
  Theory}},  \href{http://arxiv.org/abs/1703.05448}{{\tt 1703.05448}}.

\bibitem{Ball:2021tmb}
A.~Ball, S.~Narayanan, J.~Salzer and A.~Strominger, \emph{{Perturbatively Exact
  $w_{1+\infty}$ Asymptotic Symmetry of Quantum Self-Dual Gravity}},
  \href{http://arxiv.org/abs/2111.10392}{{\tt 2111.10392}}.

\bibitem{Costello:2021bah}
K.~J. Costello, \emph{{Quantizing local holomorphic field theories on twistor
  space}},  \href{http://arxiv.org/abs/2111.08879}{{\tt 2111.08879}}.

\bibitem{Cachazo:2004kj}
F.~Cachazo, P.~Svrcek and E.~Witten, \emph{{MHV vertices and tree amplitudes in
  gauge theory}},
  \href{http://dx.doi.org/10.1088/1126-6708/2004/09/006}{\emph{JHEP} {\bf 09}
  (2004) 006}, [\href{http://arxiv.org/abs/hep-th/0403047}{{\tt
  hep-th/0403047}}].

\bibitem{Bern:1991aq}
Z.~Bern and D.~A. Kosower, \emph{{The Computation of loop amplitudes in gauge
  theories}}, \href{http://dx.doi.org/10.1016/0550-3213(92)90134-W}{\emph{Nucl.
  Phys. B} {\bf 379} (1992) 451--561}.

\bibitem{Mahlon:1993si}
G.~Mahlon, \emph{{Multi - gluon helicity amplitudes involving a quark loop}},
  \href{http://dx.doi.org/10.1103/PhysRevD.49.4438}{\emph{Phys. Rev. D} {\bf
  49} (1994) 4438--4453}, [\href{http://arxiv.org/abs/hep-ph/9312276}{{\tt
  hep-ph/9312276}}].

\bibitem{Bern:1993qk}
Z.~Bern, G.~Chalmers, L.~J. Dixon and D.~A. Kosower, \emph{{One loop N gluon
  amplitudes with maximal helicity violation via collinear limits}},
  \href{http://dx.doi.org/10.1103/PhysRevLett.72.2134}{\emph{Phys. Rev. Lett.}
  {\bf 72} (1994) 2134--2137}, [\href{http://arxiv.org/abs/hep-ph/9312333}{{\tt
  hep-ph/9312333}}].

\bibitem{Parke:1986gb}
S.~J. Parke and T.~R. Taylor, \emph{{An Amplitude for $n$ Gluon Scattering}},
  \href{http://dx.doi.org/10.1103/PhysRevLett.56.2459}{\emph{Phys. Rev. Lett.}
  {\bf 56} (1986) 2459}.

\bibitem{Bern:1993mq}
Z.~Bern, L.~J. Dixon and D.~A. Kosower, \emph{{One loop corrections to five
  gluon amplitudes}},
  \href{http://dx.doi.org/10.1103/PhysRevLett.70.2677}{\emph{Phys. Rev. Lett.}
  {\bf 70} (1993) 2677--2680}, [\href{http://arxiv.org/abs/hep-ph/9302280}{{\tt
  hep-ph/9302280}}].

\bibitem{Bern:2005ji}
Z.~Bern, L.~J. Dixon and D.~A. Kosower, \emph{{The last of the finite loop
  amplitudes in QCD}},
  \href{http://dx.doi.org/10.1103/PhysRevD.72.125003}{\emph{Phys. Rev. D} {\bf
  72} (2005) 125003}, [\href{http://arxiv.org/abs/hep-ph/0505055}{{\tt
  hep-ph/0505055}}].

\bibitem{Dixon:2004za}
L.~J. Dixon, E.~W.~N. Glover and V.~V. Khoze, \emph{{MHV rules for Higgs plus
  multi-gluon amplitudes}},
  \href{http://dx.doi.org/10.1088/1126-6708/2004/12/015}{\emph{JHEP} {\bf 12}
  (2004) 015}, [\href{http://arxiv.org/abs/hep-th/0411092}{{\tt
  hep-th/0411092}}].

\bibitem{Boels:2007qn}
R.~Boels, L.~J. Mason and D.~Skinner, \emph{{From twistor actions to MHV
  diagrams}},
  \href{http://dx.doi.org/10.1016/j.physletb.2007.02.058}{\emph{Phys. Lett. B}
  {\bf 648} (2007) 90--96}, [\href{http://arxiv.org/abs/hep-th/0702035}{{\tt
  hep-th/0702035}}].

\bibitem{Kosower:1999rx}
D.~A. Kosower and P.~Uwer, \emph{{One loop splitting amplitudes in gauge
  theory}}, \href{http://dx.doi.org/10.1016/S0550-3213(99)00583-0}{\emph{Nucl.
  Phys. B} {\bf 563} (1999) 477--505},
  [\href{http://arxiv.org/abs/hep-ph/9903515}{{\tt hep-ph/9903515}}].

\bibitem{Okubo:1978qe}
S.~Okubo, \emph{{Quartic Trace Identity for Exceptional Lie Algebras}},
  \href{http://dx.doi.org/10.1063/1.524127}{\emph{J. Math. Phys.} {\bf 20}
  (1979) 586}.

\bibitem{Fernandez:2024qnu}
V.~E.~Fern\'andez and N.~M.~Paquette,
``Associativity is enough: an all-orders 2d chiral algebra for 4d form factors,''
 \href{http://arxiv.org/abs/2412.17168}{{\tt 2412.17168}}.



\bibitem{Zeng:2023qqp}
K.~Zeng,
``Twisted Holography and Celestial Holography from Boundary Chiral Algebra,''
Commun. Math. Phys. \textbf{405}, no.1, 19 (2024)
doi:10.1007/s00220-023-04917-0
 \href{http://arxiv.org/abs/2302.06693}{{\tt 2302.06693}}.


\bibitem{Britto:2005fq}
R.~Britto, F.~Cachazo, B.~Feng and E.~Witten, \emph{{Direct proof of tree-level
  recursion relation in Yang-Mills theory}},
  \href{http://dx.doi.org/10.1103/PhysRevLett.94.181602}{\emph{Phys. Rev.
  Lett.} {\bf 94} (2005) 181602},
  [\href{http://arxiv.org/abs/hep-th/0501052}{{\tt hep-th/0501052}}].

\bibitem{bern2005shell}
Z.~Bern, L.~J. Dixon and D.~A. Kosower, \emph{On-shell recurrence relations for
  one-loop qcd amplitudes}, {\emph{Physical Review D} {\bf 71} (2005) 105013}.

\end{thebibliography}
\end{document}